%% file: integral.tex
\documentclass[]{aa}  
\usepackage{graphicx,subfigure}
\usepackage{txfonts}
\usepackage{amssymb,natbib}
\usepackage{longtable,supertabular}

\input{symbols}

\begin{document}
   \title{Multi-wavelength observations of Galactic hard X-ray sources 
discovered by \textit{INTEGRAL}. \thanks{Based on observations collected at the European Organisation for
  Astronomical Research in the Southern Hemisphere, Chile (ESO
  Programme 073.D-0339) (PI S. Chaty).}}

   \subtitle{I. 
    The nature of the companion star}

   \author{S. Chaty\inst{1} \and F. Rahoui\inst{2,1} \and C. Foellmi\inst{3} 
     \and J.\,A. Tomsick\inst{4} \and J. Rodriguez\inst{1} \and R. Walter\inst{5}
          }

   \offprints{S. Chaty}

   \institute{Laboratoire AIM, CEA/DSM - CNRS - Universit\'e Paris Diderot,
Irfu/Service d'Astrophysique, B\^at. 709, CEA-Saclay,
FR-91191 Gif-sur-Yvette Cedex, France, \email{chaty@cea.fr}
	\and ESO, Alonso de Cordova 3107, Vitacura, Casilla 19001,
	     Santiago 19, Chile
	\and Laboratoire d'Astrophysique, Observatoire de Grenoble, 
	     BP 53, F-38041 Grenoble Cedex 9, France
	\and Space Sciences Laboratory, 7 Gauss Way, University
of California, Berkeley, CA 94720-7450, USA
	\and INTEGRAL Science Data Centre, Chemin d'\'Ecogia 16, 
	     CH-1290 Versoix, Switzerland
             }

   \date{Received \today; accepted \today}

 
  \abstract 
  {(abridged) The \textit{INTEGRAL} hard X-ray observatory has revealed an
    emerging population of highly obscured X-ray binary systems
    through multi-wavelength observations.  Previous studies have
    shown that many of these sources are high-mass X-ray binaries
    hosting neutron stars orbiting around luminous and evolved
    companion stars.  }
  {To better understand this newly-discovered population, we have
    selected a sample of sources for which an accurate localisation is
    available to identify the stellar counterpart and reveal the
    nature of the companion star and of the binary system.  }
  {We performed an intensive study of a sample of thirteen
    \textit{INTEGRAL} sources, through multi-wavelength optical to NIR
    photometric and spectroscopic observations, using EMMI and SofI
    instruments at the ESO NTT telescope. We performed accurate
    astrometry and identified candidate counterparts for which we give
    the optical and NIR magnitudes.  We detected many spectral lines
    allowing us to determine the spectral type of the companion star.
    We fitted with stellar black bodies the mid-infrared to optical
    spectral energy distributions of these sources.  From the spectral
    analysis and SED fitting we identified the nature of the companion
    stars and of the binary systems.}
  {Through spectroscopic analysis of the most likely candidates we
    found the spectral types of IGR~J16320-4751, IGR~J16358-4726,
    IGR~J16479-4514, IGR~J17252-3616, IGR~J18027-2016: They all host
    OB type supergiant companion stars, with IGR~J16358-4726 likely
    hosting an sgB[e]. Our spectra also confirm the supergiant O and B
    nature of IGR~J17391-3021 and IGR~J19140+0951. From SED fitting we
    found that IGR~J16418-4532 is a (likely OB supergiant) HMXB,
    IGR~J16393-4643 a (likely BIV-V star) HMXB, and IGR~J18483-0311 a
    likely HMXB system.  Through accurate astrometry, we rejected the
    proposed counterparts of IGR~J17091-3624 and IGR~J17597-2201, and
    we discovered two new candidate counterparts for each source, both
    suggesting an LMXB from SED fitting.  We confirm the AGN nature of
    IGR\,J16558-5203. Finally, we report that NIR fields of four
    sources of our sample exhibit large-scale regions of absorption.}


\authorrunning{S. Chaty et al.}
\titlerunning{Optical/NIR observations revealing the obscured {\it INTEGRAL} 
binary systems}

   \maketitle
%

\section{Introduction}

The hard X-ray {\it INTEGRAL} observatory was launched on the 17th of
October 2002, and since then, it has performed a detailed survey of
the Galactic plane.  The ISGRI detector on the IBIS imager
\citep{lebrun:2003} has discovered many new hard X-ray
sources\footnote{An updated list of these sources is maintained at
  {\rm http://isdc.unige.ch/$\sim$rodrigue/html/igrsources.html}.},
including binary systems, pulsars and AGNs, all these so-called IGR
sources being reported in \cite{bird:2007} and
\cite{bodaghee:2007}. One of the most important achievements of the
{\it INTEGRAL} observatory to date is that it is revealing hard X-ray
sources which were not easily detected in earlier soft X-ray
(typically $\leq 10\ \mathrm{keV}$) observations, bringing to light a
previously hidden part of a population of highly obscured high-energy
binary systems in our Galaxy. These objects share characteristics that
have rarely been seen (see \citeauthor{dean:2005}
\citeyear{dean:2005}).  They are high-mass X-ray binaries (HMXBs)
hosting a neutron star orbiting an OB companion star, in some cases a
supergiant star (see e.g. \citeauthor{filliatre:2004}
\citeyear{filliatre:2004} and \citeauthor{pellizza:2006}
\citeyear{pellizza:2006}), some of them possibly being long-period
X-ray pulsars.  Many of these new sources are highly absorbed,
exhibiting column densities higher than about $10^{23} \cmmoinsdeux$,
and are concentrated in directions tangential to Galactic arms, for
instance the Norma arm (see \citeauthor{chaty:2005a}
\citeyear{chaty:2005a}, \citeauthor{tomsick:2004a}
\citeyear{tomsick:2004a} and \citeauthor{walter:2004b}
\citeyear{walter:2004b}), the richest arm of our Galaxy in high-mass
star forming regions.  This short-lived population hosts the likely
progenitors of extremely compact binary objects, which are good
candidates of gravitational wave emitters, and might constitute a key
sample in the understanding of the evolution of high-energy binary
systems.

Among these HMXBs hosting an OB supergiant companion star, two
classes, which might overlap, appear \citep{chaty:2006c}. The first
class is constituted of intrinsically highly obscured hard X-ray
sources, exhibiting a huge local extinction.  The most extreme example
of these sources is the highly absorbed source IGR\,J16318-4848
\citep{filliatre:2004}.  The second class exhibits fast and transient
outbursts, with peak fluxes of the order of $10^{-9} \ergcms$ in the
$20-40$\,keV band, and lasting only a few hours. This last
characteristic is very unusual among HMXBs. For this reason they are
called Supergiant Fast X-ray Transients (SFXTs,
\citeauthor{negueruela:2006a}, \citeyear{negueruela:2006a}).  These
SFXTs have faint quiescent emission, and their hard X-ray spectra
require a black hole or neutron star accretor.  Among these sources,
IGR\,J17544-2619 \citep{pellizza:2006} seems to be the archetype of
this new class of HMXBs, with long quiescent periods (Zurita Heras \&
Chaty in prep.).

Even if the {\it INTEGRAL} observatory can provide a localisation that
is accurate above 10\,keV ($\sim 2\amin$), it is not accurate enough
to pinpoint the source at other wavelengths, which is necessary to
reveal the nature of these sources.  So the first step in the study of
these sources is to look for an accurate localisation of the hard
X-ray sources by X-ray satellites such as {\it XMM-Newton}, {\it
  Swift} or {\it Chandra}.  While {\it XMM-Newton} and {\it Swift}
provide positions good enough to restrict the list of possible
counterparts to a small number, only {\it Chandra} gives unique
identifications in most cases (see for instance the multi-wavelength
study of four {\it INTEGRAL} sources in the direction of the Norma arm
by \citeauthor{tomsick:2006a} \citeyear{tomsick:2006a}, via {\it
  Chandra} localisation).  Once the source position is known to better
than several arcseconds, the hunt for the optical counterpart of the
source can begin.  However, a difficulty persists, due to the high
level of interstellar absorption in this region of the Galaxy, close
to the Galactic plane.  Near-infrared (NIR) observations, thanks to
accurate astrometry, and photometric and spectroscopic analysis, allow
for the nature of these sources to be revealed, constraining the
spectral type of the companion star and the type of the binary system.

Finally, mid-infrared (MIR) observations are required to understand
why these sources exhibit a strong local absorption, by characterizing
the nature of the absorbing matter, and determining whether it is made of
cold gas or dust, or anything else.  Based on the optical/NIR
observations we report here, we have performed MIR observations with
the VISIR instrument on ESO VLT/UT3. These observations, focusing on
the characterization of the presence, temperature, extension and
composition of the absorbing material constituting the circumstellar
medium enshrouding the obscured sources, are described in a companion
paper by \cite{rahoui:2008}.

Here we report on an intensive multi-wavelength study of a sample of
13 {\it INTEGRAL} sources belonging to both obscured and SFXT classes,
for which accurate X-ray localisations are available, aimed at
identifying their counterparts and constraining the nature of the
companion star and of the binary system.  We first describe the ESO
optical/NIR photometric and spectroscopic observations, and build the
spectral energy distributions (SEDs) in Section
\ref{section:observations}. We then review hard X-ray properties of
each source, and report the results of our optical/NIR observations in
Section \ref{section:results}.  We give general results and discuss
them in Section \ref{section:discussion}, and provide our conclusions
in Section \ref{section:conclusions}.


\section{Observations} \label{section:observations}

The multi-wavelength observations that we describe here are
based on astrometry, photometry and spectroscopy on 13 {\it
  INTEGRAL} sources indicated in Table \ref{table:sources}.  They were
performed at the European Southern Observatory (ESO, Chile), in 2
domains: in optical ($0.4-0.8 \microns$) with the EMMI instrument and
in NIR ($1-2.5 \microns$) with the SofI instrument, both with the 3.5m
New Technology Telescope (NTT) at La Silla Observatory.  Our optical
and NIR observations were carried out as part of the programme ESO \#
073.D-0339, in the visitor mode\footnote{The reduced data are
  available for retrieval at {\rm http://wikimbad.org}.}.

     \subsection{Optical observations}

     On 2004 July 10 between UT 0.0 and 11.0 we obtained optical
     photometry in $B$, $V$ $R$, $I$ and $Z$ bands of the sources
     given in Table \ref{table:optical} with the spectro-imager EMMI,
     installed on the NTT.  We used the large field of EMMI's
     detector, with the images binned by a factor of 2, giving an
     image scale of $0\asecp332$\,/pixel and a field of view of
     $9\aminp0 \times 10\aminp0$.  The photometric observations were
     performed with an integration time between 1 and 30\,s for each
     exposure, as reported in Table \ref{table:optical}. We observed
     five photometric standard stars of the optical standard star
     catalogue of \cite{landolt:1992}: PG~1633+099, PG~1633+099A;
     PG~1633+099B; PG~1633+099C and PG~1633+099D.

     We also carried out optical spectroscopy on 2004 July 9, taking
     30 spectra of the source IGR~J17391-3021 with the low-resolution
     grism \#1 of EMMI, with a slit of $1\asecp0$ providing a
     resolution of about 350, and a spectral range between 4370 and
     $10270~\AA$ with a dispersion of about $7.3~\AA$.  Each
     individual spectrum had an exposure time between 60 and 120\,s,
     giving a total integration time between 720 and 1440\,s, as
     reported in Table \ref{table:spectra}.  We observed the
     spectro-photometric standard star LTT~7379 taken with a similar
     airmass to flux calibrate the optical spectra.

     Finally, through internal Director's Discretionary Time (DDT) we
     obtained long-slit low-resolution spectra of 2 sources --IGR
     J17252-3616 and IGR J18027-2016-- with the spectro-imager EFOSC2
     installed on the 3.6~m telescope of La Silla Observatory.


     \subsection{NIR observations}

     We performed NIR photometry in $J$, $H$ and $K_{s}$ bands of the
     sources given in Table \ref{table:sources} on 2004 July 08-11
     with the spectro-imager SofI, installed on the NTT. We used the
     large field of SofI's detector, giving an image scale of
     $0\asecp288$\,/pixel and a field of view of $4\aminp92
     \times 4\aminp92$.  The photometric observations were obtained by
     repeating a set of images for each filter with 9 different
     30$\asec$ offset positions including the targets, following the
     standard jitter procedure allowing us to cleanly subtract the sky
     emission in NIR.  The integration time varied between 10 and
     50\,s for each individual exposure, giving a total exposure time
     between 108 and 450\,s.  The NIR photometry results are given in
     Table \ref{table:infrared}.  We observed three photometric
     standard stars of the faint NIR standard star catalogue of
     \cite{persson:1998}: sj9157, sj9172 and sj9181.

     We also carried out NIR spectroscopy with SofI between 0.9 and
     $2.5 \microns$, taking 12 spectra using alternatively the
     low-resolution Blue and Red grisms respectively, half of them
     with the $1\asecp0$ slit on the source and the other half with an
     offset of 30$\asec$, in order to subtract the NIR sky
     emission. Each individual spectrum has an exposure time from
     60 to 180\,s, giving a total integration time between 720 and
     2160\,s in each grism, as reported in Table \ref{table:spectra}.

     \subsection{Data reduction}

     We used the \rm{IRAF} (Image Reduction and Analysis Facility)
     suite to perform data reduction, carrying out standard procedures
     of optical and NIR image reduction, including flat-fielding and
     NIR sky subtraction.

\begin{figure*}
	\centering
\subfigure[IGR\,J16320-4751 (3$\asec$ {\it XMM-Newton})]{
          \label{fig:igrj16320}
          \includegraphics[angle=0,width=5.5cm]{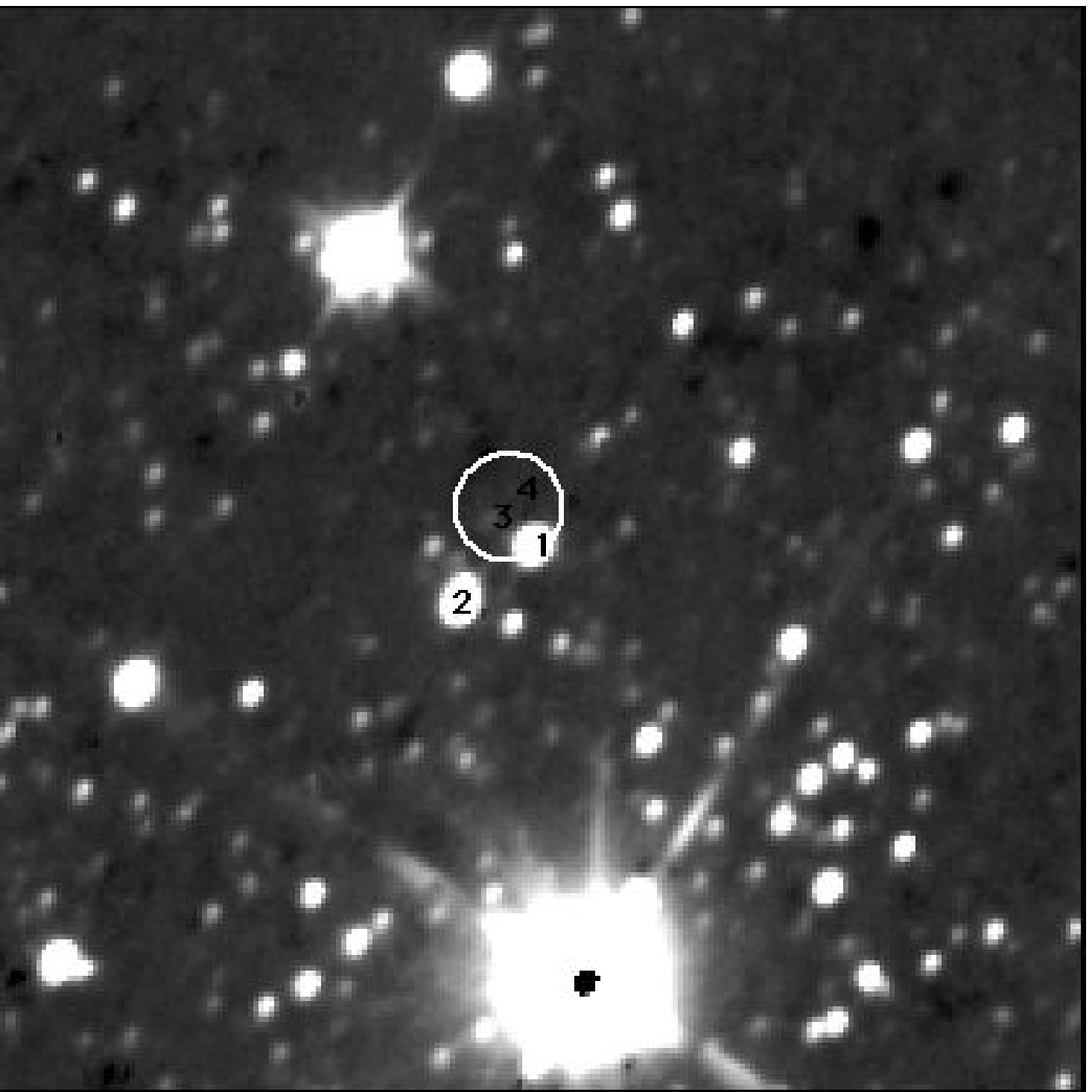}}
\subfigure[IGR\,J16358-4726 (0.6$\asec$ {\it Chandra})]{
          \label{fig:igrj16358}
          \includegraphics[angle=0,width=5.5cm]{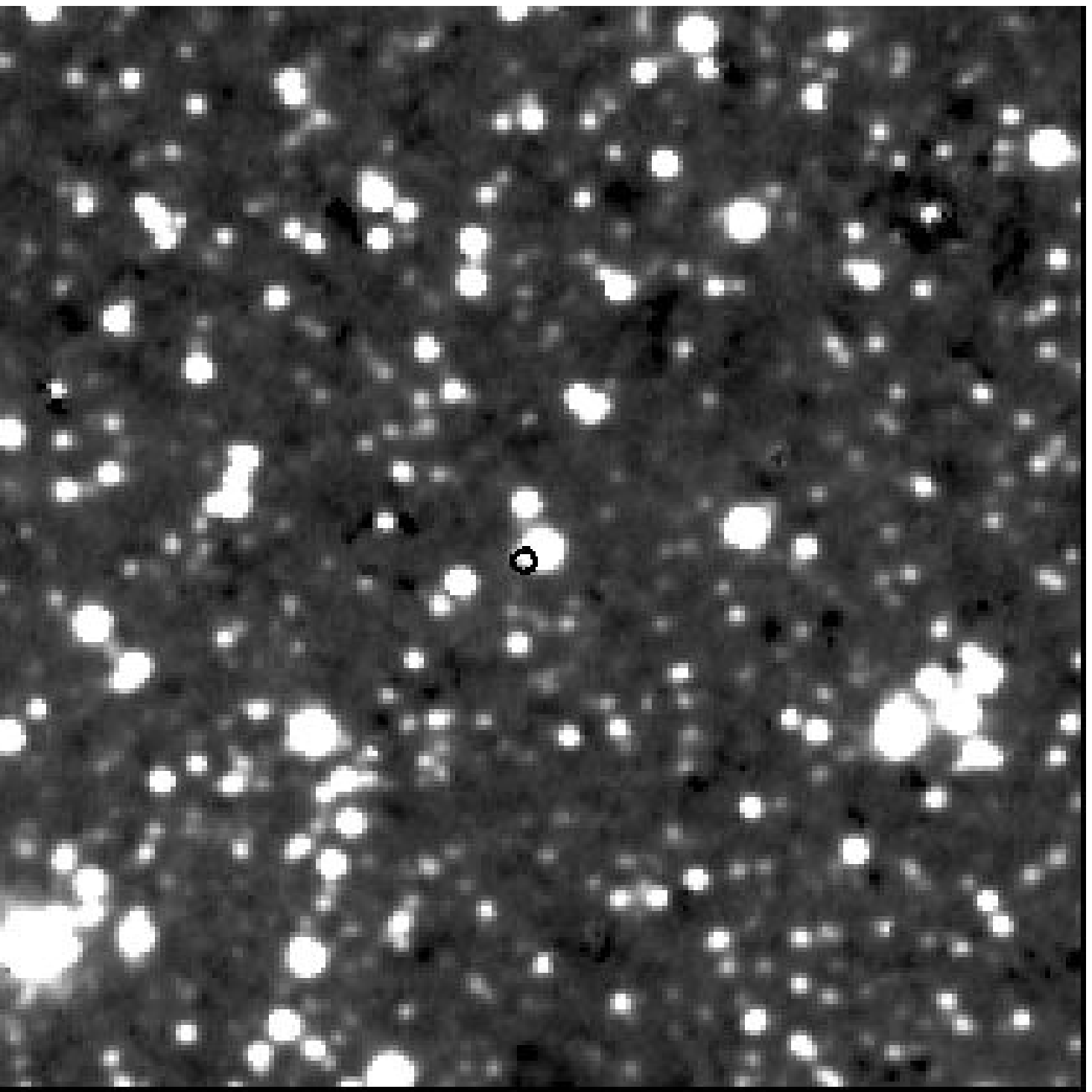}}
\subfigure[IGR\,J16393-4643 (4$\asec$ {\it XMM-Newton})]{
          \label{fig:igrj16393}
          \includegraphics[angle=0,width=5.5cm]{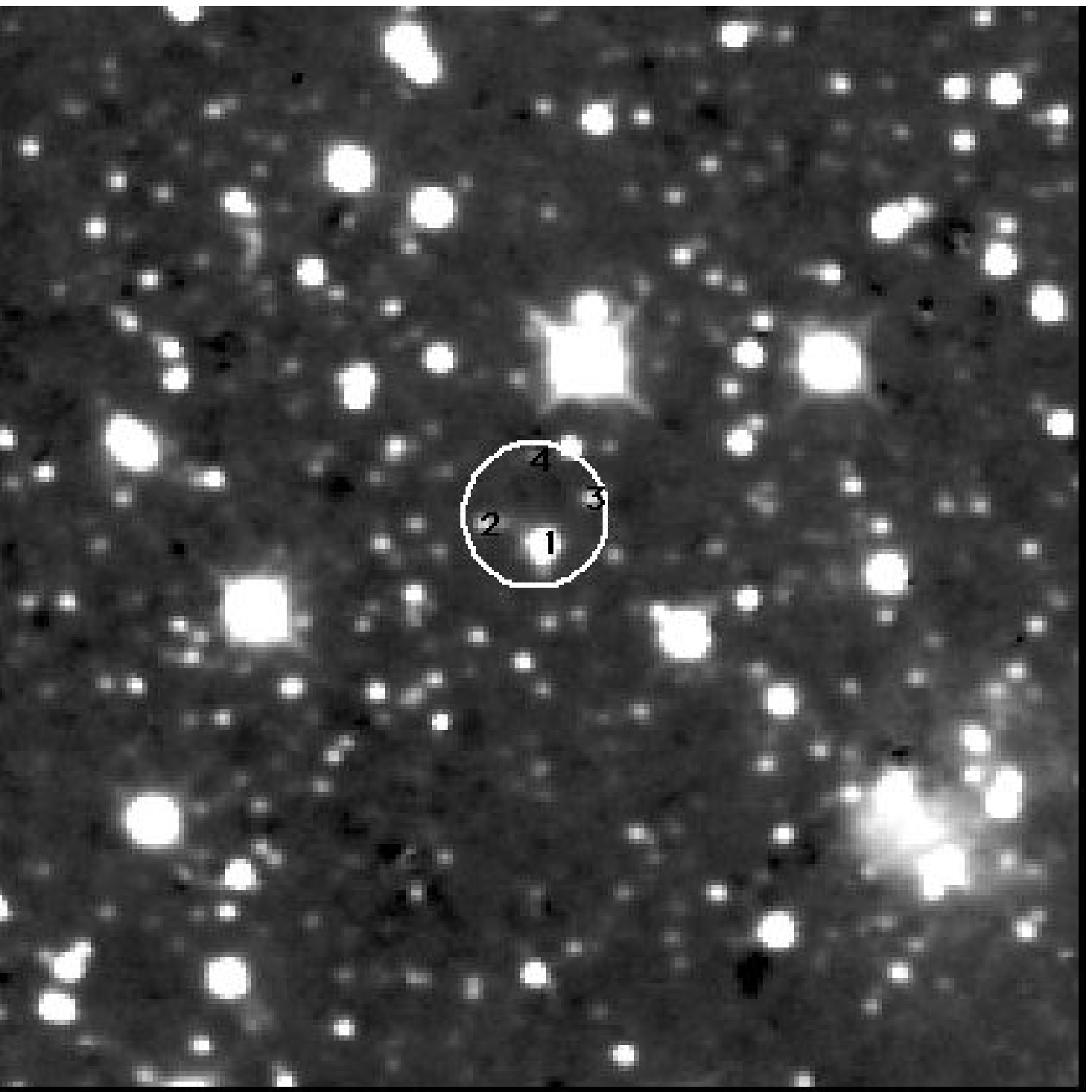}}
\subfigure[IGR\,J16418-4532 (4$\asec$ {\it XMM-Newton})]{
          \label{fig:igrj16418}
          \includegraphics[angle=0,width=5.5cm]{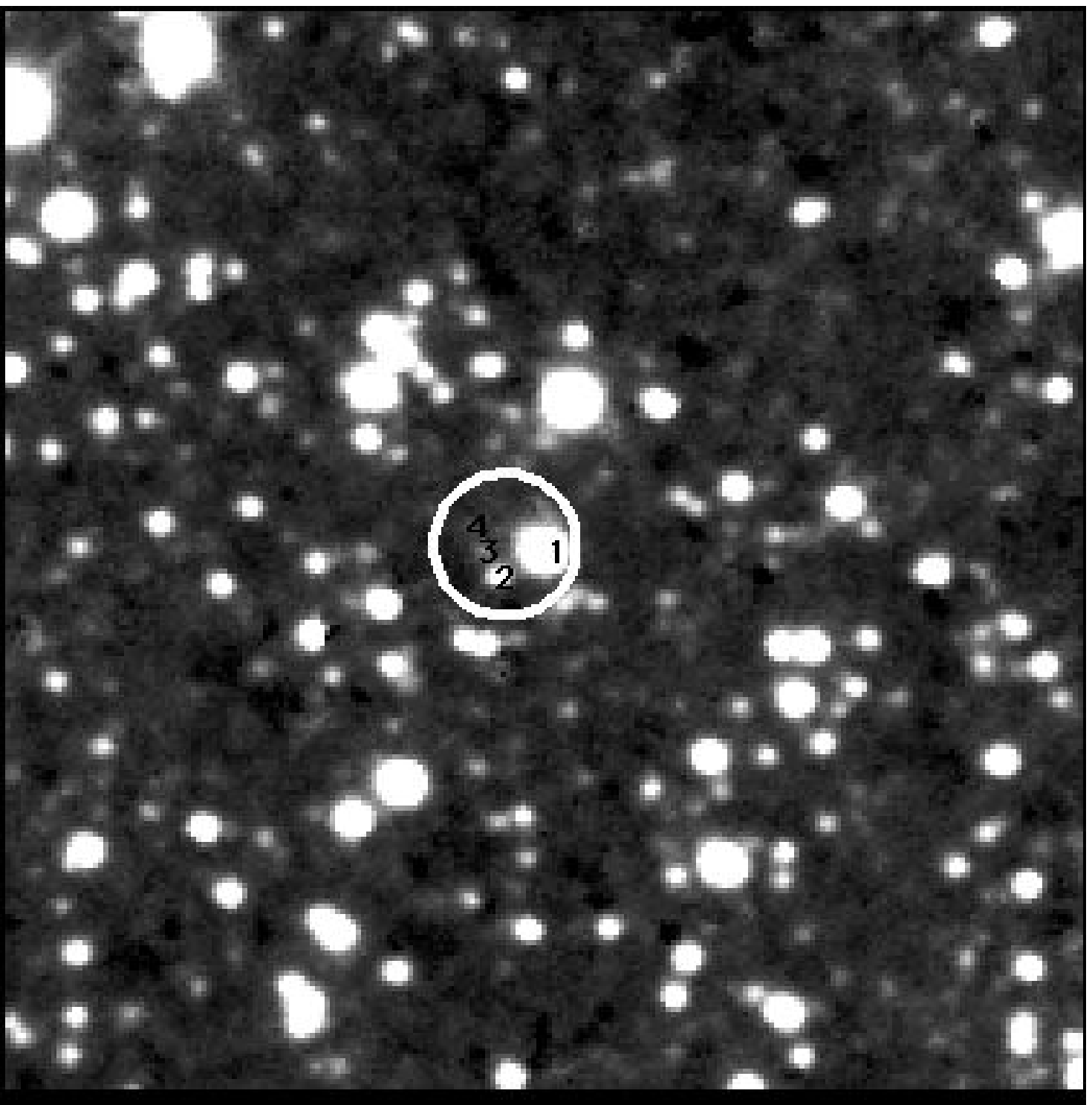}}
\subfigure[IGR\,J16479-4514 (4$\asec$ {\it XMM-Newton})]{
          \label{fig:igrj16479}
          \includegraphics[angle=0,width=5.5cm]{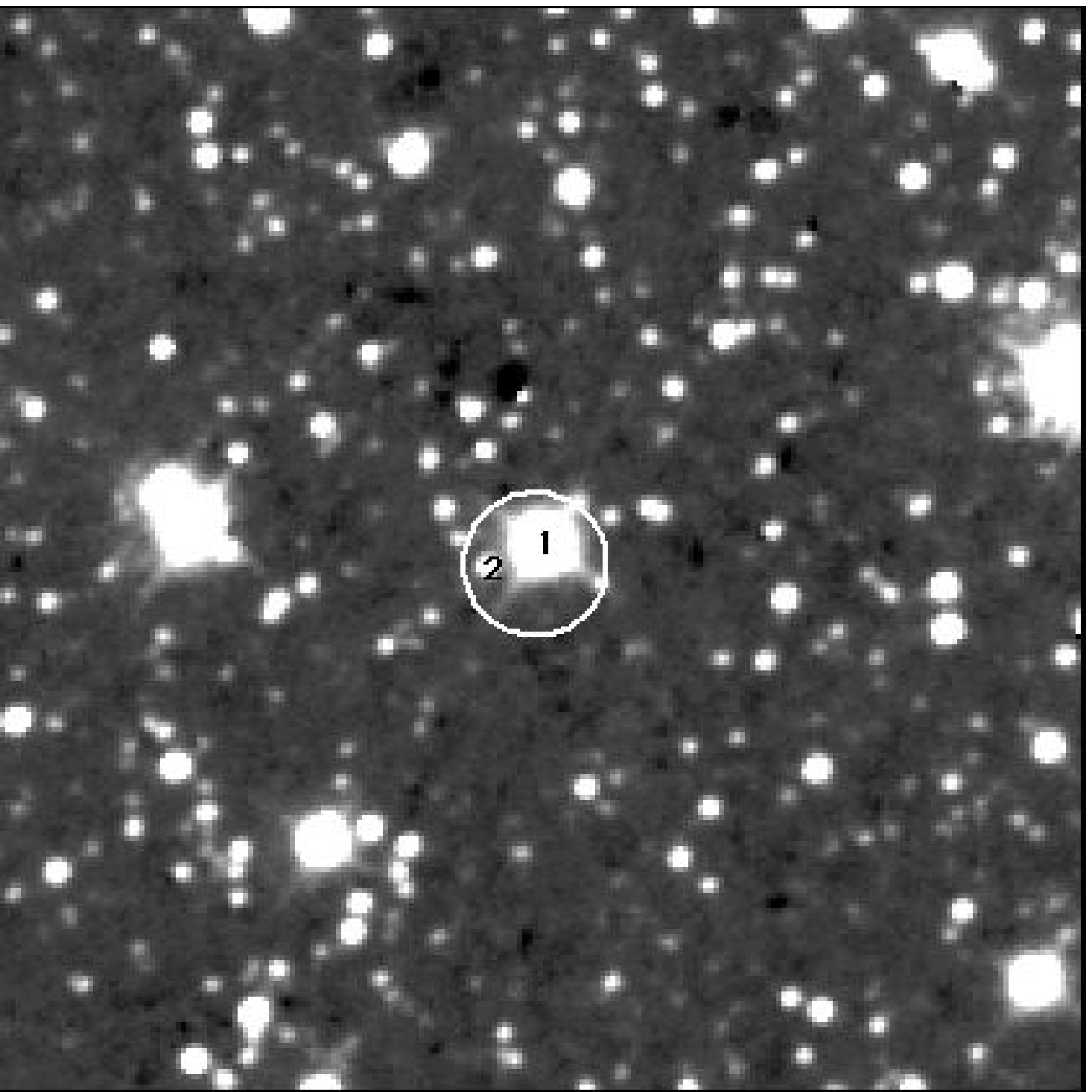}}
\subfigure[IGR\,J16558-5203 ($3\asecp52$ {\it Swift})]{
          \label{fig:igrj16558}
          \includegraphics[angle=0,width=5.5cm]{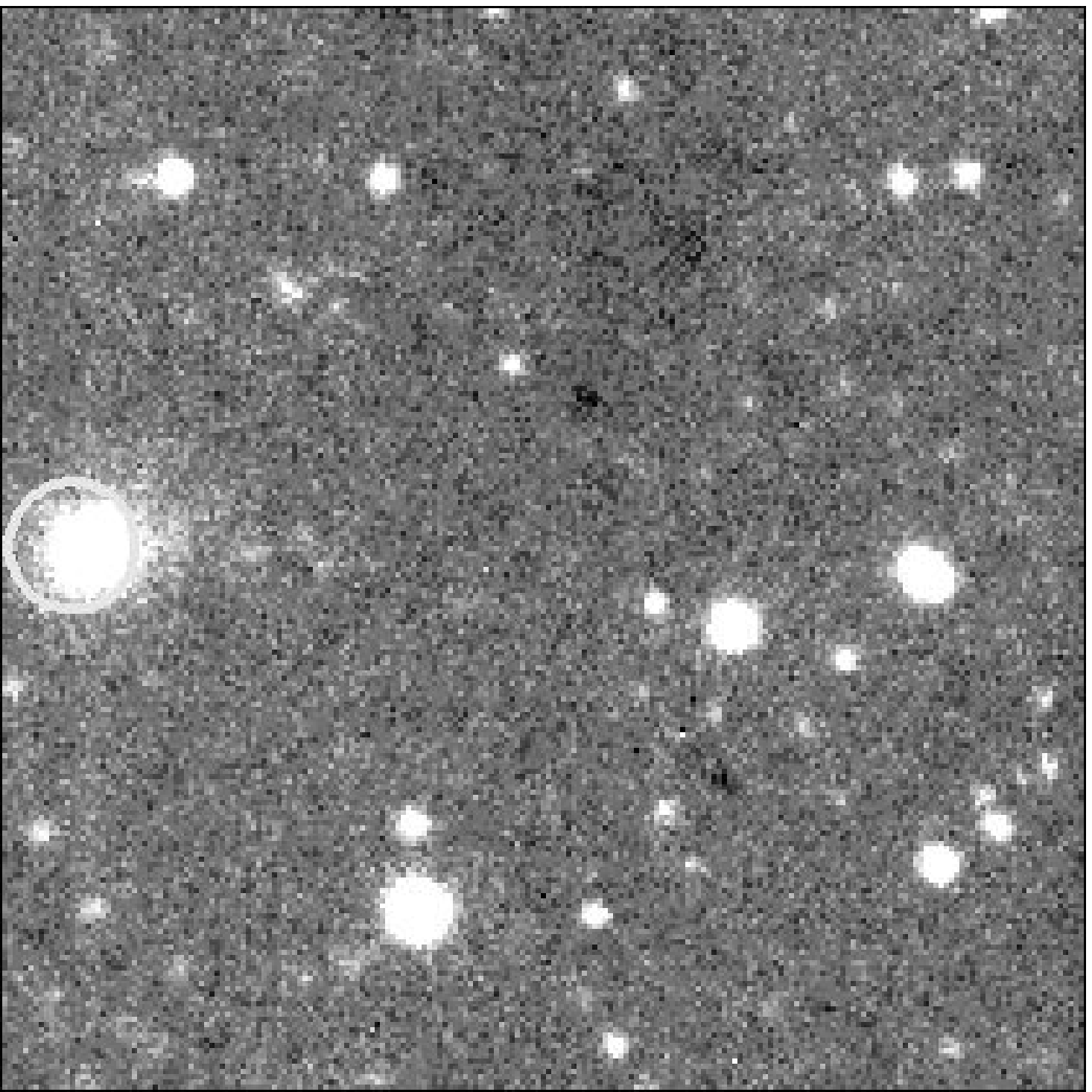}}
\subfigure[IGR\,J17091-3624 (3.6$\asec$ {\it Swift})]{
          \label{fig:igrj17091}
          \includegraphics[angle=0,width=5.5cm]{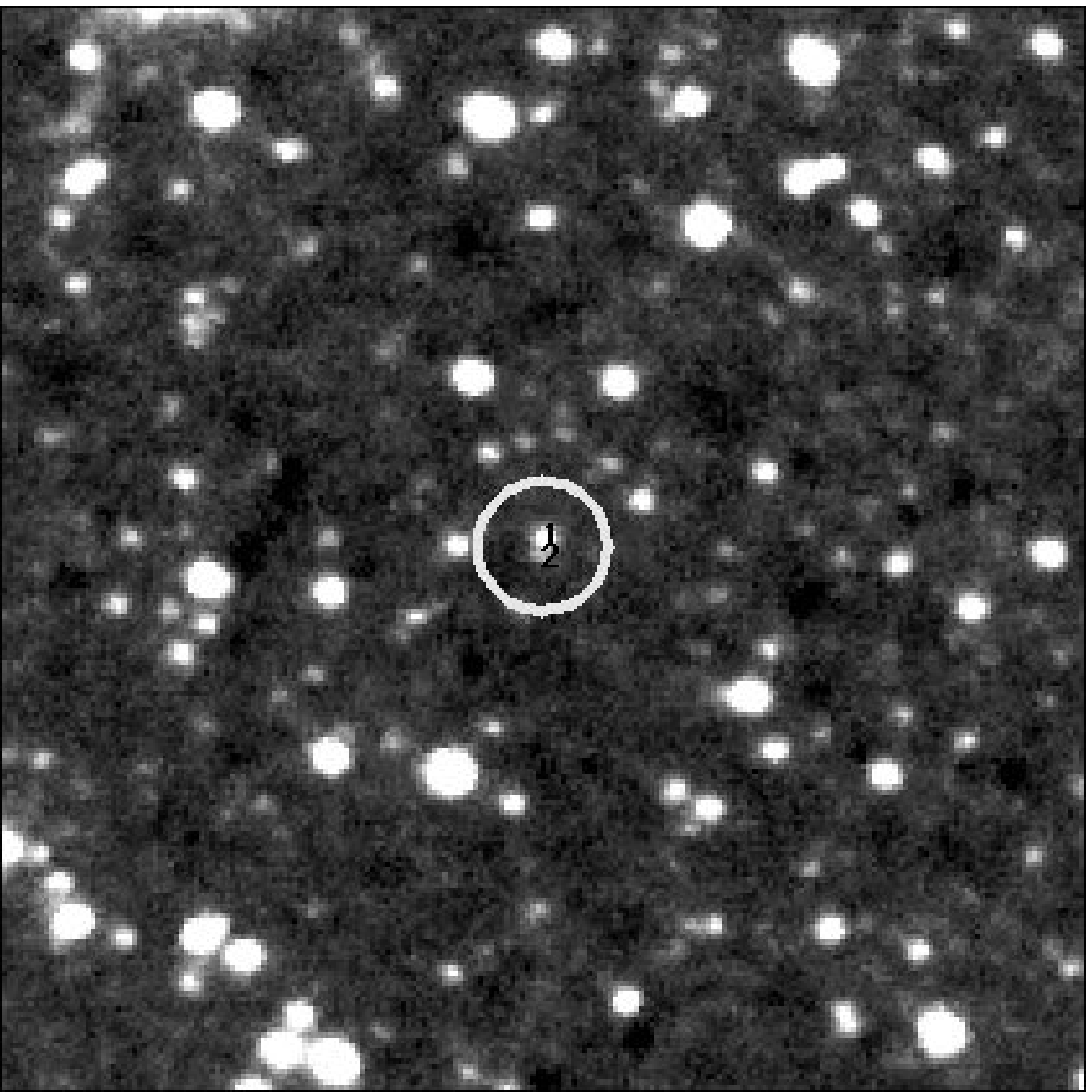}}
\subfigure[IGR\,J17252-3616 (4$\asec$ {\it XMM-Newton})]{
          \label{fig:igrj17252}
          \includegraphics[angle=0,width=5.5cm]{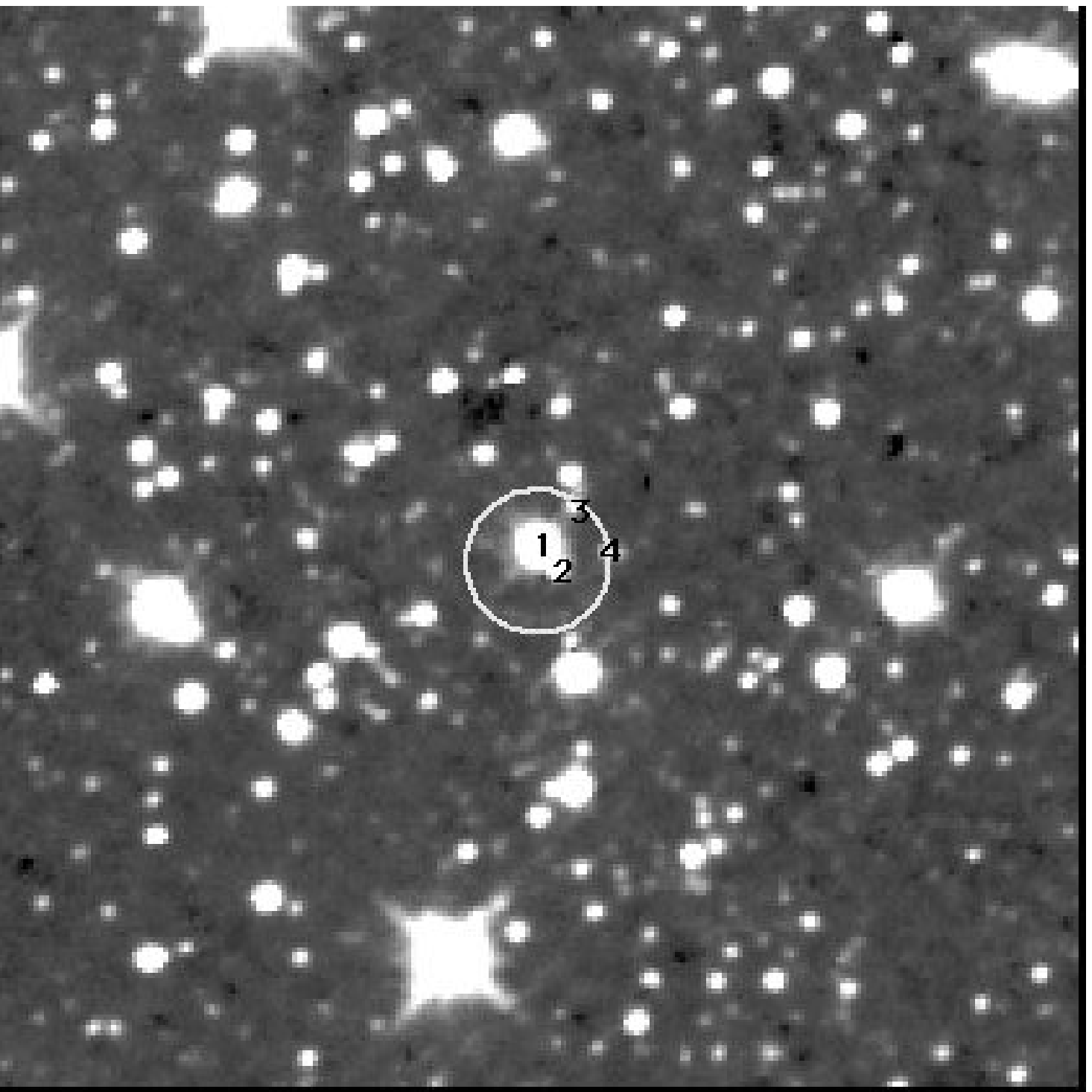}}
\subfigure[IGR\,J17391-3021 (1$\asec$ {\it Chandra})]{
          \label{fig:igrj17391}
          \includegraphics[angle=0,width=5.5cm]{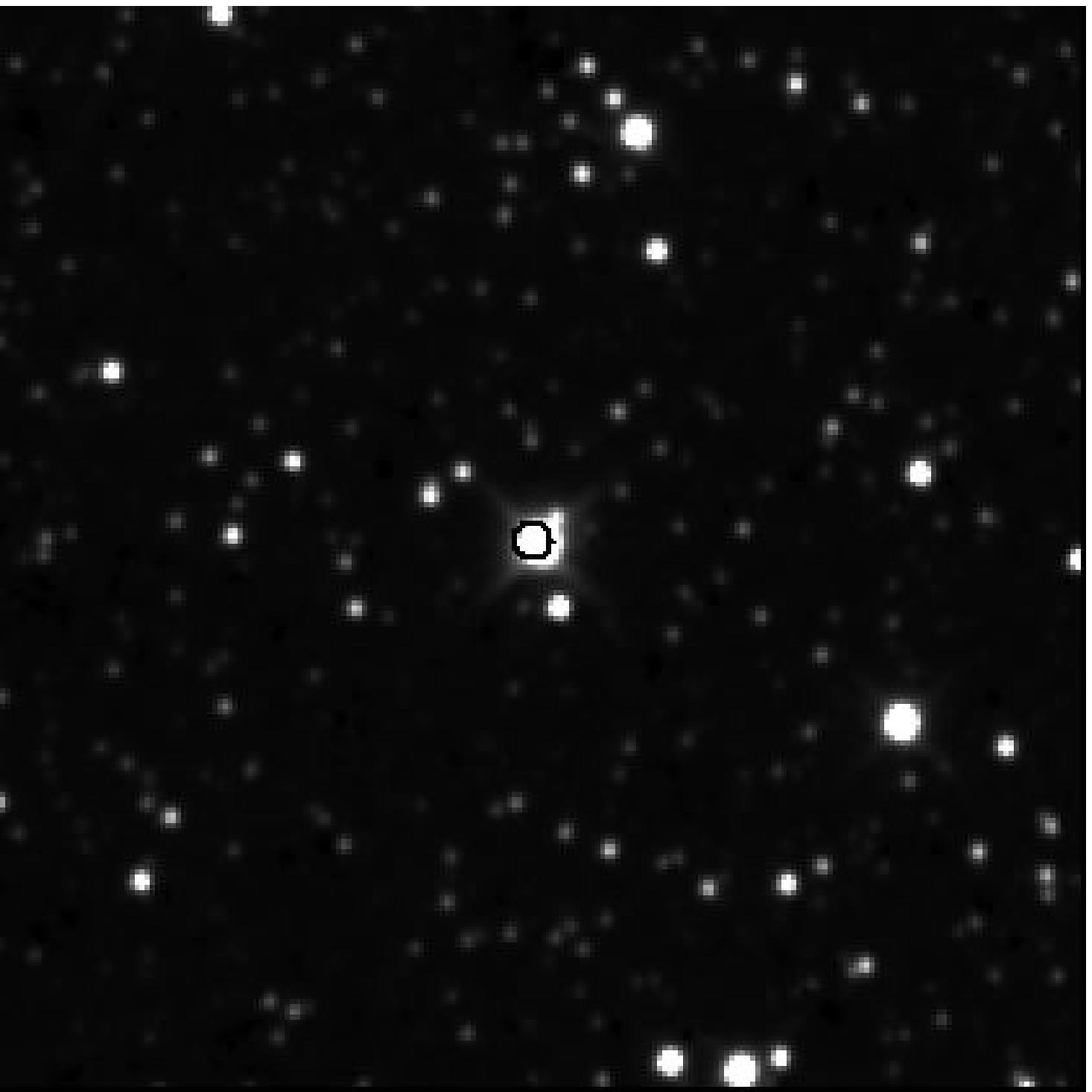}}
	\caption[]{\label{figure:smallfields} Finding charts of the
studied {\it INTEGRAL} sources, observed at ESO NTT telescope in the
infrared K$_{\rm S}$ band (2.2 $\microns$). Size: 1$\amin$x1$\amin$; North is to
the top and East to the left.  They are all centered apart from
IGR\,J16558-5203, because the source was caught at the edge of the
CCD. We overplot the most accurate localisation available to date.
%
%
%
%
}
\end{figure*}

\begin{figure*}
\subfigure[IGR\,J17597-2201 (4$\asec$ {\it XMM-Newton})]{
          \label{fig:igrj17597}
          \includegraphics[angle=0,width=5.5cm]{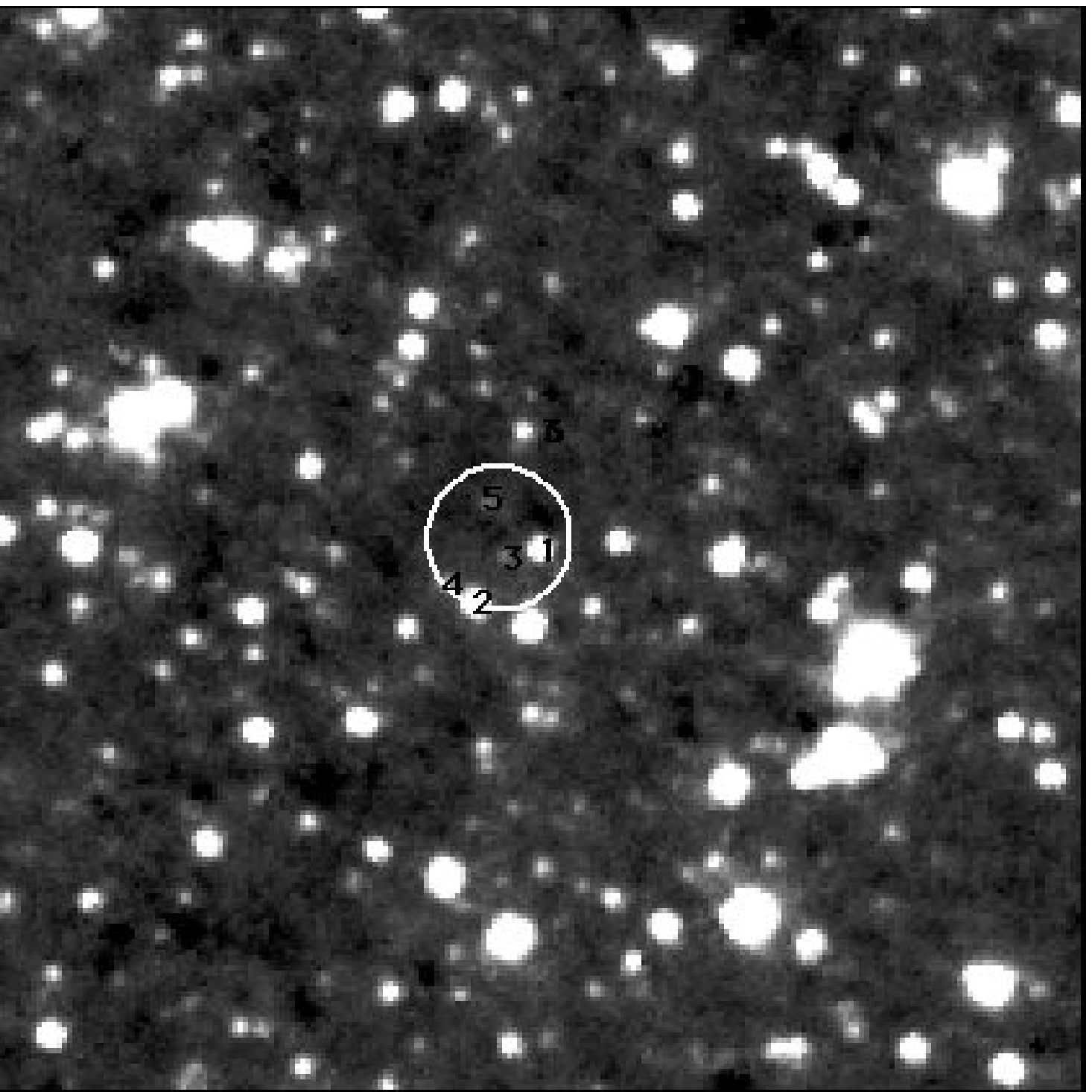}}
\subfigure[IGR\,J18027-2016 (4$\asec$ {\it XMM-Newton})]{
          \label{fig:igrj18027}
          \includegraphics[angle=0,width=5.5cm]{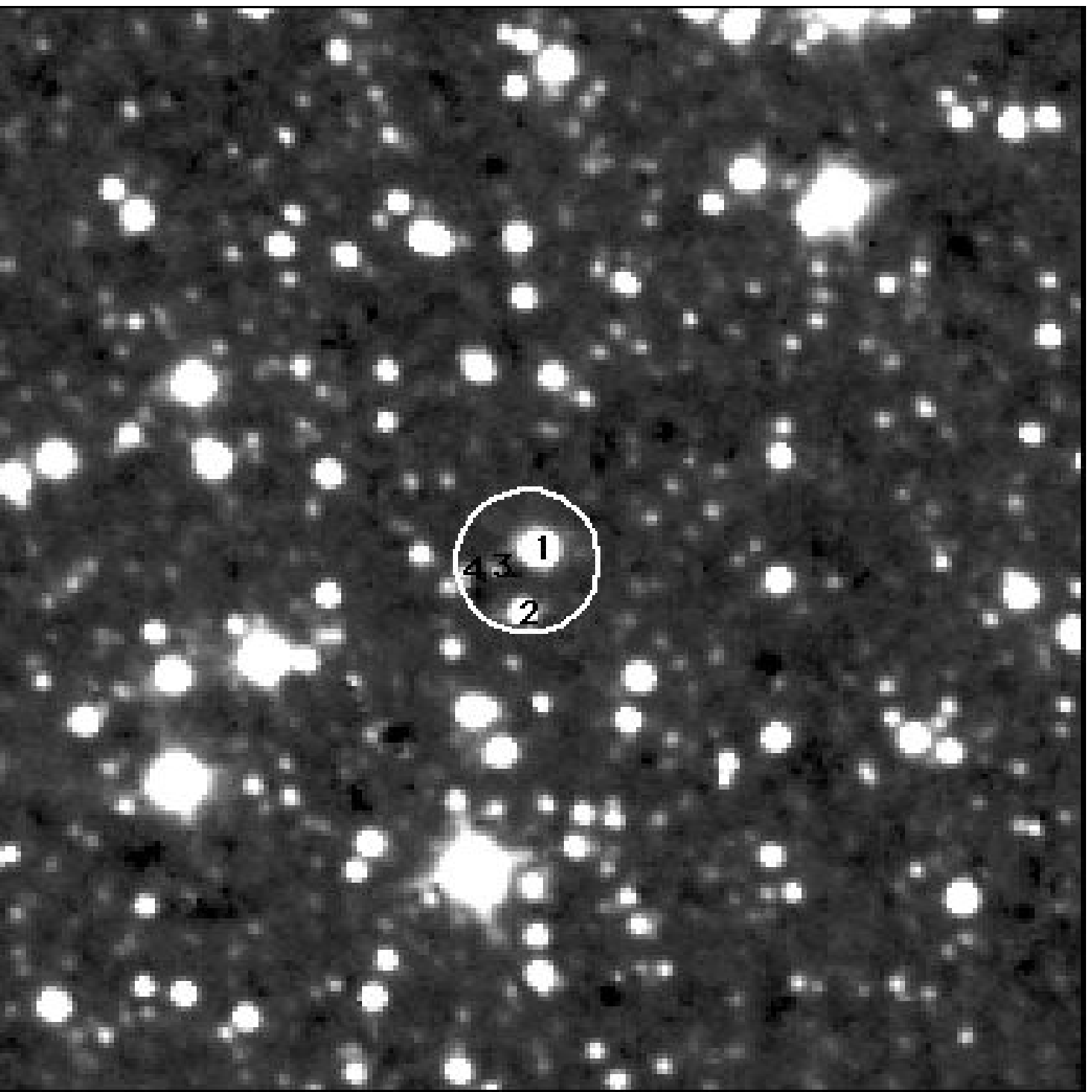}}
\subfigure[IGR\,J18483-0311 (3.3$\asec$ {\it Swift})]{
          \label{fig:igrj18483}
          \includegraphics[angle=0,width=5.5cm]{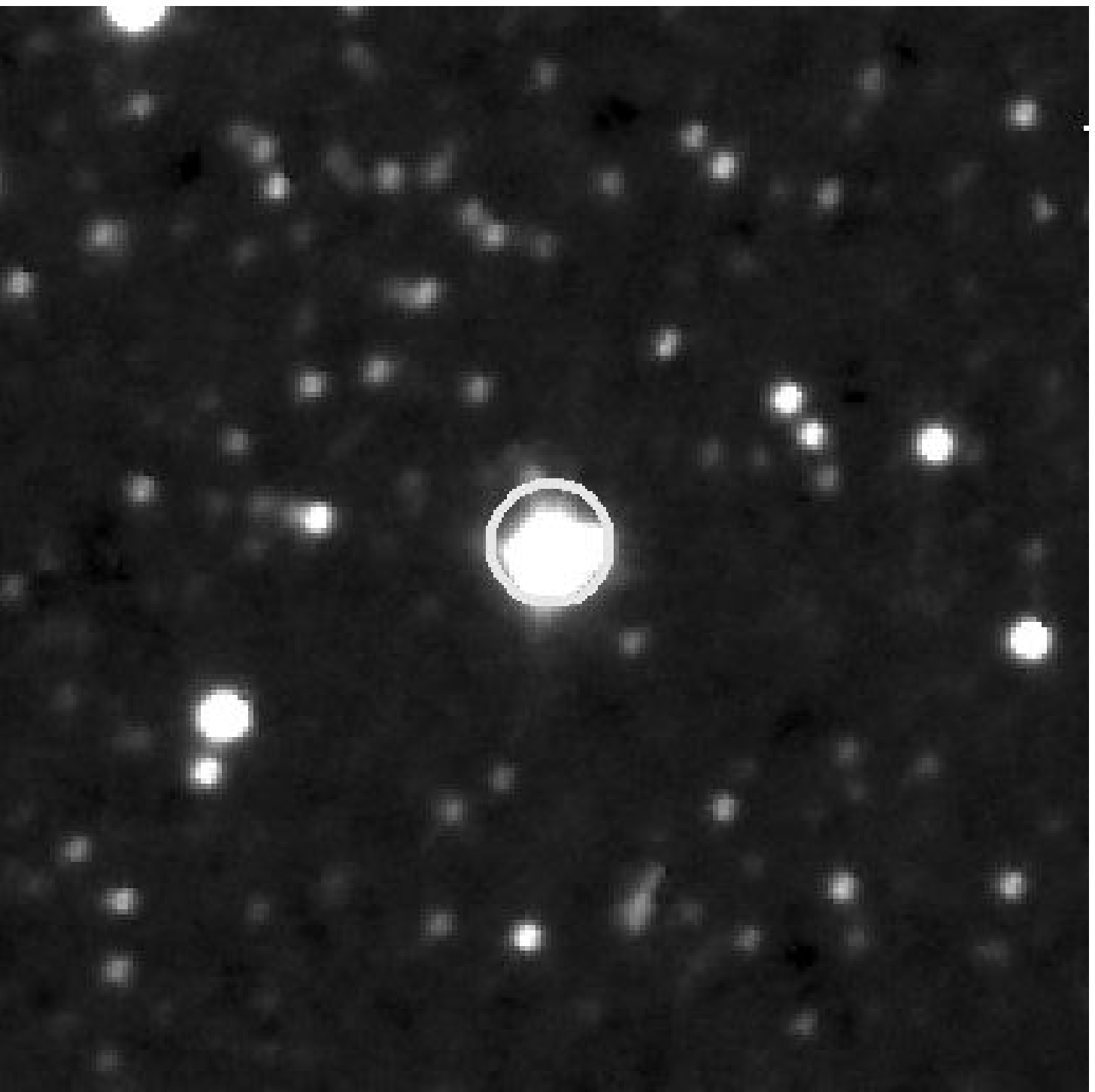}}
\subfigure[IGR\,J19140+0951 (0.6$\asec$ {\it Chandra})]{
          \label{fig:igrj19140}
	  \includegraphics[angle=0,width=5.5cm]{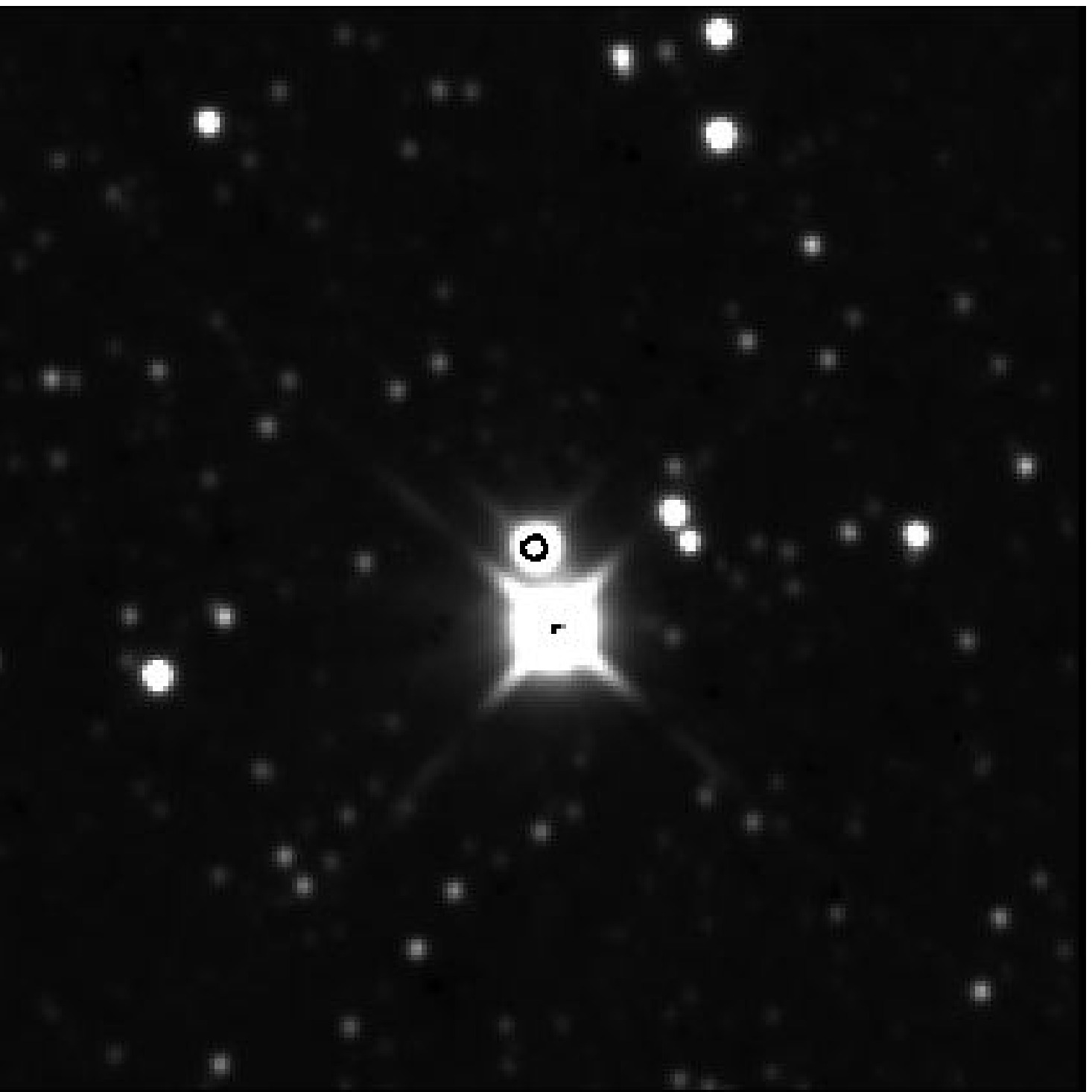}}
	\caption[]{\label{figure:smallfieldsbis} 
Figure \ref{figure:smallfields} cont'd: finding chart of the
studied {\it INTEGRAL} sources.
}
\end{figure*}

     We performed accurate astrometry on each entire SofI $4\aminp92
     \times 4\aminp92$ field, using all stars from the 2MASS catalogue
     present in this field\footnote{ The pixel size of 2MASS is
       $2\asecp0$ and the position reconstruction error is $\sim
       0\asecp2$.} (amounting to $\sim 1000$ 2MASS objects). The rms
     of astrometry fit is always less than $0\asecp5$, and we obtained
     a pixel scale in x,y axis of $-0\asecp28783$ and
     $0\asecp28801$\,/pixel respectively.  The finding charts
     including the results of our astrometry are shown in Figures
     \ref{figure:smallfields} and \ref{figure:smallfieldsbis} for all
     sources of our sample.

     We carried out aperture photometry, and we then transformed
     instrumental magnitudes into apparent magnitudes using the
     standard relation: $mag_{app} = mag_{inst} - Zp - ext \times AM$
     where $mag_{app}$ and $mag_{inst}$ are the apparent and
     instrumental magnitudes, Zp is the zero-point, $ext$ the
     extinction and $AM$ the airmass. We used for the extinction
     $ext_\mathrm{J} = 0.06$, $ext_\mathrm{H} = 0.04$ and
     $ext_\mathrm{Ks} = 0.10$ typical of the La Silla observatory.
     The log of the observations and the results of photometry are
     given in Tables \ref{table:optical} and \ref{table:infrared} for
     the optical and NIR respectively.

     We analyzed the optical spectra using standard {\rm IRAF} tasks,
     subtracting bias and correcting for flat field, and we used the
     {\it IRAF noao.twodspec} package in order to extract spectra and
     perform wavelength and flux calibrations.  The optical spectra
     were reduced and flux-calibrated in ``F-lambda'' units -
     $\ergcmsang$.

     NIR spectra were reduced using {\rm IRAF} by flat-fielding,
     correcting the geometrical distortion using the arc frame,
     shifting the individual images using the jitter offsets,
     combining these images and finally extracting the spectra. The
     analysis of SofI spectroscopic data, and more precisely the sky
     subtraction, was difficult due to a variable sky, mainly in the
     red part of the blue grism, causing some wave patterns.  The
     target spectra were then corrected for the telluric lines using a
     median of various standard stars observed with the same
     configuration during the corresponding nights.  All spectra,
     optical and NIR, are finally shifted to the heliocentric rest
     frame.


     \subsection{SEDs} \label{section:seds}

\begin{figure*}
	\centering
	\includegraphics[angle=-90,width=5.9cm]{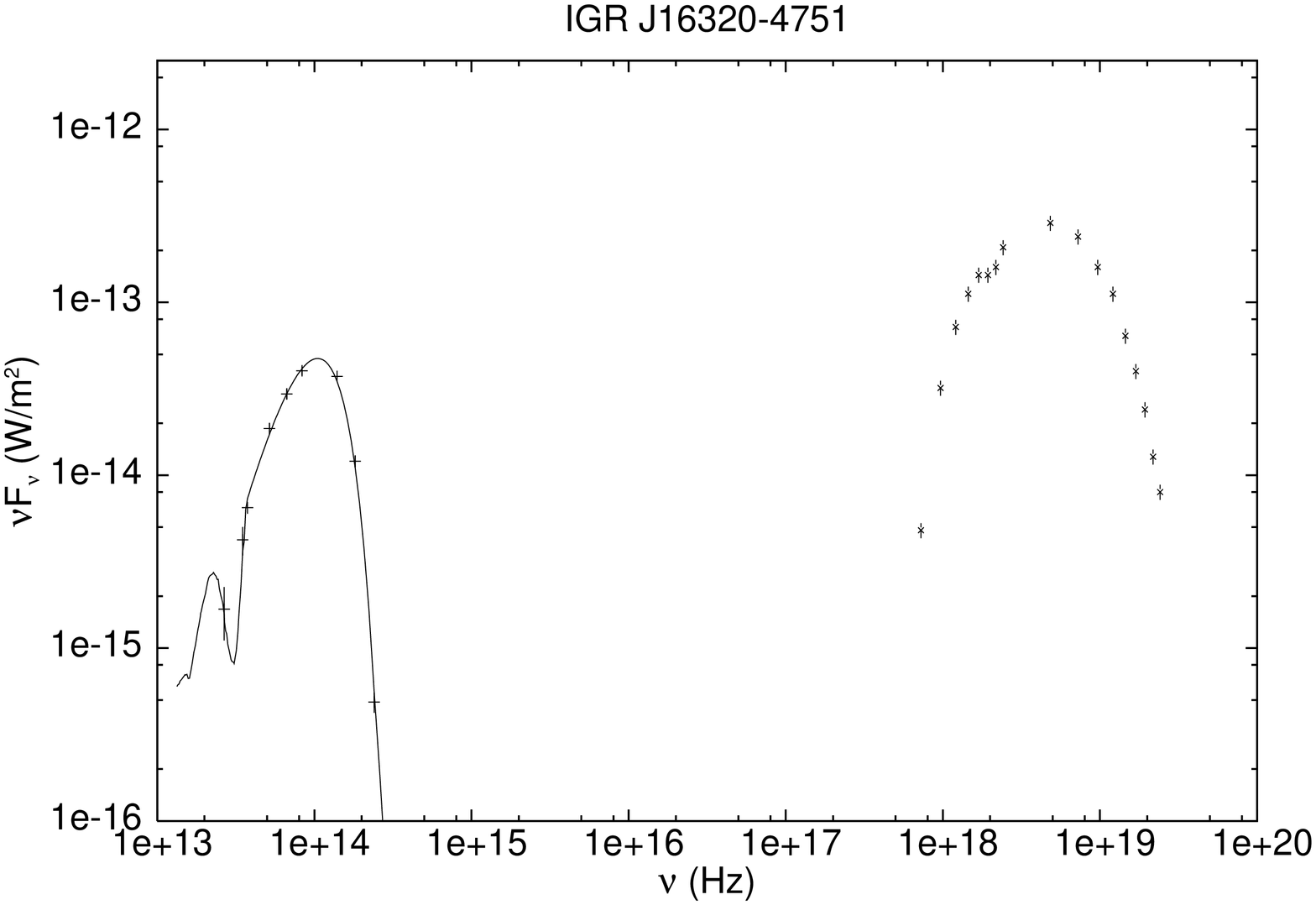}
	\includegraphics[angle=-90,width=5.9cm]{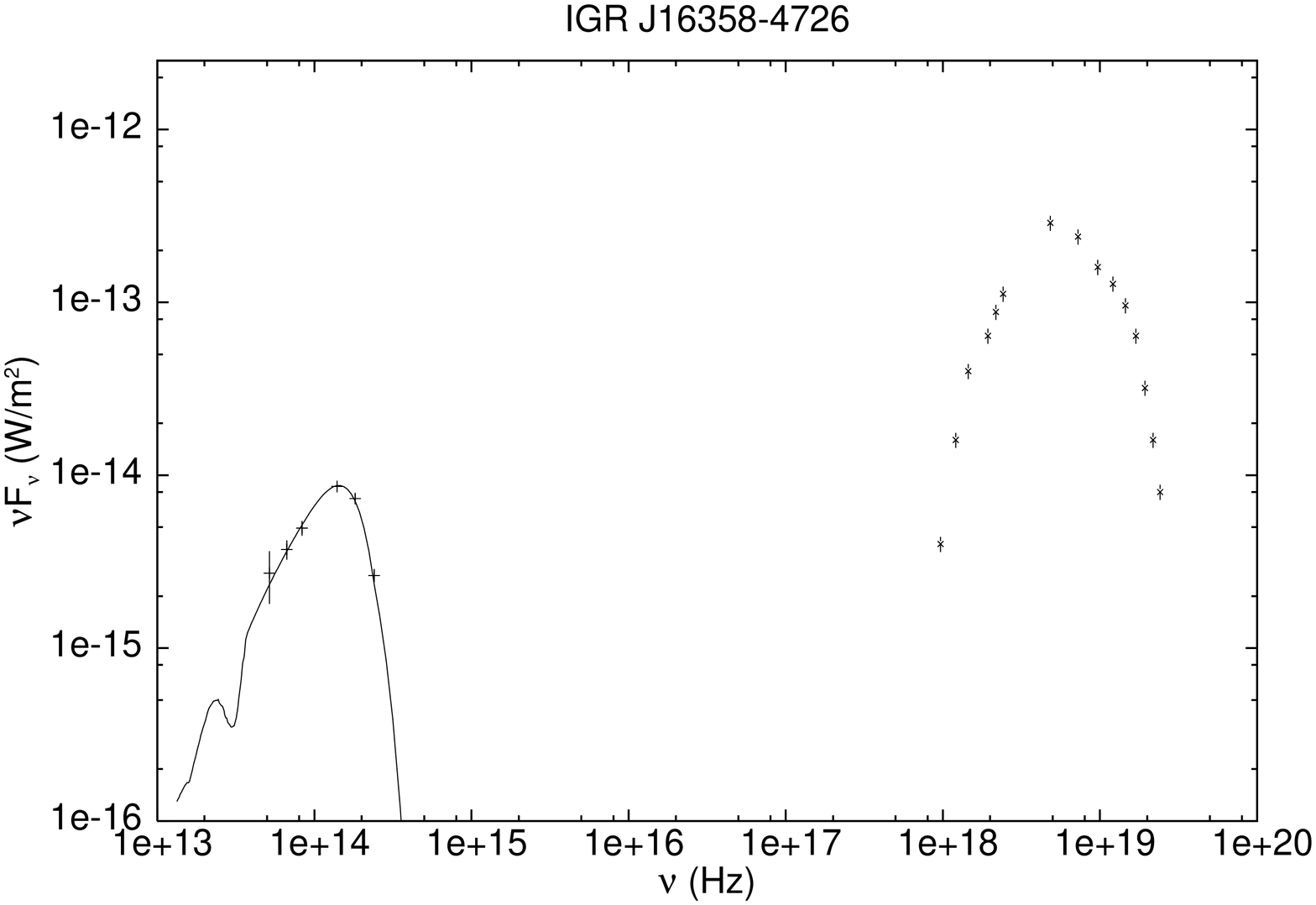}
	\includegraphics[angle=-90,width=5.9cm]{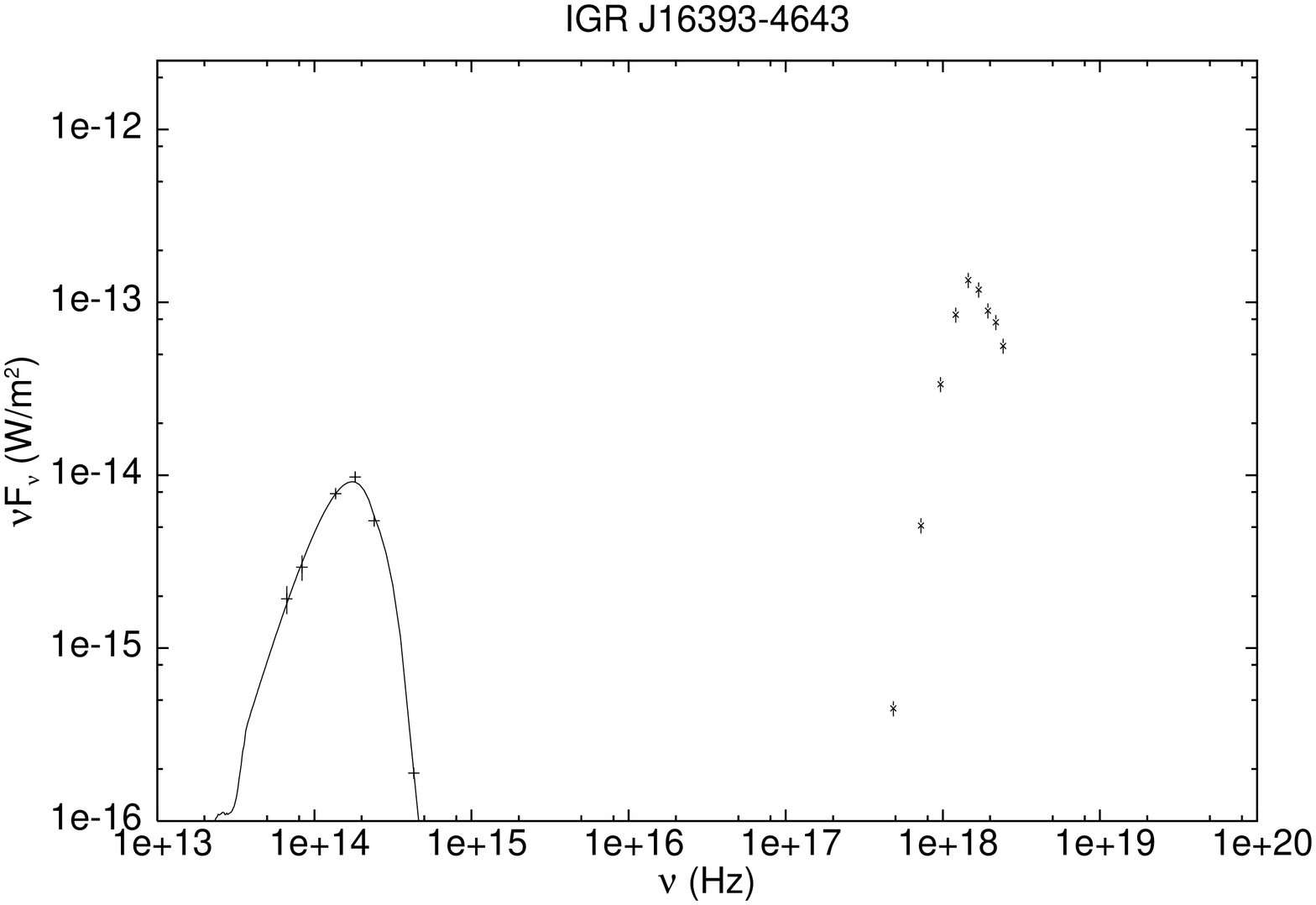}
	\includegraphics[angle=-90,width=5.9cm]{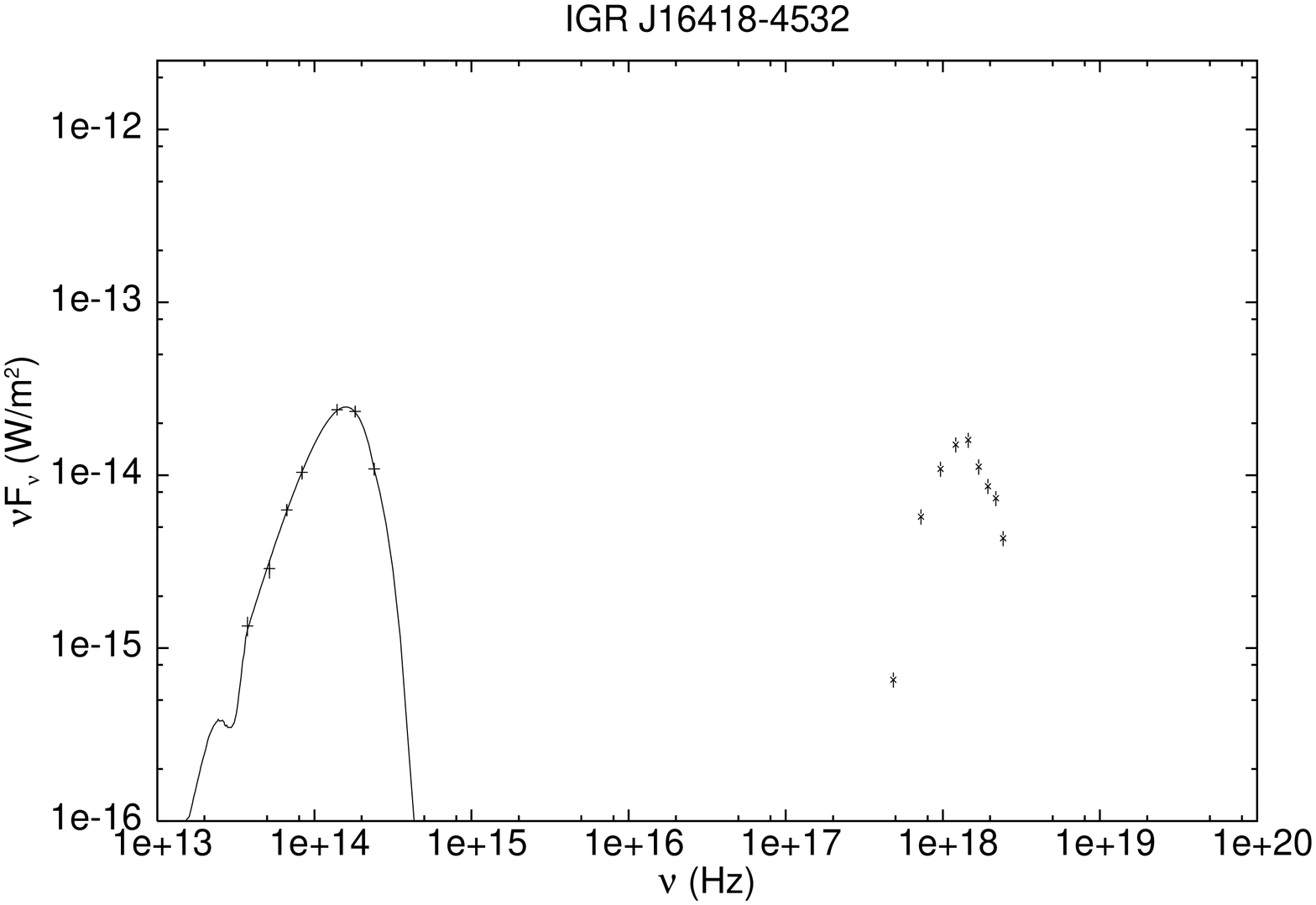}
	\includegraphics[angle=-90,width=5.9cm]{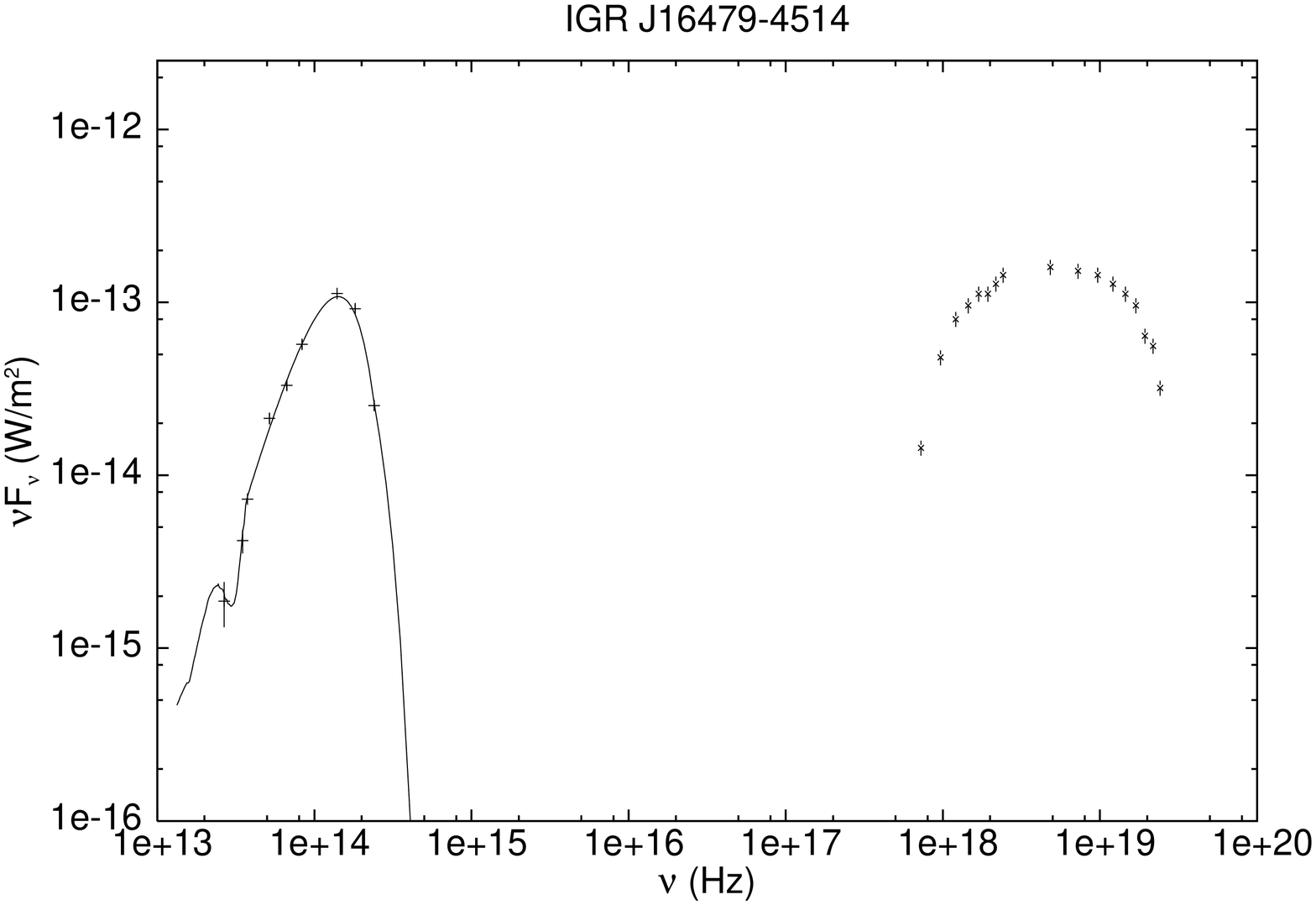}
	\includegraphics[angle=-90,width=5.9cm]{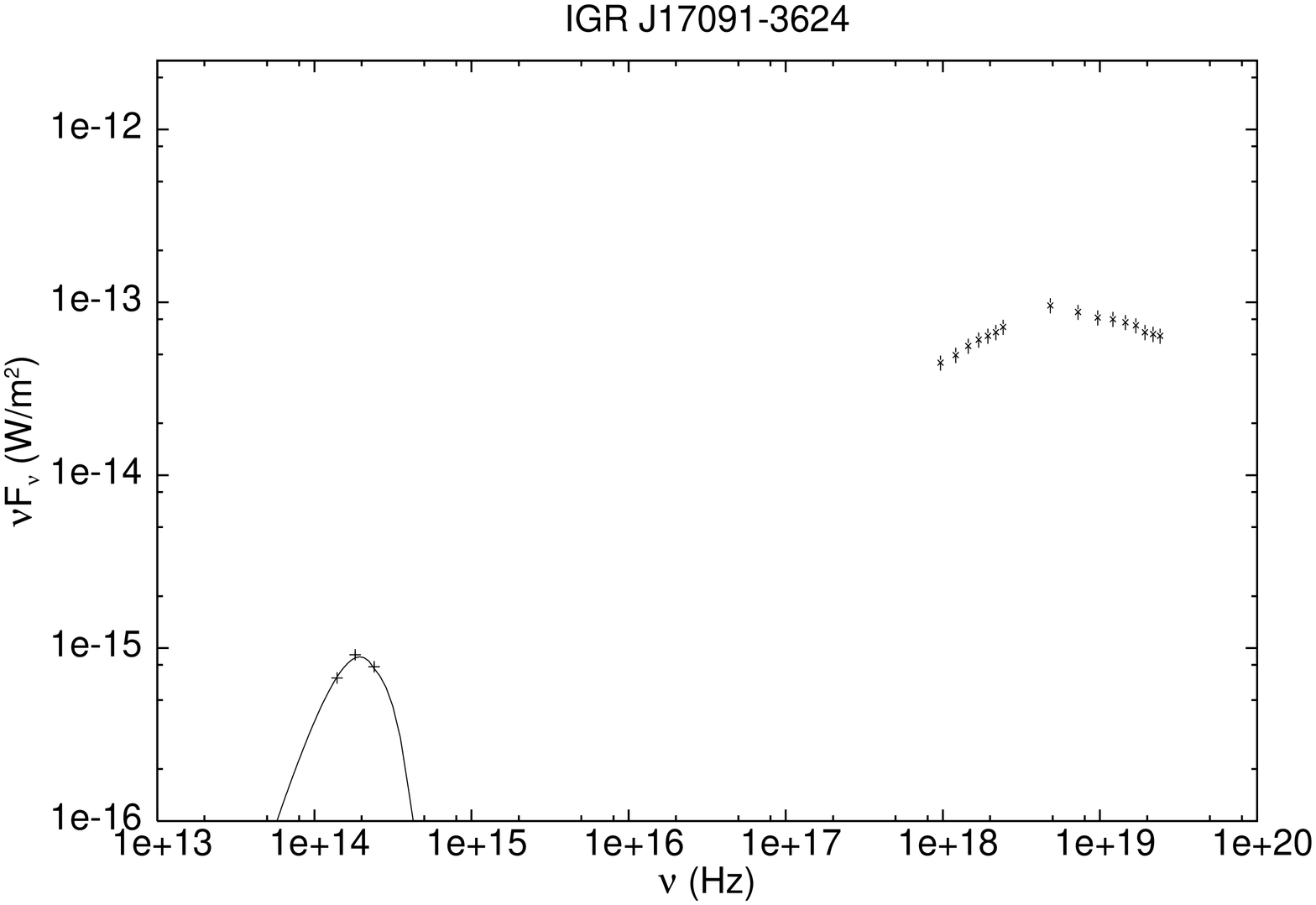}
	\includegraphics[angle=-90,width=5.9cm]{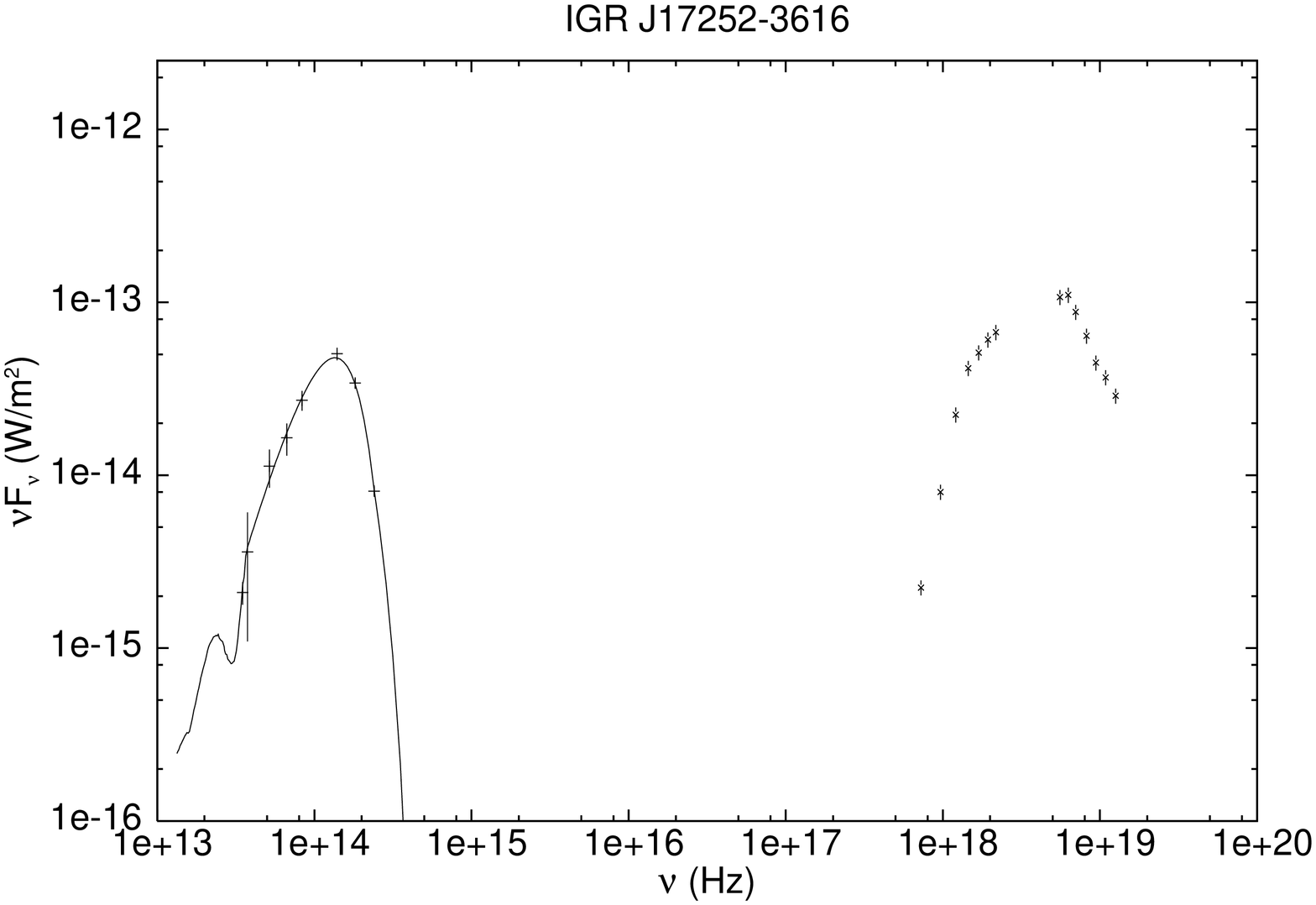}
	\includegraphics[angle=-90,width=5.9cm]{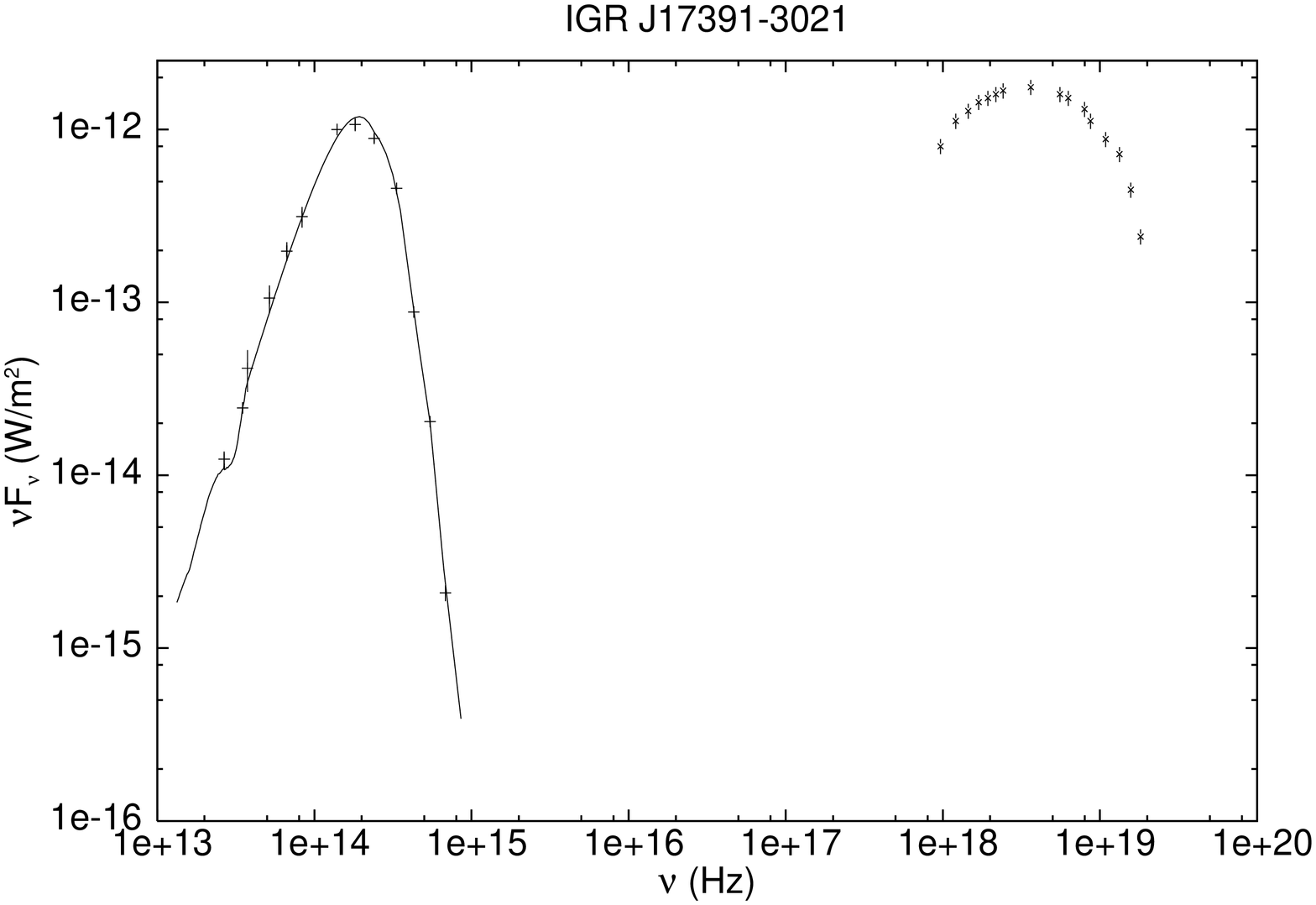}
	\includegraphics[angle=-90,width=5.9cm]{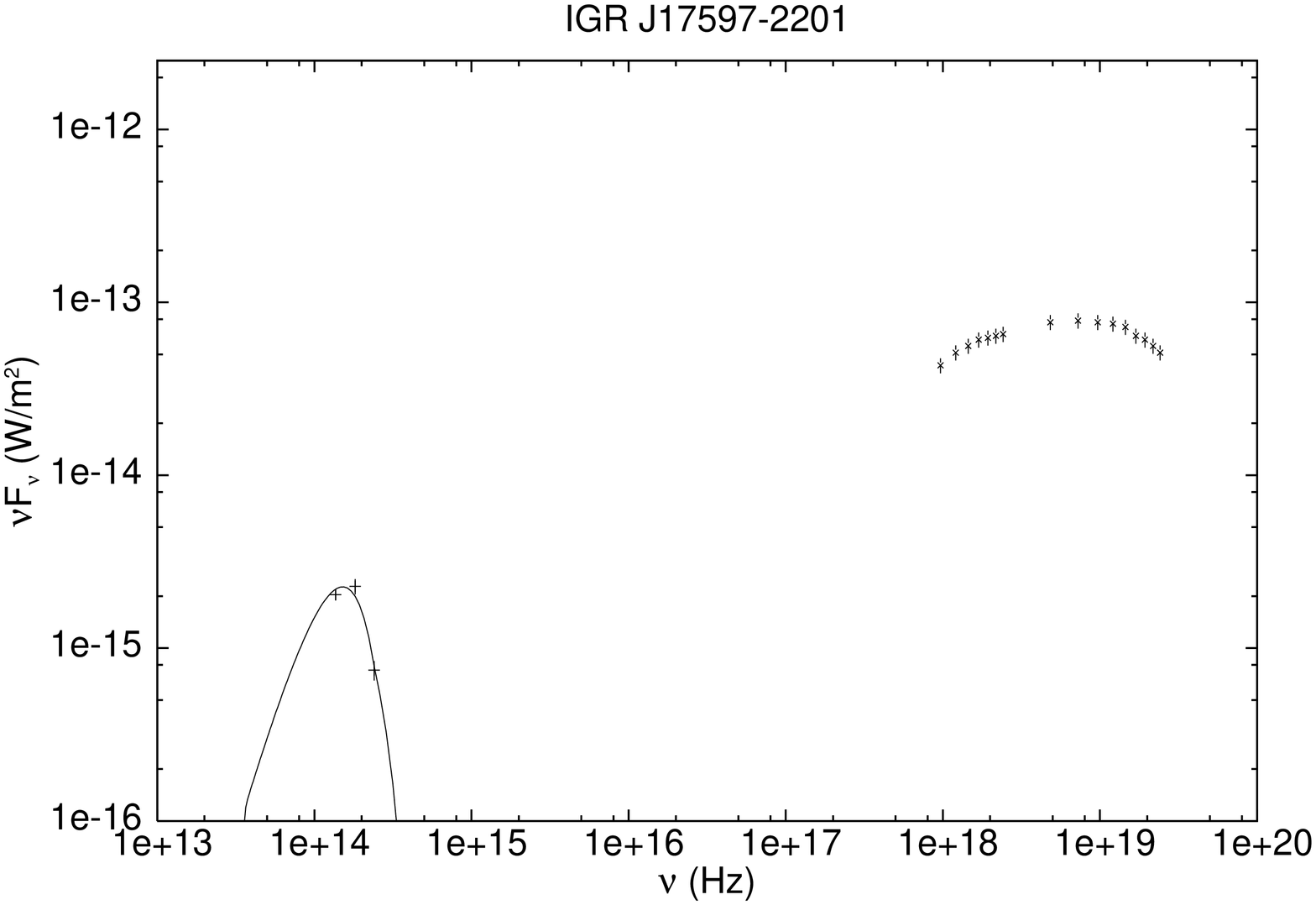}
	\includegraphics[angle=-90,width=5.9cm]{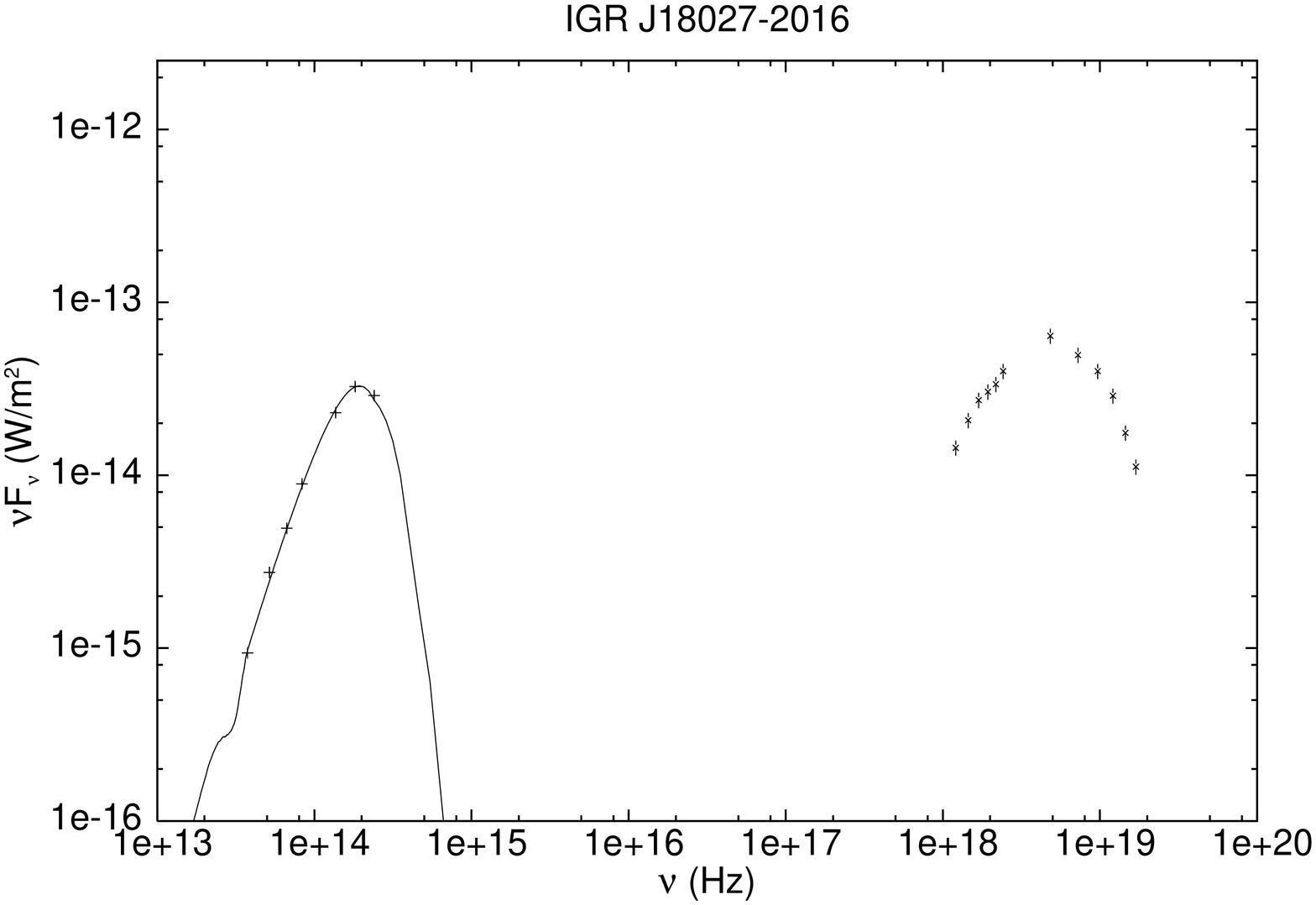}
	\includegraphics[angle=-90,width=5.9cm]{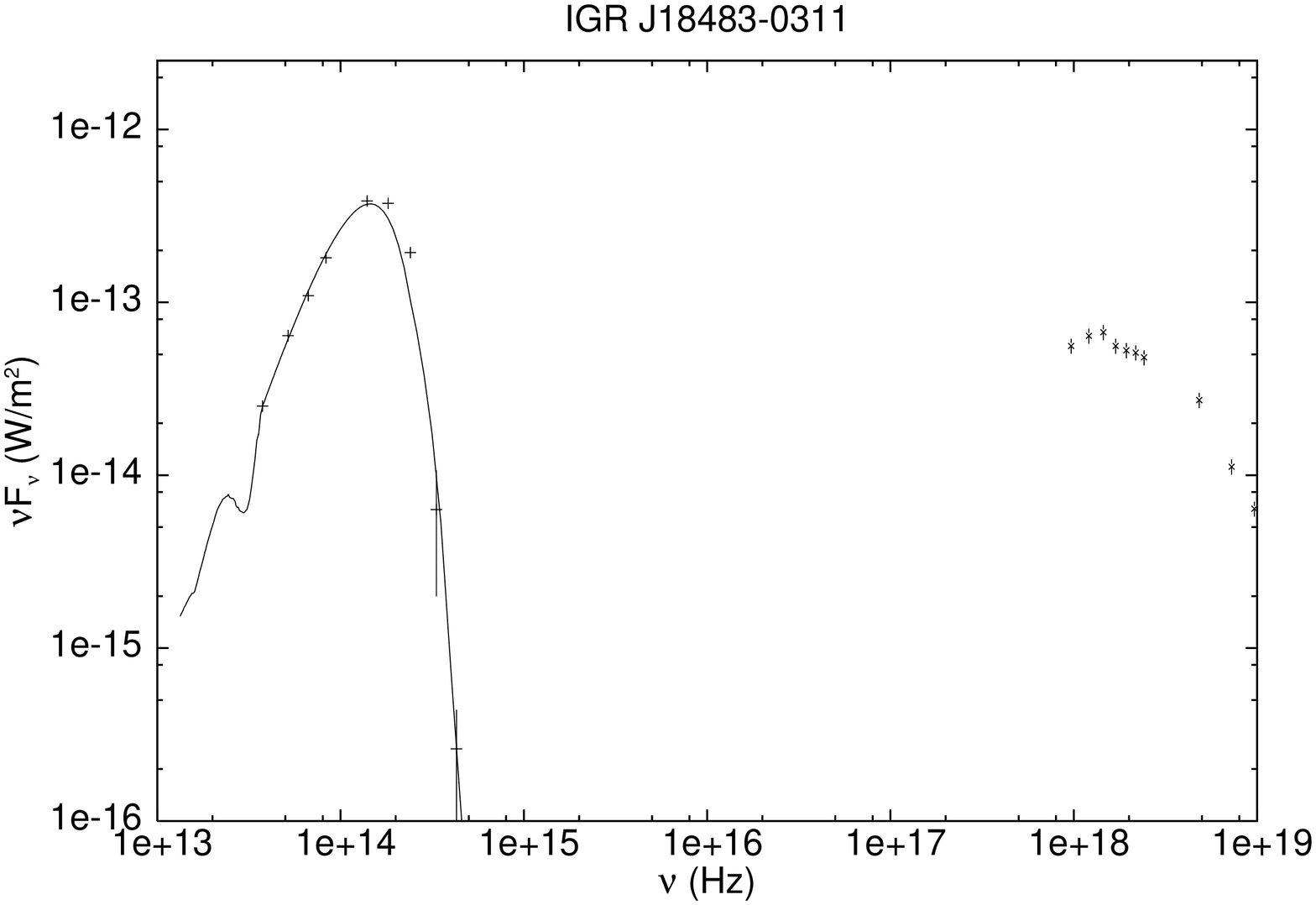}
	\includegraphics[angle=-90,width=5.9cm]{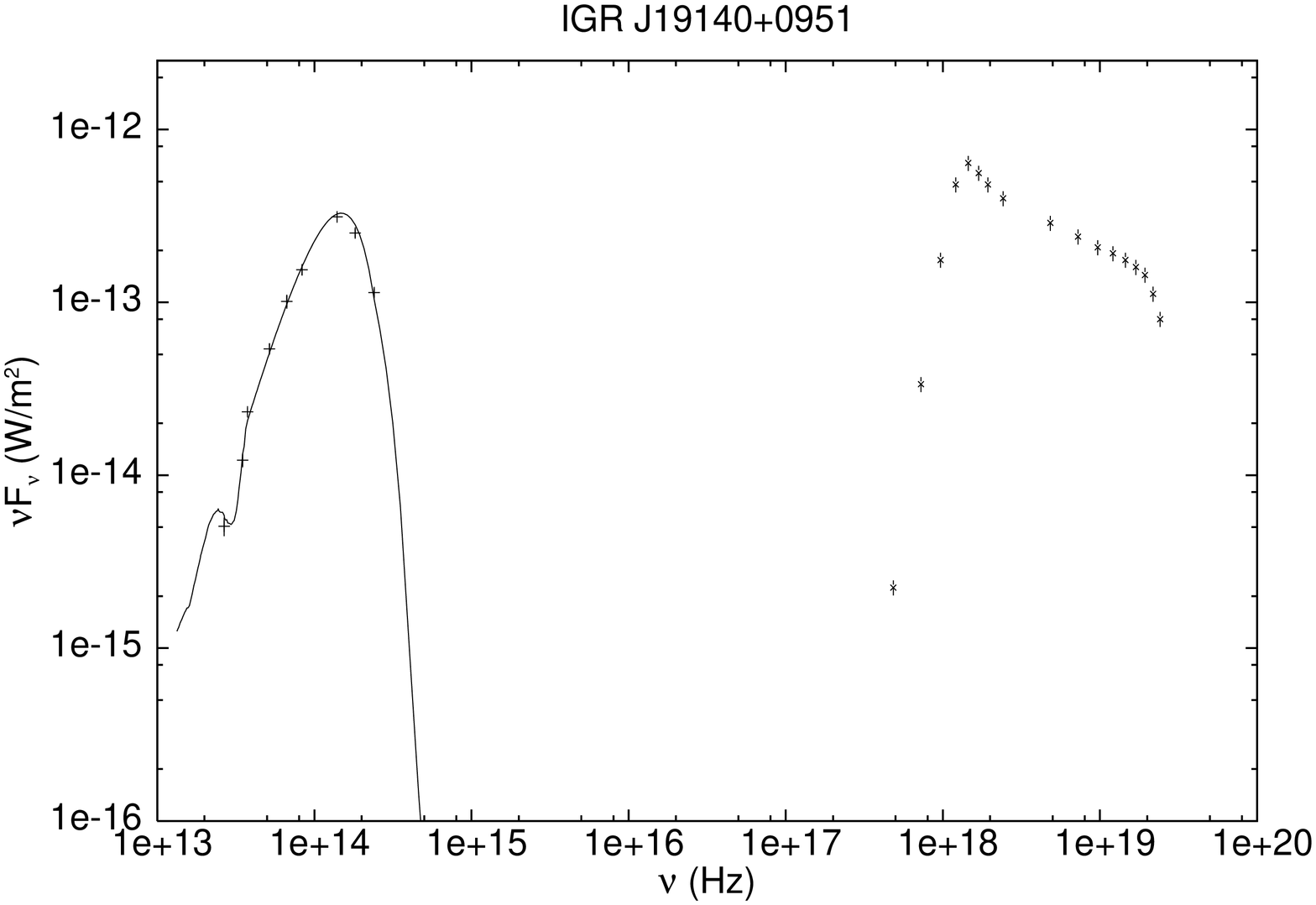}
	\caption[]{\label{figure:SEDs} SEDs of all these {\it
INTEGRAL} sources, showing the observations from hard X-rays to MIR
wavelengths. In each case, we overplot the black body emission
representing the stellar spectral type of the companion star, which is
given in Table \ref{table:results} for the sources IGR~J16393-4643,
IGR~J16418-4532, IGR~J17091-3624, IGR~J17597-2201, IGR~J18027-2016 and
IGR~J18483-0535, and in \cite{rahoui:2008} for the remaining
sources. See Section \ref{section:seds} for more details on the X-ray data.
From top to bottom, and left to right: IGR\,J16320-4751,
IGR\,J16358-4726, IGR\,J16393-4643, IGR\,J16418-4532,
IGR\,J16479-4514,
IGR\,J17091-3624,
IGR\,J17252-3616, 
IGR\,J17391-3021, 
IGR\,J17597-2201,
IGR\,J18027-2016, 
IGR\,J18483-0311
and IGR\,J19140+0951. 
}
\end{figure*}

Once the most likely counterpart was identified through astrometry,
photometry and spectroscopy when available, we were able to build the
optical/NIR SEDs of all sources of our sample, shown in Figure
\ref{figure:SEDs}. While these SEDs are mainly based on optical and
NIR photometry from this paper, we also add MIR observations ($5-20
\microns$) obtained with the VISIR instrument on Melipal, the 8~m
third Unit Telescope (UT3) of the ESO Very Large Telescope (VLT) at
Paranal Observatory, reported in the companion paper by
\cite{rahoui:2008}. We also put MIR data taken from the
{\it Spitzer} GLIMPSE survey, reported in \cite{rahoui:2008}, and in
Table \ref{table:glimpse} for the sources not included in this
companion paper. We include in these SEDs data from
X-ray observations taken with {\it INTEGRAL}/IBIS for all the sources, 
{\it XMM} for IGR\,J16418-4532, IGR\,J17252-3616 and IGR\,J19140+0951,
{\it RXTE} for IGR\,J16358-4726, IGR\,J17091-3624, IGR\,J17597-2201 and
IGR\,J17391-3021, 
{\it ASCA} for IGR\,J16320-4751, IGR\,J16393-4643 and IGR\,J16479-4514,
{\it BeppoSAX} for IGR\,J18027-2016, 
and finally {\it INTEGRAL}/JEM-X for IGR\,J18483-0311.

We fitted the optical to MIR observing data with a black body emission
reproducing the stellar emission. The broad band SEDs of these sources
were modelled using an absorbed black body emission component,
representing the companion star emission ($D_{\ast}$ and $R_{\ast}$ are
respectively the distance and radius of the star):

      \begin{equation}
	\lambda{F(\lambda)}\,=\,\frac{2\pi{h}{c}^2}{{D_{\ast}}^2{\lambda}^4}10^{-{0.4A_\lambda}}\left[\frac{{R_\ast}^2}{e^{\frac{hc}{{\lambda}k{T}_\ast}}-1}\right]\,\,\,\,\,\,\,\textrm{in W.m}^{-2}
      \end{equation}

The free parameters of the fits were the absorption in the V\,band
A$_\textrm{v}$, the companion star black body temperature T$_\ast$ and
its $\frac{\textrm{R}_\ast}{\textrm{D}_{\ast}}$ ratio.  The fits were
performed using a $\chi^2$ minimization.

Best-fitting parameters for individual sources, as well as
corresponding $\chi^2$ and 90\%--confidence ranges of parameters, are
reported in Table \ref{table:fitsseds} for the sources
IGR\,J16393-4643, IGR\,J16418-4532, IGR\,J17091-3624, IGR J17597-2201,
IGR\,J18027-2016 and IGR\,J18483-0311, and in \cite{rahoui:2008} for
the remaining sources.  The IGR\,J16558-5203 SED has not been fitted,
because of its AGN nature. In this Table we also give the interstellar
extinction in magnitude, $A_i$, obtained from the neutral hydrogen
column density $\nhone$, and the X-ray extinction in magnitude, $A_X$,
obtained from X-ray observations.  Both extinctions were converted to
magnitudes using the conversion between $\nh$ and $A_v$ given by
\cite{predehl:1995}. $\nhone$ has been computed using the $\nh$ tool
from HEASARC\footnote{\rm
http://heasarc.gsfc.nasa.gov/cgi-bin/Tools/w3nh/w3nh.pl}
\citep{dickey:1990}. Since $\nhone$ is the total galactic column
density along the line of sight, it is likely overestimated compared
to the real value at the distance of the sources.  

We overplot on the SEDs of Figure \ref{figure:SEDs} the
best-fitted model to our observations.  The dip seen at $\sim
4\times10^{13}$\,Hz is due to silicate absorption present in our
extinction model (more details on this model are given in
\citeauthor{rahoui:2008} \citeyear{rahoui:2008}).

\begin{table*}
\caption{Sample of sources. Name and
coordinates of the sources: position (RA, DEC, J2000.0), galactic
longitude and latitude (l,b), uncertainty (Unc. in arcmin) and
reference (Ref1) of the discovery of the source by {\it INTEGRAL}; and
position (RA, DEC, J2000.0), uncertainty (Unc.) and reference (Ref2)
of the most accurate position by the satellite indicated in
parenthesis. 
References are b:
\cite{bodaghee:2006}, c: \cite{chernyakova:2003}, h:
\cite{hannikainen:2003}, i: \cite{intzand:2006}, ke:
\cite{kennea:2007}, ko: \cite{kouveliotou:2003}, ku:
\cite{kuulkers:2003},
l3: \cite{lutovinov:2003c}, ma: \cite{malizia:2004}, mo:
\cite{molkov:2003}, 
re2: \cite{revnivtsev:2003a}, re4: \cite{revnivtsev:2004}, ro:
\cite{rodriguez:2006}, sg: \cite{sguera:2007}, sm: \cite{smith:2006},
st: \cite{stephen:2005}, su: \cite{sunyaev:2003a}, t3:
\cite{tomsick:2003}, t4: \cite{tomsick:2004b}, w4: \cite{walter:2004a},
w6: \cite{walter:2006}, z: \cite{zurita-heras:2006}.}
\label{table:sources}      
\centering          
\begin{tabular}{c c c c c c c c c c c}     
\hline\hline       
Source & RA & DEC & l & b & Unc. & Ref1 & RA & DEC & Unc. & Ref2 \\
\hline
%
%
%
%
IGR~J16320-4751 & 248.006 & -47.875 & 336.3 & 0.169 & 0.4 & t3 & $16\hour 32\min 01\secp9$ & $-47\adeg 52\amin27\asec$ & $3\asec$ ({\it XMM}) & ro \\
IGR\,J16358-4726 & 248.976 & -47.425 & 337.01 & -0.007 & 0.8 & re2 & $16\hour 35\min 53\secp8$ & $-47\adeg 25\amin41\asecp1$ & $0\asecp6$ ({\it Chandra}) & ko \\
IGR\,J16393-4643 & 249.775 & -46.706 & 338.015 & 0.100 & 0.7 & ma & $16\hour 39\min 05\secp4$ & $-46\adeg 42\amin12\asec$ & $4\asec$ ({\it XMM}) & b \\
IGR\,J16418-4532 & 250.468 & -45.548 & 339.19 & 0.489 & 1.0 & t4 & $16\hour 41\min 51\secp0$ & $-45\adeg 32\amin25\asec$ & $4\asec$ ({\it XMM}) & w6 \\
IGR\,J16479-4514 & 252.015 & -45.216 & 340.16 & 0.124 & 1.4 & mo & $16\hour 48\min 06\secp6$ & $-45\adeg 12\amin08\asec$ & $4\asec$ ({\it XMM}) & w6 \\
%
IGR\,J16558-5203 & 254.010 & -52.062 & 335.687 & -05.493 & 2.0 & w4 & $16\hour 56\min 05\secp73$ & $-52\adeg 03\amin41\asecp18$ & $3\asecp52$ ({\it Swift}) & st \\
%
%
IGR\,J17091-3624 & 257.280 & -36.407 & 349.5 & 2.2 & 0.5 & ku & $17\hour 09\min 07\secp6$ & $-36\adeg 24\amin 24\asecp9$ & $3\asecp6$ ({\it Swift}) & ke \\
%
%
IGR\,J17252-3616 & 261.299 & -36.282 & 351.5 & -0.354 & 0.5 & w4 & $17\hour 25\min 11\secp4$ & $-36\adeg 16\amin58\asecp6$ & $4\asec$ ({\it XMM}) & z \\
IGR\,J17391-3021 & 264.800 & -30.349 & 358.07 & 0.445 & 1.2 & su & $17\hour 39\min 11\secp58$ & $-30\adeg 20\amin37\asecp6$ & $\sim1\asec$ ({\it Chandra}) & sm \\
IGR\,J17597-2201 & 269.935 & -22.026 & 7.581 & 0.775 & 0.6 & l3 & $17\hour 59\min 45\secp7$ & $-22\adeg 01\amin39\asec$ & $4\asec$ ({\it XMM}) & w6 \\
%
%
IGR\,J18027-2016 & 270.661 & -20.304 & 9.418 & 1.044 & 0.7 & re4 & $18\hour 02\min 42\secp0$ & $-20\adeg 17\amin18\asec$ & $4\asec$ ({\it XMM}) & w6 \\
%
%
IGR\,J18483-0311 & 282.068 & -3.171 & 29.760 & -0.744 & 0.8 & c & $18\hour48\min17\secp17$ & $-03\adeg10\amin15\asecp54$ & $3.3\asec$ ({\it Swift}) & sg \\
%
%
IGR\,J19140+0951 & 288.516 & 9.878 & 44.30 & -0.469 & 0.5 & h & $19\hour 14\min 4\secp232$ & $+09\adeg 52\amin58\asecp29$ & $0\asecp6$ ({\it Chandra}) & i \\
\hline                  
\end{tabular}
\end{table*}

%
%
\begin{table*}
\caption{Results in Optical.  We indicate the name of the source, the
 date and UT time of the observations, the airmass (AM),
the exposure time in seconds (ET) and the B, V, R, I
 and Z magnitudes. Z-band magnitudes are instrumental.}
\label{table:optical}      
\centering          
\begin{tabular}{c c c c c c c c c}     
\hline\hline       
Source & Date & AM & ET & B & V & R & I & Z \\ 
\hline                    
%
%
%
\hline
IGR\,J16320-4751 C1 & 2004-07-10T03:04 & 1.1 & 30 & $>22.79\pm0.27$ & $>24.70\pm0.46$ & $>22.10\pm0.26$ & $>22.31\pm0.25$ & $>21.67\pm0.42$ \\
IGR\,J16320-4751 C2 & 2004-07-10T03:04 & 1.1 & 30 & $18.97\pm0.04$ & $16.62\pm0.02$ & $15.32\pm0.01$ & $13.23\pm0.02$ & $13.781\pm0.003$ \\
\hline
IGR\,J16358-4726 & 2004-07-10T04:40 & 1.2 & 30 & $>23.42\pm0.28$ & $>23.67\pm0.33$ & $23.75\pm0.34$ & $20.49\pm0.10$ & $18.59\pm0.08$ \\
IGR\,J16393-4643 & 2004-07-10T06:56 & 1.8 & 30 & $>24.97\pm0.80$ & $21.53\pm0.13$ & $19.62\pm0.05$ & $17.92\pm0.05$ & $16.99\pm0.02$ \\
%
%
IGR\,J17391-3021 & 2004-07-10T06:22 & 1.3 & 1 & $>21.16\pm0.71$ & $14.97\pm0.02$ & $12.94\pm0.02$ & $11.28\pm0.02$ & $10.458\pm0.001$ \\
\hline                  
\end{tabular}
\end{table*}

%
\begin{table*}
\caption{Log of optical and NIR spectra. We indicate the name of the
 source, the telescope used, the date and UT time of the observations,
 the airmass (AM), the exposure time in seconds in optical and NIR
 (blue and red grisms respectively). All spectra were obtained at
 ESO/NTT with EMMI and SofI instruments, except the optical spectra of
 IGR\,J17252-3616 and IGR\,J18027-2016 obtained at ESO/3.6m with
 EFOSC2.}
\label{table:spectra}      
\centering          
\begin{tabular}{c c c c c c c}     
\hline\hline       
Source & Tel & Date & AM & optical & NIR blue grism & NIR red grism \\ 
\hline                    
IGR\,J16320-4751 C1 & NTT  & 2004-07-08T23:18 & 1.255 & - & 2160 & 2160 \\
IGR\,J16320-4751 C2 & NTT  & 2004-07-10T03:36 & 1.112   & 1440 & - & - \\
''                   & NTT  & 2004-07-08T23:18 & 1.255 & - & 2160 & 2160 \\
IGR\,J16358-4726    & NTT  & 2004-07-09T02:06 & 1.052 & - & 2160 & 2160 \\
%
%
IGR\,J16479-4514    & NTT  & 2004-07-10T23:35 & 1.222 & - & 720 & 720 \\
IGR\,J17252-3616    & 3.6m & 2005-10-01T01:49 & 1.683 & 4800 & - & - \\
''                  & NTT  & 2004-07-11T01:08 & 1.083 & - & 720 & 720 \\
IGR\,J17391-3021    & NTT  & 2004-07-10T05:36 & 1.177 & 720 & - & - \\
IGR\,J17391-3021    & NTT  & 2004-07-09T04:24 & 1.038   & -  & 1080 & 1080 \\
IGR\,J18027-2016    & 3.6m & 2005-09-30T02:38 & 2.006 & 1800 & - & - \\
''                  & NTT  & 2004-07-11T02:34 & 1.036 & -    & 720 & 720 \\
IGR\,J19140+0951    & NTT  & 2004-07-11T04:38 & 1.289 & -    & 720 & 720 \\
\hline                  
\end{tabular}
\end{table*}

\begin{table*}
\caption{Results in NIR. We indicate the name of the source (C1, C2,
etc. indicate the different candidates as labeled in the finding
charts of Figure \ref{figure:smallfields}), the date and UT time of
the observations, the airmass, the exposure time (Exptime) in seconds, and
the J, H and K$_{\rm S}$ magnitudes.
} 
\label{table:infrared}      
\centering          
\begin{tabular}{l c c c c c c}     
\hline\hline       
Source & Date & Airmass & Exptime & J & H & K$_{\rm S}$ \\           
 & & & (s) & $1.25 \microns$ & $1.65 \microns$ & $2.2 \microns$ \\    
\hline      
%
%
%
%
%
%
IGR\,J16320-4751 C1& 2004-07-08T23:05 & 1.3 & 108 & 17.24$\pm$0.11 & 13.28$\pm$0.04 & 11.21$\pm$0.05 \\
IGR\,J16320-4751 C2& 2004-07-08T23:05 & 1.3 & 108 & 12.30$\pm$0.03 & 11.41$\pm$0.03 & 11.05$\pm$0.05 \\
IGR\,J16320-4751 C3& 2004-07-08T23:05 & 1.3 & 108 & 18.86$\pm$0.20 & 17.70$\pm$0.17 & 16.85$\pm$0.15 \\
IGR\,J16320-4751 C4& 2004-07-08T23:05 & 1.3 & 108 & - & - & 18.21$\pm$0.33 \\
\hline
IGR\,J16358-4726 & 2004-07-09T01:43 & 1.1 & 450 & 15.46$\pm$0.04 & 13.57$\pm$0.03 & 12.60$\pm$0.05 \\
\hline
IGR\,J16393-4643 C1 & 2004-07-09T06:25 & 1.5 & 108 & 14.62$\pm$0.03 & 13.26$\pm$0.03 & 12.67$\pm$0.05 \\
IGR\,J16393-4643 C2 & 2004-07-09T06:25 & 1.5 & 108 & 16.25$\pm$0.08 & 15.46$\pm$0.09 & 14.88$\pm$0.10 \\
IGR\,J16393-4643 C3 & 2004-07-09T06:25 & 1.5 & 108 & - & 17.87$\pm$0.97 & 15.47$\pm$0.24 \\
IGR\,J16393-4643 C4 & 2004-07-09T06:25 & 1.5 & 108 & 16.24$\pm$0.10 & 14.79$\pm$0.07 & 14.68$\pm$0.13 \\
\hline
IGR\,J16418-4532 C1 & 2004-07-11T05:55 & 1.4 & 108 & 14.03$\pm$0.03 & 16.62$\pm$0.03 & 11.61$\pm$0.05 \\
IGR\,J16418-4532 C2 & 2004-07-11T05:55 & 1.4 & 108 & 16.48$\pm$0.09 & 19.44$\pm$0.26 & 14.76$\pm$0.09 \\
IGR\,J16418-4532 C3 & 2004-07-11T05:55 & 1.4 & 108 & 17.48$\pm$0.17 & 20.60$\pm$0.51 & 16.12$\pm$0.27 \\
IGR\,J16418-4532 C4 & 2004-07-11T05:55 & 1.4 & 108 & 18.06$\pm$0.17 & 21.05$\pm$0.51 & 16.96$\pm$0.30 \\
\hline
IGR\,J16479-4514 C1 & 2004-07-10T23:21 & 1.3 & 108 & 13.06$\pm$0.02 & 10.92$\pm$0.02 & 9.79$\pm$0.05 \\
IGR\,J16479-4514 C2 & 2004-07-10T23:21 & 1.3 & 108 & 16.22$\pm$0.07 & 15.53$\pm$0.08 & 15.02$\pm$0.10 \\
\hline
IGR\,J16558-5203 & 2004-07-11T06:18 & 1.5 & 108 & 12.82$\pm$0.03 & 11.66$\pm$0.03 & 10.52$\pm$0.05 \\
\hline
IGR\,J17091-3624 C1 & 2004-07-11T06:35 & 1.5 & 108 & 16.73$\pm$0.05 & 15.83$\pm$0.05 & 15.36$\pm$0.07 \\
IGR\,J17091-3624 C2 & 2004-07-11T06:35 & 1.5 & 108 & 18.19$\pm$0.11 & 17.12$\pm$0.12 & 16.65$\pm$0.14 \\
%
\hline
%
%
%
IGR\,J17252-3616 C1 & 2004-07-11T00:39 & 1.2 & 108 & 14.19$\pm$0.02 & 11.90$\pm$0.03 & 10.67$\pm$0.05 \\
IGR\,J17252-3616 C2 & 2004-07-11T00:39 & 1.2 & 108 & 17.40$\pm$0.10 & 15.13$\pm$0.05 & 13.92$\pm$0.07 \\
IGR\,J17252-3616 C3 & 2004-07-11T00:39 & 1.2 & 108 & -          & 15.00$\pm$0.06 & 13.82$\pm$0.09 \\
IGR\,J17252-3616 C4 & 2004-07-11T00:39 & 1.2 & 108 & -          & 18.61 1.17 & 15.17$\pm$0.13 \\
\hline
IGR\,J17391-3021 & 2004-07-09T04:08 & 1.0 & 108 & 9.09$\pm$0.02 & 8.70$\pm$0.02 & 8.16$\pm$0.05 \\
\hline
IGR\,J17597-2201 C1 & 2004-07-11T01:36 & 1.1 & 108 & 16.78$\pm$0.10 & 14.84$\pm$0.07 & 14.13$\pm$0.08 \\
IGR\,J17597-2201 C2 & 2004-07-11T01:36 & 1.1 & 108 & 16.13$\pm$0.08 & 14.24$\pm$0.06 & 13.49$\pm$0.07 \\
IGR\,J17597-2201 C3 & 2004-07-11T01:36 & 1.1 & 108 & 18.50$\pm$0.23 & 17.36$\pm$0.28 & 17.28$\pm$0.44 \\
IGR\,J17597-2201 C4 & 2004-07-11T01:36 & 1.1 & 108 & 17.80$\pm$0.13 & 16.62$\pm$0.12 & 16.30$\pm$0.22 \\
IGR\,J17597-2201 C5 & 2004-07-11T01:36 & 1.1 & 108 & 19.19$\pm$0.25 & 17.89$\pm$0.20 & 17.22$\pm$0.36 \\
IGR\,J17597-2201 C6 & 2004-07-11T01:36 & 1.1 & 108 & 17.62$\pm$0.25 & 15.59$\pm$0.12 & 14.68$\pm$0.11 \\
\hline
%
%
%
IGR\,J18027-2016 C1 & 2004-07-11T02:08 & 1.1 & 180 & 12.81$\pm$0.02 & 11.95$\pm$0.03 & 11.50$\pm$0.05 \\
IGR\,J18027-2016 C2 & 2004-07-11T02:08 & 1.1 & 180 & 13.97$\pm$0.03 & 13.3$\pm$0.04 & 13.14$\pm$0.06 \\
IGR\,J18027-2016 C3 & 2004-07-11T02:08 & 1.1 & 180 & 17.61$\pm$0.22 & 16.75$\pm$0.23 & 16.08$\pm$0.29 \\
IGR\,J18027-2016 C4 & 2004-07-11T02:08 & 1.1 & 180 & 16.88$\pm$0.14 & 15.65$\pm$0.15 & 15.20$\pm$0.24 \\
\hline
%
%
%
IGR\,J18483-0311 & 2004-07-11T07:44 & 1.7 & 180 & 10.77$\pm$0.03 & 9.21$\pm$0.03 & 8.39$\pm$0.04 \\
\hline
%
%
%
IGR\,J19140+0951 & 2004-07-11T04:26 & 1.3 & 108 & 11.32$\pm$0.02 & 9.73$\pm$0.02 & 8.84$\pm$0.04 \\
\hline                  
\end{tabular}
\end{table*}
%

   \begin{table*}
	\caption{\label{table:glimpse}  GLIMPSE fluxes (in mJy) 
	  for the sources IGR~J16393-4643, IGR~J16418-4532, 
	  IGR~J18027-2016 and IGR~J18483-0535. 
	  The fluxes for the other
	sources are given in \cite{rahoui:2008}.}
	$$
  \begin{array}{c c c c c}
     \hline
     \hline
     \textrm{Sources}&3.6\,\mu\textrm{m}&4.5\,\mu\textrm{m}&5.8\,\mu\textrm{m}&8\,\mu\textrm{m}\\
     \hline
     \textrm{IGR~J16393-4643}&3.53\pm0.52&2.89\pm0.47& - & - \\
     \hline
     \textrm{IGR~J16418-4532}&12.46\pm0.90&9.45\pm0.58&5.57\pm0.58&3.58\pm0.41\\
     \hline
     \textrm{IGR~J18027-2016}&10.70\pm0.28&7.40\pm0.18&5.30\pm0.26&2.50\pm0.05\\
     \hline
     \textrm{IGR~J18483-0535}&217.00\pm8.90&164.00\pm7.20&124.00\pm5.50&67.00\pm2.10\\
     \hline
   \end{array}
   $$
     \end{table*} 

   \begin{table*}
	\caption{\label{table:fitsseds} Summary of parameters we used
	  to fit the SEDs of the sources. We give their
	  name, the interstellar extinction in magnitudes
	  $\textrm{A}_\textrm{i}$, the X-ray extinction of the source
	  in magnitudes $\textrm{A}_\textrm{x}$ and the
	  parameters of the fit: the extinction in the optical
	  $\textrm{A}_\textrm{v}$, the temperature $\textrm{T}_\ast$
	  and the $\frac{\textrm{R}_{\ast}}{\textrm{D}_{\ast}}$ ratio
	  of the companion (more details on these parameters are
	  given in the text). 
	  The 90\%-confidence ranges of 
	  these parameters are given in parenthesis.
	  We also give the reduced $\chi^2$. The parameters for the other
	sources are given in \cite{rahoui:2008}.}
	$$
	\begin{array}{c c c c c c c}
	  \hline
	  \hline
	  \textrm{Sources}&\textrm{A}_\textrm{i}&\textrm{A}_\textrm{x}&\textrm{A}_\textrm{v}&\textrm{T}_{\ast} (K)&\frac{\textrm{R}_{\ast}}{\textrm{D}_{\ast}}&\chi^2/\textrm{dof}\\
	  \hline
	  \textrm{IGR~J16393-4643}&11.71&133.61&11.5(10.9-11.8)&24400(12800-34200)&2.21(1.81-3.13)\times10^{-11}&3.25/2\\
	  \hline
	  \textrm{IGR~J16418-4532}&10.05&53.45&14.5(13.1-14.9)&32800(10600-36000)&3.77(2.64-6.85)\times10^{-11}&1.5/4\\
	  \hline
	  \textrm{IGR~J17091-3624}&4.13&5.34&6.8(2.0-9.6)&6900(3000,34300)&1.15(0.50-2.00)\times10^{-11}&1.005/1\\
	  \hline
	  \textrm{IGR~J17597-2201}&6.251&24.05&16.1(13.1-17.1)&31700(6500-36000)&1.28(1.30-3.00)\times10^{-11}&2.9/1\\
	  \hline
	  \textrm{IGR~J18027-2016}&5.56&48.42&8.8(8-9.1)&20800(12800-32200)&3.7(2.8-4.76)\times10^{-11}&6.00/4\\
	  \hline
	  \textrm{IGR~J18483-0535}&8.66&148.1&17.4(16.9-18.3)&22500(16400-36000)&2.15(1.75-2.75)\times10^{-10}&10/5\\
	  \hline
	\end{array}
	$$
      \end{table*}      

\section{Results on individual sources} \label{section:results}

All the sources studied in this paper were discovered with the
IBIS/ISGRI detector onboard the {\it INTEGRAL} observatory.  The
sample of 13 sources, along with their position, uncertainty, and
references about their discovery and position are given in Table
\ref{table:sources}. We present in the following our results on each
source, for which we followed the same strategy. We first observed the
field in optical and NIR, we performed accurate astrometry, and we
derived the photometry of the candidates inside the X-ray satellite
error circle. We then analyzed the optical/NIR spectrum of the most
likely candidate, when available. We give the results of the
optical-MIR SED fitting.

     \subsection{IGR\,J16320-4751}

IGR\,J16320-4751 was discovered on 2003 February by
\cite{tomsick:2003} at the position $\mathrm{RA} = 16\hour32\minp0$, Decl. =
$-47\adeg51\amin$ (equinox J2000.0; uncertainty $2\amin$). Follow-up
{\it XMM-Newton} observations localized the source at
$16\hour32\min01\secp9$ $-47\adeg52\amin27\asec$ with 3$\asec$
accuracy (\citeauthor{rodriguez:2003} \citeyear{rodriguez:2003};
\citeauthor{rodriguez:2006} \citeyear{rodriguez:2006}). It is a
heavily absorbed variable source with
$\nh\sim2.1\times10^{23}\ \cmmoinsdeux$, and a hard X-ray spectrum
fitted by an absorbed power-law, with $\Gamma\sim1.6$
\citep{rodriguez:2006}.  Soft X-ray pulsations have been detected from
this source at a period of P$\sim1309\pm40$~s with {\it XMM-Newton}
and P$\sim1295\pm50$~s with {\it ASCA} observations, these pulsations
being the signature of an X-ray pulsar \citep{lutovinov:2005b}.  An
orbital period of $8.96\pm0.01\ \mathrm{days}$ was found from a {\it Swift}/BAT
lightcurve extending from 2004 December 21 to 2005 September 17
\citep{corbet:2005}, and of $8.99\pm0.05\ \mathrm{days}$ with {\it INTEGRAL}
\citep{walter:2006}. Putting the spin and orbital
periods of this source on a Corbet diagram \citep{corbet:1986} 
suggests a supergiant HMXB nature. IGR\,J16320-4751 might have
been persistent for at least 8 years, since this source is the
rediscovery of a previously known {\it ASCA} source AX\,J1631.9-4752.

We performed accurate astrometry of the field (rms of fit=$0\asecp49$), and overplot the 3$\asec$ {\it XMM-Newton} error circle
in the finding chart of Figure \ref{figure:smallfields}.  We give the
infrared magnitudes of all the candidate counterparts in Table
\ref{table:infrared}. Two candidate counterparts had been proposed for
this source \citep{rodriguez:2003}, however the {\it XMM-Newton}
3$\asec$ error circle made the ambiguity disappear, accurately
localizing the candidate labeled 1 in Figure \ref{figure:smallfields},
and making it the most likely counterpart (2MASS\,J16320215-4752289;
\citeauthor{rodriguez:2006} \citeyear{rodriguez:2006}). This result is
in agreement with \cite{negueruela:2007} rejecting candidate 2 on the
basis of 2MASS photometry. In addition, Candidate 1 is much more
absorbed than Candidate 2, as can be seen from the optical and NIR
magnitudes given for both candidate counterparts in Tables
\ref{table:optical} and \ref{table:infrared}, since
Candidate 1 is invisible in the optical, but becomes as bright as
Candidate 2 in the K$_{\rm S}$ band. There are at
least two faint candidate counterparts (labeled 3 and 4) inside the
error circle, therefore it would be useful to obtain a more accurate
position to unambiguously pinpoint the correct counterpart.  However,
the faintness of Candidates 3 and 4 tends to rule them out as
counterparts, and in the following we will consider
Candidate 1 as the most likely counterpart of this source.

NIR spectra of candidate 1 of IGR\,J16320-4751 are shown in Figure
\ref{figure:nirspec1}. We report the detected lines in Table
\ref{table:spectralines}. There are only a few lines visible in the
blue NIR spectrum, probably because it is very faint and absorbed.
The red NIR spectrum exhibits a very red continuum, and the presence
of absorption and emission lines: the Pa(7-3) emission line, the
Brackett series with P-Cygni profiles between 1.5 and $2.17 \microns$,
and $\ion{He}{i}$ at $2.166 \microns$ (perhaps with P-Cygni profile).
The presence of these narrow and deep Paschen and $\ion{He}{i}$ lines,
associated with P-Cygni profiles, are typical of early-type stars, and
more precisely of luminous supergiant OB stars
(\citeauthor{caron:2003} \citeyear{caron:2003},
\citeauthor{munari:1999} \citeyear{munari:1999}), which is therefore
the likely spectral type of the companion star. Furthermore, the
presence of a wide Br $\gamma$ emission line constrains the spectral
type to a O supergiant or even O hypergiant \citep{hanson:2005}.  We also
took optical and NIR spectra of Candidate 2: they do not exhibit any
emission lines, and seem typical of a late-type star. Such a spectral
type would therefore be hard to reconcile with the wind accretion hard
X-ray spectra and the localisation of this source in the Corbet
diagram. Therefore both astrometry and spectroscopy allow us to
exclude Candidate 2 as a candidate counterpart.  These results
strengthen Candidate 1 as the real counterpart of
IGR\,J16320-4751. From the nature of the companion star and of the
compact object, we derive that this source belongs to the very
obscured supergiant HMXB class, hosting a neutron star. This result is
also in agreement with the fit of its SED, computed in
\cite{rahoui:2008} and shown in Figure \ref{figure:SEDs}.

     \subsection{IGR\,J16358-4726}


IGR\,J16358-4726 was discovered on 2003 March 19 by
\cite{revnivtsev:2003a} at the position (RA DEC J2000.0) = ($16\hour35\minp8$,
$-47\adeg26\amin$, $1\aminp5$ uncertainty).
This hard X-ray source was observed for 25700\,s
serendipitously with {\it Chandra} during a scheduled observation of
SGR 1627-41 on 2003 March 24.  {\it Chandra} localized the source at
($16\hour35\min53\secp8$, $-47\adeg25\amin41\asecp1$) with $0\asecp6$
accuracy \citep{kouveliotou:2003}. We point out however that the
uncertainty may be somewhat underestimated since the source was
$9\aminp7$ from the {\it Chandra} aimpoint so that the point-spread
function is significantly broadened.  This source is a transient
source, its hard X-ray spectrum being well fitted with a heavily
absorbed power-law: $\Gamma\sim0.5$ and
$\nh\sim3.3\,10^{23}\cmmoinsdeux$, with the presence of
Fe\,K$\alpha$ emission line \citep{patel:2007}. By performing detailed
spectral and timing analysis of this source using multi-satellite
archival observations, \cite{patel:2007} have detected 
$5880\pm50$\,s periodic variations, 
which could be due either to the spin of a neutron
star, or to an orbital period, and they identified a $94$~s spin-up in
8 days, corresponding to a mean spin period derivative of
$1.9\times10^{-4}$~s/s, pointing to a neutron star origin. Assuming
that this spin up is due to accretion, they estimate the source
magnetic field to be between $10^{13}$ and $10^{15}$\,G,
suggesting that the compact object might be a magnetar. These
observations suggest that this source is an X-ray pulsar at a distance
of $\sim 6-8$\,kpc. No radio emission has been detected.

A 2MASS counterpart has been suggested by \cite{kouveliotou:2003}
based on the accurate {\it Chandra} position
(2MASS\,J16355369-4725398, with J=15.41, H=13.44,
K=12.59).
We performed accurate astrometry of the field (rms of
fit=$0\asecp45$), and overplot the $0\asecp6$ {\it Chandra} error
circle, as shown in Figure \ref{figure:smallfields}. The 2MASS
counterpart is at 1.2$\asec$ from the centre, therefore outside the
error circle.  On the other hand, this 2MASS counterpart might be a
blended object, since it shows an extension towards the east, right at
the position of the error circle. A better spatial resolution would
allow us to confirm or not whether the 2MASS candidate is the real
counterpart of this source. However, its brightness favors this
candidate, and in the following we consider it as the candidate
counterpart. Optical and NIR magnitudes of this 2MASS candidate are
given in Tables \ref{table:optical} and \ref{table:infrared}
respectively.

The NIR spectra of IGR\,J16358-4726 are shown in Figure
\ref{figure:nirspec1}.  We report the detected lines in Table
\ref{table:spectralines}. The NIR spectrum is very faint and extremely
absorbed, however we detect some lines even in the blue part of its
spectrum, mainly $\ion{He}{ii}$ emission lines.  The red part of the NIR
spectrum exhibits a red continuum, and the presence of absorption and
emission lines, with tentative P-Cygni profiles: the H Brackett series
with P-Cygni profiles between 1.5 and $2.2 \microns$, and $\ion{He}{i}$ and
$\ion{He}{ii}$ absorption lines.  The presence of these lines, associated with
P-Cygni profiles, are typical of an OB supergiant star, which is
therefore probably the spectral type of the companion star. In this
case it would be a supergiant HMXB.  In addtion, we clearly detect the
forbidden [$\ion{Fe}{ii}$] line at $2.22\microns$ (and tentatively the allowed $\ion{Fe}{ii}$
line at $1.98\microns$), suggesting that the companion star is a
sgB[e] star.  Furthermore, it is interesting to note that
\cite{rahoui:2008} propose, using an independent method of SED
fitting, that the companion might be a sgB[e] star. Our result would
therefore be in agreement with the fit of its SED, shown in Figure
\ref{figure:SEDs}.

     \subsection{IGR\,J16393-4643}

IGR\,J16393-4643 was discovered by \cite{malizia:2004} at the position
(RA, DEC, J2000.0) = ($16\hour39\minp3, -46\adeg43\amin$) ($2\amin$
uncertainty).  The improved position from {\it XMM-Newton}/EPIC is (RA
DEC, J2000.0) = ($16\hour39\min05\secp4, -46\adeg42\amin12\asec$)
($4\asec$ uncertainty) which is compatible with that of 2MASS
J16390535-4642137 \citep{bodaghee:2006}.  It is a persistent,
heavily-absorbed ($\nh = 2.5 \times 10^{23} \cmmoinsdeux$), and hard
($\Gamma=1.3\pm1.0$) wind-accreting pulsar, and a pulse period of
$912.0\pm0.1\ \mathrm{s}$ was discovered in the ISGRI and EPIC light curves,
characteristic of a spin period of an X-ray pulsar
\citep{bodaghee:2006}.
The high column density and hard spectral index suggest that
IGR\,J16393-4643 is an HMXB. This source exhibits large variations in
intensity, and shows X-ray lines.  An orbital period of
$3.6875\pm0.0006$~days has been detected in {\it Rossi-XTE} data,
implying a mass function of $6.5\pm1.1\Msol$ (or up to $14\Msol$ if
the orbit is eccentric; \citeauthor{thompson:2006}
\citeyear{thompson:2006}), this lower limit on the mass confirming
that the system is an HMXB.

We performed accurate astrometry of the field (rms of
fit=$0\asecp54$).  Inside the error circle, there is the 2MASS
candidate counterpart proposed by \cite{bodaghee:2006} (2MASS
J16390535-4642137), and in addition there are 3 more candidates,
labeled 1 to 4 in Figure \ref{figure:smallfields}.  They are located
at 1.7, 3.1, 3.4 and $3\asecp3$ respectively from the centre of the
error circle.  Optical and NIR magnitudes of these candidates are
given in Tables \ref{table:optical} and \ref{table:infrared}
respectively. A more precise localisation of the hard X-ray source is
therefore necessary to know which is the real counterpart of the
source, however Candidate 1 seems to be the most likely counterpart
based on the astrometry and on its NIR brightness.

We fitted its SED with the model described in Section
\ref{section:seds}, taking the optical and NIR magnitudes of Candidate
1. We obtain a stellar temperature $T_\mathrm{\ast} = 24400\
\mathrm{K}$, typical of a B spectral type companion star. The other
parameters are given in Table \ref{table:fitsseds}, and the SED is
shown in Figure \ref{figure:SEDs}. By taking the $R_\ast/D_\ast$
($=2.21\times10^{-11}$) minimizing $\chi^2$, and assuming a radius
$R_\mathrm{\ast} = 10\ \Rsol$ typical of a BIV-V spectral type
companion star, we derive a distance of $D_\ast = 10.6\ \kpc$.  On the
other hand, by taking the minimum radius of a B supergiant star,
i.e. at least $20 \Rsol$, we derive a distance of nearly $20.4\kpc$,
probably too large to be plausible. The fit therefore is entirely
consistent with its HMXB nature and favours a BIV-V companion star.

     \subsection{IGR\,J16418-4532}

     IGR\,J16418-4532 was discovered on 2003 February 1-5 at the
     position (RA DEC J2000.0) = ($16\hour41\minp8$,
     $-45\adeg32\amin$, $2\amin$ uncertainty), towards the Norma
     region \citep{tomsick:2004b}.  {\it XMM-Newton} localized the
     source at ($16\hour41\min51\secp0, -45\adeg32\amin25\asec$) with
     4$\asec$ accuracy \citep{walter:2006}.  {\it XMM-Newton}
     observations have shown that it is a heavily absorbed X-ray
     pulsar exhibiting a column density of $\nh\sim 1.0\times10^{23}
     \cmmoinsdeux$, a peak-flux of $\sim\,$80 mCrab (20-30 keV), and a
     pulse period of $1246\pm100$~s \citep{walter:2006}.  This source
     is an SFXT candidate, as proposed by \cite{sguera:2006} using
     {\it INTEGRAL} observations.  A $3.75\ \mathrm{day}$ modulation
     was found in {\it Rossi-XTE}/ASM and {\it Swift}/BAT lightcurves,
     with a possible total eclipse, which would suggest either a high
     binary inclination, or the presence of a supergiant companion
     star \citep{corbet:2006}.  The latter case would be consistent
     with the position of this object in the Corbet diagram.

     We performed the astrometry of this source with 1139 2MASS stars
     (rms of fit=$0\asecp38$).  The field is shown in Figure
     \ref{figure:smallfields}.  The centre of the {\it XMM-Newton}
     error circle is $2\asecp2$ away from a bright 2MASS source
     (2MASS~J16415078-4532253, labelled Candidate 1), as pointed out
     by \cite{walter:2006}, however there are at least three more
     possible counterparts in the $4\asec$ {\it XMM-Newton} error
     circle, not present in the 2MASS catalogue, at 1.9, 1.3 and
     $2\asecp1$ respectively from the centre of the error
     circle. These additional candidate counterparts are labeled 2 to
     4 respectively. We give their NIR magnitudes in
     Table \ref{table:infrared}.
     From the brightness in NIR we favor Candidate 1 as the candidate
     counterpart.

     The fit of its SED, shown in Figure \ref{figure:SEDs}, and taking
     optical and NIR magnitudes as described in Section
     \ref{section:seds}, gives a stellar temperature $T_\mathrm{\ast}
     = 32800\ \mathrm{K}$, suggesting an OB spectral type companion
     star. The other parameters of the fit are given in Table
     \ref{table:fitsseds}. The $R_\ast/D_\ast$ ratio which minimizes
     $\chi^2$ ($=3.77\times10^{-11}$) allows us to derive a minimal
     distance of 13\,kpc for a supergiant, which is plausible. This
     source is therefore an HMXB, and could be a supergiant HMXB,
     consistent with its position in the Corbet diagram. We point out
     however its membership of the SFXT class is uncertain, based on
     its X-ray behaviour (Zurita Heras \& Chaty in prep.).

     \subsection{IGR\,J16479-4514}

     IGR\,J16479-4514 was discovered on 2003 August 8-9 at the
     position (RA DEC J2000.0) = ($16\hour47\minp9, -45\adeg14\amin$),
     uncertainty $\sim 3\amin$, by \cite{molkov:2003}. {\it
       XMM-Newton} observations localized the source at
     ($16\hour48\min06\secp6, -45\adeg12\amin08\asec$) with a $4\asec$
     accuracy \citep{walter:2006}. These {\it XMM-Newton} observations
     have shown a column density of $\nh = 0.77-1.2 \times 10^{23}\
     \cmmoinsdeux$. This source has recurrent outbursts,
     making it a fast transient with a peak-flux of $\sim\,$120 mCrab
     (20-60 keV) (\citeauthor{sguera:2005} \citeyear{sguera:2005};
     \citeauthor{sguera:2006} \citeyear{sguera:2006}).  There is an IR
     source IRAS\,16441-4506 in the error circle.

     We performed accurate astrometry of the field (rms of
     fit$=0\asecp41$), shown in Figure \ref{figure:smallfields}.
     There is a bright J\,=\,12.9, H\,=\,10.8 and K\,=\,9.8 2MASS
     source (2MASS\,J16480656-4512068) $1\asecp1$ away from the centre
     of the {\it XMM-Newton} error circle, which is the closest 2MASS
     source, suggested by \cite{walter:2006} as being a candidate
     counterpart.  In addition, we find another candidate counterpart
     inside the error circle, at $2\asecp6$ from the centre of the
     error circle, labeled 2 in Figure \ref{figure:smallfields}.  The
     NIR magnitudes of both candidate counterparts are given in Table
     \ref{table:infrared}.  However, on the basis of the astrometry
     (close to the error circle centre) and the photometry (NIR
     brightness) we favour Candidate 1 as the candidate counterpart.

     NIR spectra of Candidate 1 are shown in Figure
     \ref{figure:nirspec1}.  We report the detected lines in Table
     \ref{table:spectralines}.  We find emission of H (Brackett
     series) between $1.5-2.17\microns$, and also $\ion{He}{i}$,
     $\ion{He}{ii}$ and $\ion{Fe}{ii}$ emission lines.  These NIR
     spectra are typical of a supergiant OB star, therefore
     strengthening Candidate 1 as the likely counterpart of this
     source, the companion star of IGR~J16479-4514 having an OB
     spectral type. Furthermore, the presence of a Br $\gamma$
     emission line and $\ion{He}{i}$ at $2.11\microns$ absorption line
     constrains the spectral type to a late O or peculiar early B supergiant
     \citep{hanson:2005}.  This source would therefore be a supergiant
     HMXB system, probably belonging to the SFXT class. Its SED,
     consistent with the HMXB nature of the system, is shown in Figure
     \ref{figure:SEDs}.

     \subsection{IGR\,J16558-5203}

IGR\,J16558-5203 was discovered at the position (RA DEC J2000.0) =
($16\hour55\minp8, -52\adeg03$) with a 2$\amin$ uncertainty
\citep{walter:2004a}.  This source has a {\it ROSAT} counterpart (1RXS
J165605.6-520345) at the position ($16\hour56\min05\secp60,
-52\adeg03\amin45\asecp5$), which allowed them to reduce the uncertainty to
8$\asec$ \citep{stephen:2005}. Recently, the position has been refined
by {\it Swift} observations at ($16\hour56\min05\secp73,
-52\adeg03\amin41\asecp18$) with a $3\asecp52$ error circle radius
\citep{malizia:2007}.

We performed accurate astrometry of the field, shown in Figure
\ref{figure:smallfields}, and find that there is a bright 2MASS object
inside the $3\asecp52$ {\it Swift} error circle.  The NIR magnitudes
of this object are given in Table \ref{table:infrared}.  This object
is clearly extended on the NIR images, suggesting an extragalactic
source.  This is in agreement with the result from
\cite{masetti:2006}, who showed that this object is a Seyfert 1.2 AGN
at a redshift of 0.054, exhibiting H$\beta$ and $\ion{O}{iii}$ emission lines.

     \subsection{IGR\,J17091-3624}


IGR\,J17091-3624 was discovered at the position (RA DEC J2000.0) =
($17\hour09\minp1, -36\adeg24\amin38\asec$), uncertainty $\sim3\amin$
\citep{kuulkers:2003}. Follow-up analysis of archival data from the TTM
telescope aboard the KVANT module of the MIR orbital station revealed that
this source had been detected in several observations performed on
Oct.1-10, 1994, at the position (RA DEC J2000.0) =
($17\hour09\min06\sec, -36\adeg24\amin07\asec$), error radius about
0.8$\amin$.  This TTM position is within 0.6$\amin$ of the {\it
INTEGRAL} position \citep{revnivtsev:2003b}. It is also coincident
with the SAX source 1SAX J1709-36. A variable radio counterpart has
been detected in follow-up radio observations at the position
($17\hour09\min02\secp3\pm0.4, -36\adeg23\amin33\asec$), giving an
error circle of 5$\asec$ radius \citep{rupen:2003a}. Since this source
exhibited some radio emission it has been classified as a
galactic X-ray binary, and a candidate microquasar, probably hosting a
black hole.  It exhibits a large variability in X-rays, and it is
uncertain if this source is a Be/X-ray binary or an LMXB.
\cite{capitanio:2006} confirm the low absorption
($\nh \leq 1\times10^{22}\ \cmmoinsdeux$) and Comptonised spectrum of this
source.  \cite{negueruela:2007} proposed the 2MASS\,J17090199-3623260
source as a candidate counterpart, based on photometric
catalogues. They report that this object is a late F8\,V companion
star.  However, a more accurate position has been obtained by
\cite{kennea:2007}: (RA DEC J2000.0) = ($17\hour09\min07\secp6,
-36\adeg24\amin24\asecp9$), error radius about $3\asecp6$, which
excludes the association of the high energy source with the radio
source.

We have performed accurate astrometry of the field, shown in Figure
\ref{figure:smallfields}, and we overplot the $3\asecp6$ {\it Swift} error
circle.  We find that the claimed counterpart 2MASS\,J17090199-3623260
of \cite{negueruela:2007} is well outside this error circle; we
can therefore reject it. We report the
discovery of two blended candidate counterparts 
inside the error circle, labeled 1
and 2 in Figure \ref{figure:smallfields} respectively, located at
$0\asecp5$ and $0\asecp4$ from the centre of the error circle.  We
give the NIR magnitudes of these two candidate counterparts in
Table \ref{table:infrared}.  We need a more precise position 
in order to pinpoint the real counterpart of this source, but
we favour Candidate 1 as the likely counterpart, on the basis of its
NIR brightness.

With only three infrared magnitudes for the two candidate counterparts
labeled C1 and C2, the fit of their SEDs is not very accurate, with
the model described in Section \ref{section:seds}. In addition, there
is no counterpart of this source in the Spitzer GLIMPSE survey (3.6
and $4.5 \microns$). We can therefore not firmly conclude the
nature of this binary system.  However, it is interesting to note that
the $R_\ast/D_\ast$ 90\% best fit values are very low
($=1.15\times10^{-11}$). From these values we can compute the range of
maximal radius that the massive star would have if it was at the
maximal distance inside our Galaxy, at $D_\ast=30\kpc$: we find
$R_{\ast max}=[6.7-12.8]\Rsol$. Since the maximal radius of the
companion star would be $\sim 13\Rsol$, this star cannot be either a
supergiant, or a giant. Alternatively, it could be a main sequence
early-type star, but then located very far away in our Galaxy.  We
therefore conclude that this object is more probably an LMXB system in
the Galactic bulge.

%
%
%
%

     \subsection{IGR\,J17252-3616}

     IGR\,J17252-3616 was discovered on 2004 February 9 in the
     Galactic bulge region, at the position ($17\hour25\minp2,
     -36\adeg16\amin$) by \cite{walter:2004a}.  Observations performed
     by {\it XMM-Newton} on 2004 March 21 by \cite{zurita-heras:2006}
     localized the source at ($17\hour25\min11\secp4,
     -36\adeg16\amin58\asecp6$) with 4$\asec$ accuracy.  This source
     is a rediscovery of the {\it EXOSAT} source EXO\,1722-363, based
     on similar timing and spectral properties.  It has been shown to
     be a heavily absorbed
     ($\nh\sim1.5\times10^{23}\cmmoinsdeux$) and persistent
     source, exhibiting apparent total eclipses, and a hard X-ray
     spectrum with either an absorbed Compton (kT$\sim$5.5 keV and
     $\tau\sim7.8$) or a flat power law ($\Gamma\sim0.02$), a very
     large column density, and Fe\,K$\alpha$ line at 6.4\,keV.  The
     detection of a spin period of $\sim413.7$~s and an orbital period
     of $\sim\ 9.72\ \mathrm{days}$ securely classified this source as
     a binary X-ray Pulsar \citep{zurita-heras:2006}.
     \cite{thompson:2007} refined the orbital period to $P = 9.7403\
     \mathrm{days}$ with data collected over more than 7 years.  They
     also derived a mass function of the system of $11.7\ \Msol$,
     suggesting an HMXB nature.  In addition, we point out that
     putting the spin and orbital periods of this source on a Corbet
     diagram \citep{corbet:1986} allows us to suggest a supergiant
     HMXB nature.

We performed accurate astrometry of the source, and overplot the {\it
XMM-Newton} error circle, as shown in Figure \ref{figure:smallfields}.
There is a bright 2MASS counterpart, 2MASS\,J17251139-3616575
(K$_{\rm S}$=10.7, labelled Candidate 1), inside the error circle, as suggested
by \cite{walter:2006}. However, in addition to the
2MASS source, there are three fainter candidate counterparts inside
the 4$\asec$ {\it XMM-Newton} error circle, labelled 2 to 4. 
All these candidates are located at 0.7, 1.4, 3.7 and
$3\asecp9$ from the centre of the error circle.  One of
these is blended with the 2MASS counterpart. IGR\,J17252-3616 would
therefore benefit from a more accurate localisation, however based on
astrometry and NIR brightness, we favour Candidate 1 as the most
likely candidate counterpart. NIR magnitudes of these
candidate counterparts are given in Table \ref{table:infrared}.

Optical and NIR spectra of the 2MASS counterpart (Candidate 1) are
shown in Figure \ref{figure:nirspec2}. We report the detected lines in
Table \ref{table:spectralines}. We find emission lines of H (Paschen
and Brackett), $\ion{He}{i}$ and $\ion{He}{ii}$ lines with likely
P-Cygni profiles at $2.06\microns$, and perhaps $\ion{Fe}{ii}$.  From
these NIR spectra, typical of an OB star, we conclude that the
companion star of this source has an OB spectral type. Furthermore,
the presence of $\ion{He}{i}$ in absorption and with P-Cygni profile constrains the
spectral type to a B supergiant \citep{hanson:2005}.  This result is in
agreement with the suggestion by \cite{zurita-heras:2006}, based on
the presence of this bright 2MASS source in the error circle.  This
source is therefore an HMXB system, probably hosting a supergiant
companion star, in agreement with the results by \cite{thompson:2007},
and in agreement with its position in the Corbet diagram.  The SED,
consistent with the HMXB nature of the system, is shown in Figure
\ref{figure:SEDs}.

     \subsection{IGR\,J17391-3021}

IGR\,J17391-3021 was discovered on 2003 August 26 at the position (RA
DEC J2000.0) = ($17\hour39\minp1, -30\adeg21\aminp5$) with a 
3$\amin$ uncertainty
by \cite{sunyaev:2003a}.  {\it Chandra} observations
performed on 2003 October 15 localized the source at
($17\hour39\min11\secp58, -30\adeg20\amin37\asecp6$) with
$\sim1\asec$ accuracy \citep{smith:2006}.  It is a rediscovery of a
previously known {\it ASCA} and {\it Rossi-XTE} source AX\,J1739.1-3020 =
XTE\,J1739-302 \citep{smith:2004}.
\cite{negueruela:2006b} have suggested that the optical counterpart of
IGR\,J17391-3021 is a USNO A2.0 source (USNO B1.0~0596-058586) with
B=17 and R=12.9, and that the NIR counterpart is a bright 2MASS source
(2MASS\,J17391155-3020380 with J=$8.600\pm0.021$, H=$7.823\pm0.027$
and K$_{\rm S}$=$7.428\pm0.023$; \citeauthor{smith:2006}
\citeyear{smith:2006}).  They also showed that the source is highly
reddened, with a variable absorption column density between outbursts.
During both low and high level states it exhibits hard X-ray spectra.
The fast X-ray transient behaviour (typical of an SFXT) and neutron
star spectrum have been confirmed by analyzing archival
\textit{INTEGRAL} data, with flares lasting between 30~min and 3~hours
\citep{sguera:2005}, the bright ones reaching 330~mCrab
\citep{turler:2007}. \cite{negueruela:2006b} state that this source is
most likely an HMXB, however the outbursts are shorter than expected
for HMXBs or Be/NS binaries, consistent with the SFXTs.

We performed accurate astrometry of the field (rms of fit=$0\asecp47$), and overplot the {\it Chandra} error circle, as shown
in Figure \ref{figure:smallfields}. Optical and NIR magnitudes of the
candidate counterpart are given in Tables \ref{table:optical} and
\ref{table:infrared} respectively.  However the NIR
magnitudes are out of the domain of linearity of SofI, which explains the
discrepancy with the 2MASS magnitudes.  We find that this 2MASS source
is a blended source, however from the astrometry, it is clear that the
1$\asec$ {\it Chandra} error circle confirms that the 2MASS source is
the counterpart of this source.

We examine the optical and NIR spectra shown in Figure
\ref{figure:nirspec2}. We report the detected lines in Table
\ref{table:spectralines}.  In the optical spectrum, we first detect
interstellar lines at 5800, 5890-6 (Na$_\mathrm{I}$ doublet), H$\alpha$, H lines
(Paschen), $\ion{He}{i}$ and $\ion{He}{ii}$ emission lines, N$_\mathrm{I}$ lines.  The NIR spectrum is
very rich in H lines from the Paschen and Brackett series, and also in
$\ion{He}{i}$ and $\ion{He}{ii}$ emission lines, some exhibiting P-Cygni profiles, and $\ion{O}{i}$
lines.  All these lines are characteristic of early-type stars, more
precisely of a supergiant O star, in agreement with the O8Iab(f)
spectral type derived by \cite{negueruela:2006b}.  IGR\,J17391-3021 is
therefore a supergiant HMXB, 
located at a distance of $\sim 2.3\ \mathrm{kpc}$ \citep{negueruela:2006b}, and not a
Be as stated in \cite{bird:2006}.  The SED, consistent with the
HMXB nature of the system, is shown in Figure \ref{figure:SEDs}.

We can determine the column density along the line of sight, using the
equivalent width of the Na$_\mathrm{I}$ doublet=$9\ \AA$; E(B-V)=$0.25 \times
W(\AA) = 2.25\ \mags$ (using \cite{munari:1997}); $N(\ion{H}{i}+H2) = 5.8
\times 10^{21} \times E(B-V) = 1.3 \times 10^{22}\ \mathrm{atoms/cm}^2$ (using
\citeauthor{bohlin:1978} \citeyear{bohlin:1978}), the column density
is consistent with the one indicated in Table \ref{table:results}.

     \subsection{IGR\,J17597-2201}


IGR\,J17597-2201 was discovered at the position (RA DEC J2000.0) =
($17\hour59\minp7, -22\adeg01\amin$) with $\sim2\amin$ uncertainty
\citep{lutovinov:2003c}. {\it XMM-Newton} observations localized the
source at ($17\hour59\min45\secp7, -22\adeg01\amin39\asec$) with a
4$\asec$ accuracy \citep{walter:2006}.  It exhibits an absorption of
$\nh = 4.5\pm0.7\times10^{22}\cmmoinsdeux$, or
$2.70\pm0.15\times10^{22}\cmmoinsdeux$ when a partial-covering
absorber is included. This source, associated with the {\it Rossi-XTE}
source XTE\,J1759-220, is a Low-Mass X-ray Binary system, a late-type
transient type I X-ray burster, therefore hosting a neutron star, with
a dipping behavior of $\sim 30\%$ with $\sim5$~min dip duration
\citep{markwardt:2003c}.

We performed accurate astrometry of the field, shown in Figure
\ref{figure:smallfields}.  \cite{walter:2006} proposed a 2MASS
counterpart for this source (2MASS J17594556-2201435), 
however it is well outside the error
circle, towards the south-west, at 5$\asecp$1 from the centre of the
error circle.  There are one bright and two faint candidate
counterparts which are well inside the {\it XMM-Newton} error circle
(labeled 1, 3 and 5 and located respectively at 2.25, 0.95 and
$2\asecp05$ from the centre of the {\it XMM-Newton} error circle). In
addition, there are one faint and one bright candidate counterparts,
lying on the error circle (labeled 2 and 4, and located
respectively at 3.8 and $3\asecp9$ from the centre of {\it XMM-Newton}
error circle). We give the NIR magnitudes of these candidate
counterparts in Table \ref{table:infrared}. Since the previously
proposed counterpart is outside the error circle, and since we propose
new candidate counterparts, this source deserves further observations
to find which one is the right counterpart.

By fitting the SED of the brightest candidate labeled 1, which is also
the closest to the error circle centre, with the model described in
Section \ref{section:seds}, we obtain a stellar temperature
$T_\ast=31700$\,K, consistent with an O type star.  The other
parameters are given in Table \ref{table:fitsseds}, and the SED is
shown in Figure \ref{figure:SEDs}. However our
spectral fit is not accurate enough to unambiguously constrain the
spectral type of the companion star, because we have only the three
NIR magnitudes (we did not detect any MIR counterpart in the Spitzer
GLIMPSE survey). Therefore, taking into account the errors, this SED
fitting is consistent at 90\% both with an LMXB and HMXB system.
Assuming the source is a supergiant, with a minimum radius of
$20\Rsol$, and taking a mean $R_\ast/D_\ast$ of $2\times10^{-11}$, we
obtain a distance of $35.2\kpc$, which is not
plausible. Alternatively, taking a radius $R_{\ast} = 10\ \Rsol$ typical
of a B\,V companion star, we derive a distance of
$D_{\ast}=15.2$\,kpc, a more reasonable estimate although still
high. We therefore conclude that the source is most likely an LMXB,
consistent with its X-ray properties, given the detection of type I
X-ray bursts.


%
%

     \subsection{IGR\,J18027-2016}

IGR\,J18027-2016 was discovered at the position (RA DEC J2000.0) =
($18\hour02\min46\sec, -20\adeg16\aminp3$)
\citep{revnivtsev:2004}. {\it XMM-Newton} observations allowed to
localize the source at ($18\hour02\min42\secp0,
-20\adeg17\amin18\asec$) with 4$\asec$ accuracy \citep{walter:2006}.
The source was originally called IGR\,J18029-2016,
and it is associated with the SAX source SAX\,J18027-2017.  The
absorption is $\nh = 9.1\pm0.5\times10^{22}\cmmoinsdeux$
\citep{walter:2006}.  It is an X-ray pulsar, and a transient source,
an eclipsing HMXB, likely at a distance of 10\,kpc. An orbital period
of 4.57\,days and a pulse period of 139.47\,s were reported by
\cite{hill:2005}.  These authors have derived the system parameters,
with an excentricity of $e \leq 0.2$, a mass function of
f(M)$\sim17\pm5\Msol$, implying a mass of the companion of $18.8-29.3\
\Msol$.

We performed accurate astrometry of this source (rms of
fit=$0\asecp51$), showed in Figure \ref{figure:smallfields}.  There is
a 2MASS counterpart (2MASS J18024194-2017172) at $\sim 1\asec$ from
the centre of the {\it XMM-Newton} error circle, which is the
candidate counterpart proposed by \cite{walter:2006}. However there is
another bright source well inside the error circle, labeled Candidate
2. There is also at least one faint counterpart inside the error
circle.  We give the NIR magnitudes of Candidate 1 and Candidate 2 in
Table \ref{table:infrared}. Based on the proximity to the
error circle centre, and its NIR brightness, we favour Candidate 1 as
the likely candidate.

\begin{figure*}
\setlength{\unitlength}{1.0cm}
\includegraphics[angle=0,width=9.5cm]{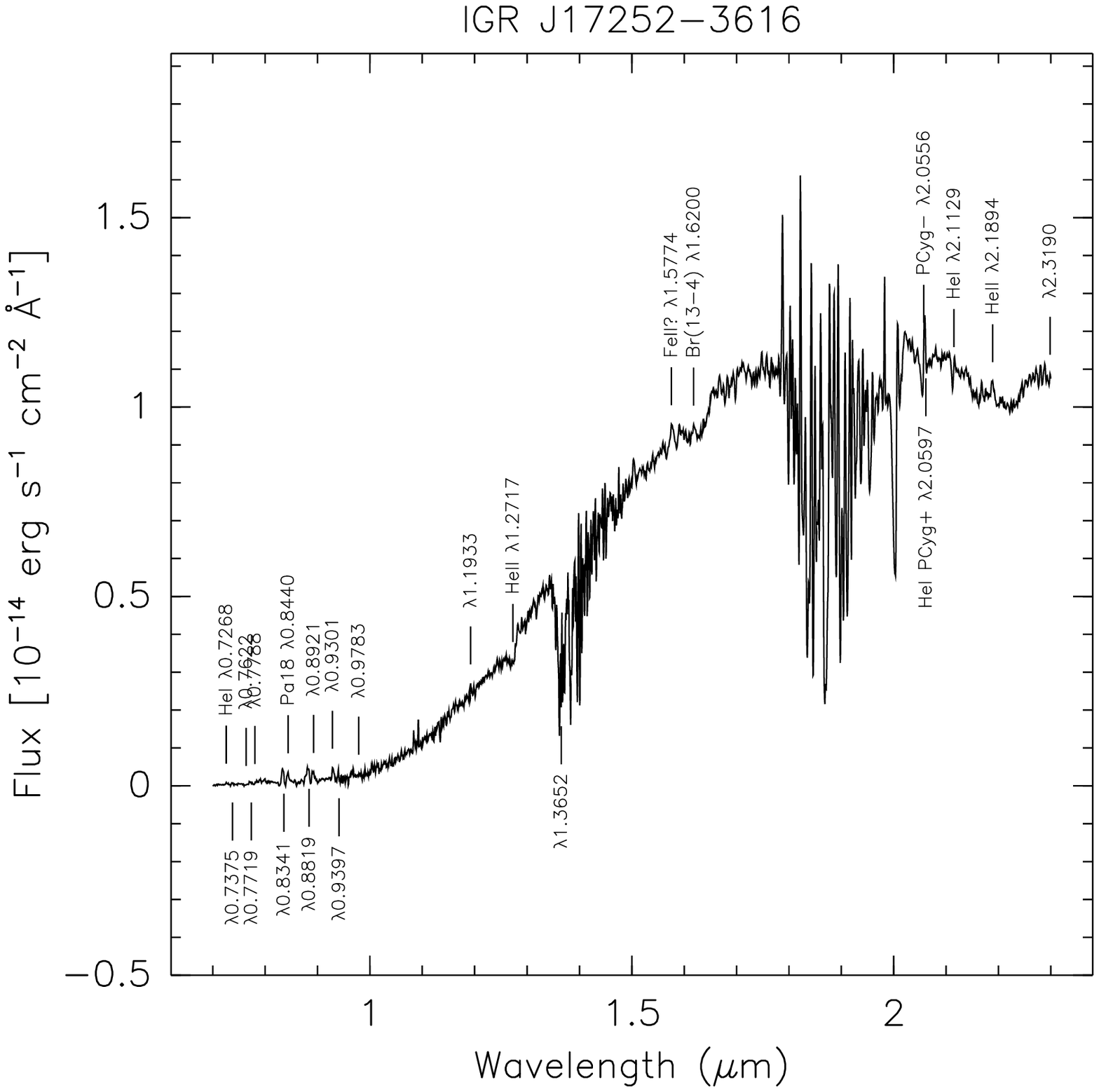}
%
\includegraphics[angle=0,width=9.5cm]{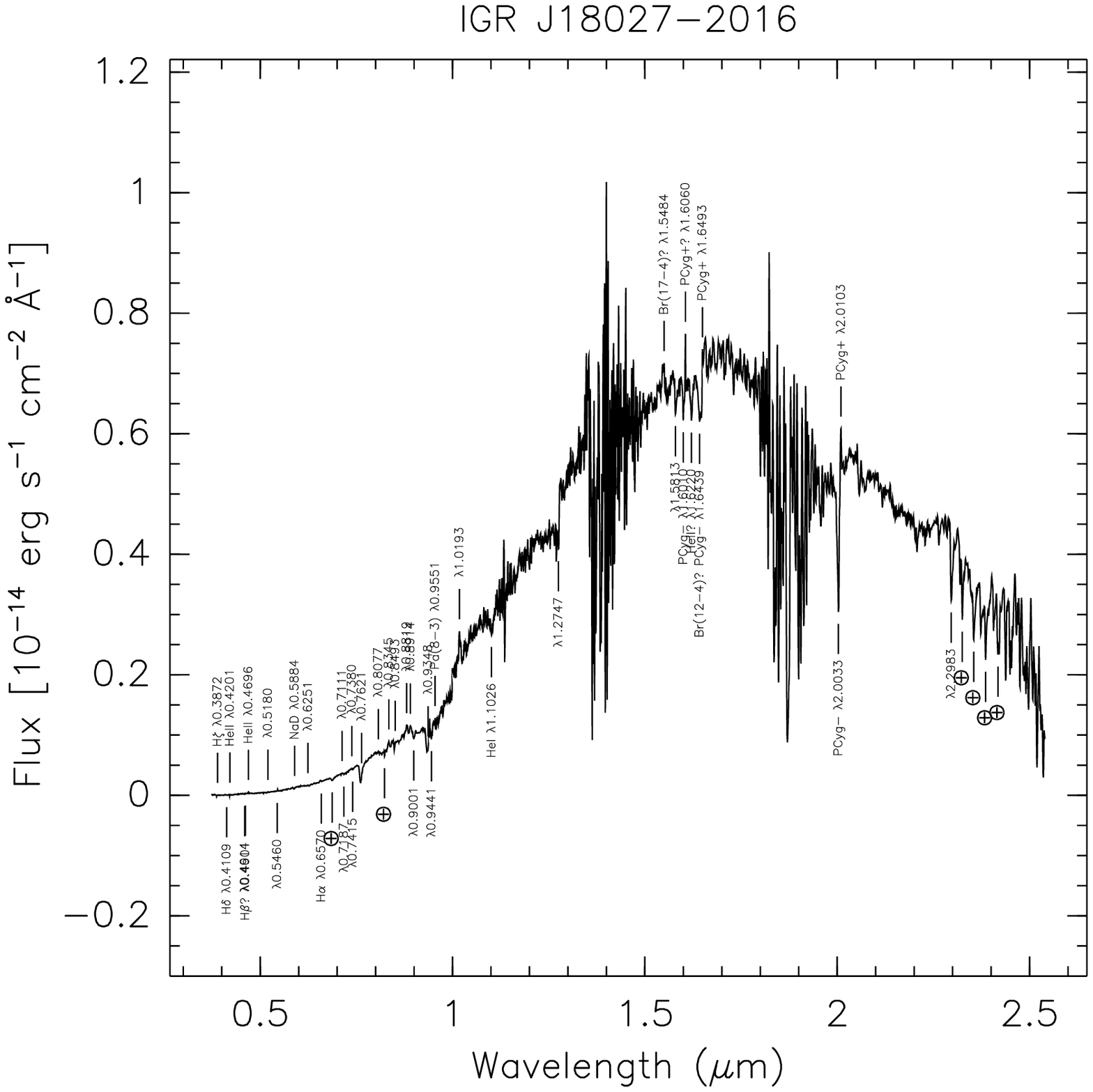}
%
%
\includegraphics[angle=0,width=9.5cm]{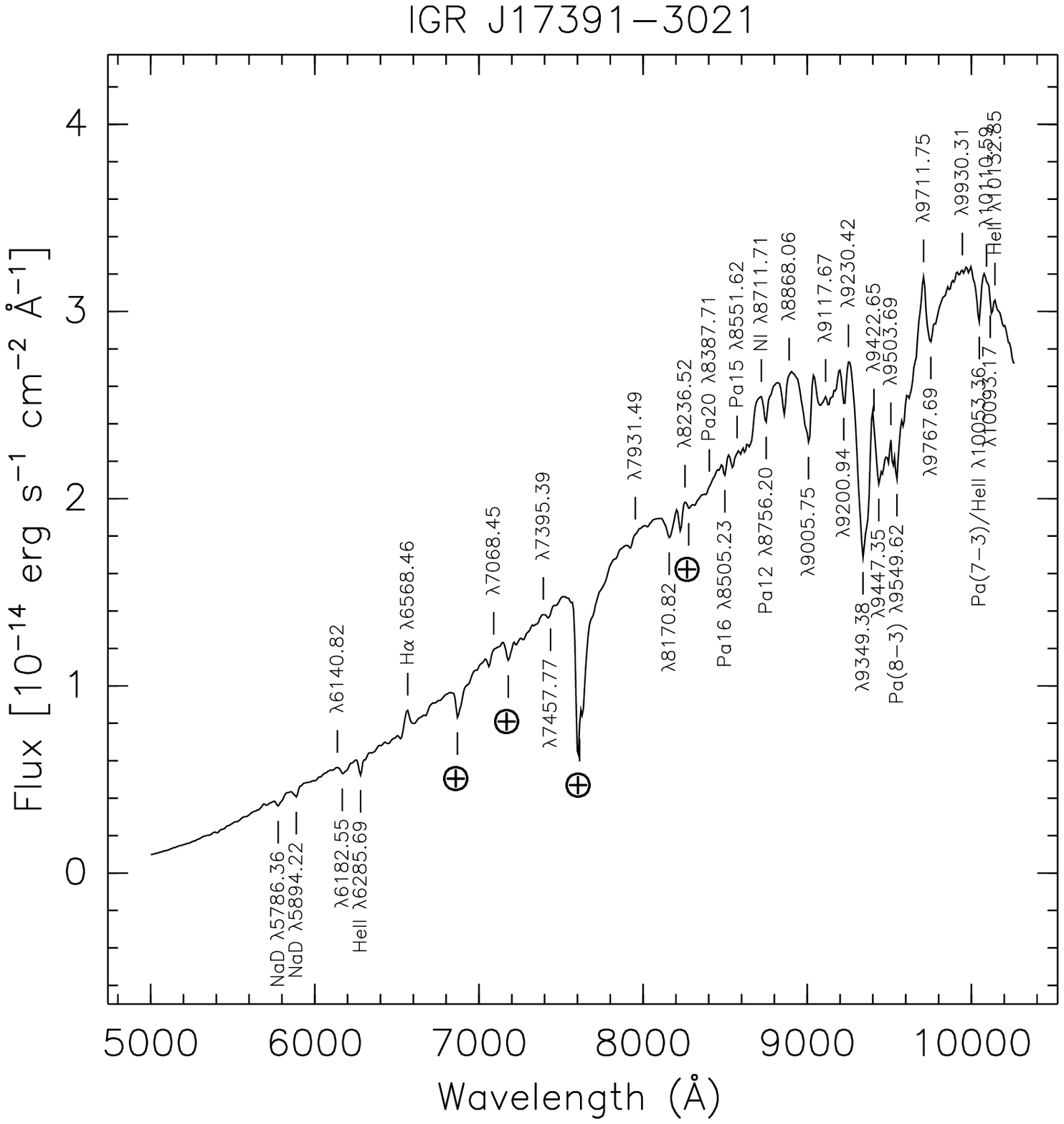}
%
\includegraphics[angle=0,width=9.5cm]{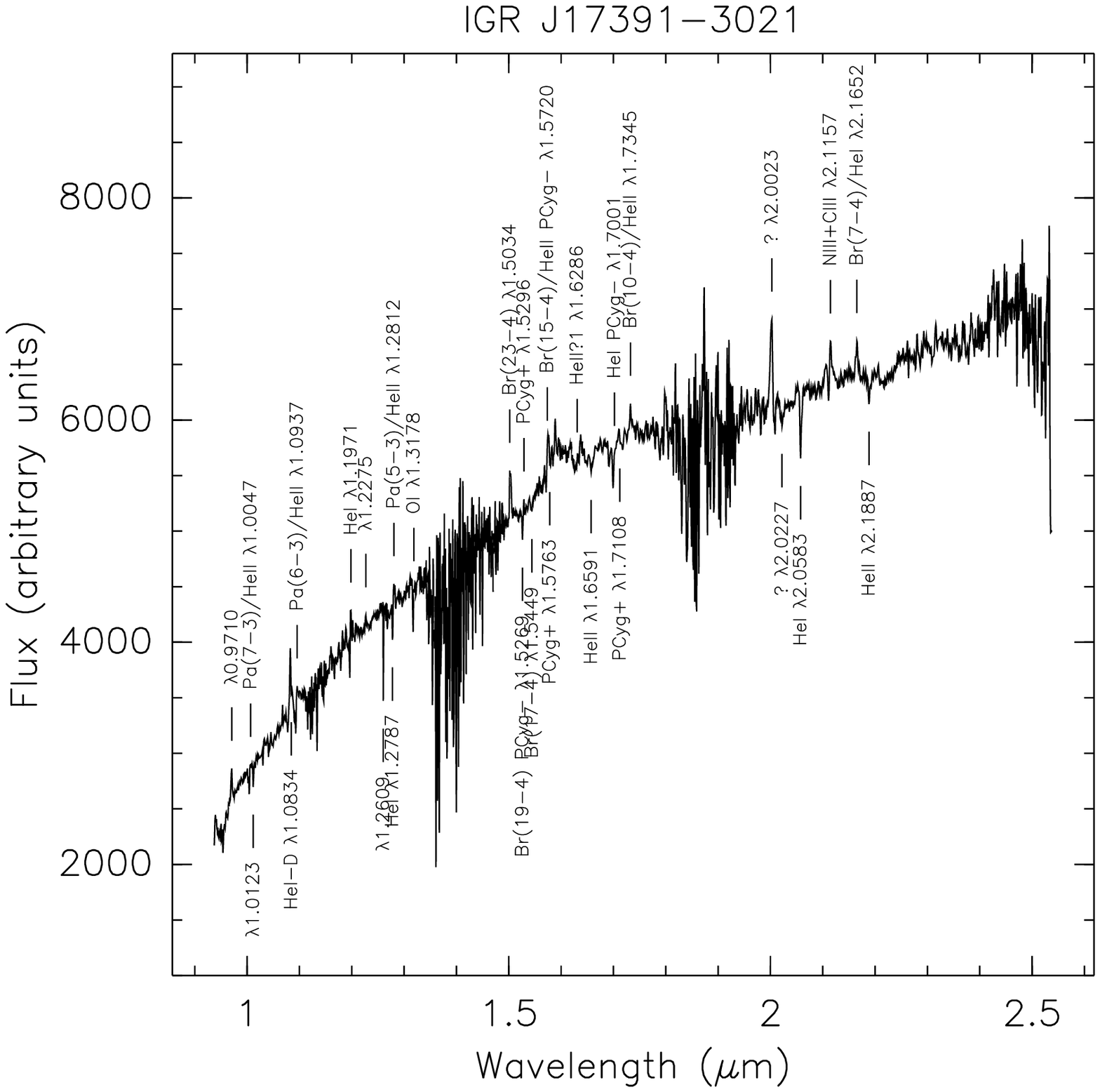}
%
\caption[]{\label{figure:nirspec2} 
Top panel: Combined flux calibrated spectra of the two sources 
IGR~J17252-3616 Candidate 1 (left) and IGR~J18027-2016 (right) respectively.  
The optical spectra were
obtained with 3.6m/EFOSC, and the blue and red grism NIR spectra with
NTT/SofI.  We combined EFOSC2 and SofI spectra by applying a scaling
factor of 18.4035 for IGR~J17252-3616 and 2.417 for IGR~J18027-2016,
the difference being due to a different calibration between EFOSC2 and
SofI, and to different exposure times.  These two factors were
obtained by dividing the flux in the overlap region between EFOSC2 and
SofI spectra. The spectra are extremely red because of the galactic
reddening. Besides the low S/N of the spectra, a number of narrow
absorption lines could be detected.  The y axis of the spectra are
given in "$\lambda F_\lambda$" units: $\ergcms$. \\ 
Bottom left panel:
IGR\,J17391-3021 flux calibrated optical spectrum, with the y axis
given in "$\lambda F_\lambda$" units: $\ergcms$.  Bottom right panel:
combined blue and red grism NIR spectra of IGR\,J17391-3021, with the
y axis in arbitrary units.
%
%
}
\end{figure*}

The optical and NIR spectra of Candidate 1 are shown in Figure
\ref{figure:nirspec2}. We report the detected lines in Table
\ref{table:spectralines}. In the optical spectrum we detect hydrogen
($\alpha$ to $\zeta$) and $\ion{He}{ii}$ emission lines, and in the NIR spectrum we
detect H (Paschen and Brackett series), $\ion{He}{i}$ and $\ion{He}{ii}$ emission lines,
some tentatively exhibiting P-Cygni profiles.  These NIR spectra are
typical of a supergiant OB star, which is therefore likely the
spectral type of the companion star: IGR\,J18027-2016 is therefore a supergiant
HMXB system.
%
Fitting the data with the model described in Section
\ref{section:seds}, we obtain a stellar temperature $T_\ast = 20800\
\mathrm{K}$, which is typical of a B supergiant star. The other
parameters are given in Table \ref{table:fitsseds}, and the SED is
shown in Figure \ref{figure:SEDs}. By taking the $R_\ast/D_\ast$
minimizing $\chi^2$ ($3.7\times10^{-11}$), and assuming a typical
radius of a B supergiant star, i.e. $R_\ast=20\Rsol$, we derive a
distance of $D_\ast = 11.9\ \mathrm{kpc}$. We therefore favour the supergiant
nature of the companion star, with a B spectral type, in agreement
with the results derived from the spectra.

%
%

     \subsection{IGR\,J18483-0311}

     IGR\,J18483-0311 was discovered at the position (RA DEC J2000.0)
     = ($18\hour48\minp3, -03\adeg11\amin$), with uncertainty
     $\sim2\amin$ \citep{chernyakova:2003}.  The {\it Swift}
     observations allowed \cite{sguera:2007} to refine the position of
     the source to ($18\hour48\min17\secp17,
     -03\adeg10\amin15\asecp54)$, uncertainty $3\asecp3$.  The {\it
       Swift} refined position allowed them to identify an optical
     counterpart from the USNO-B1.0 and 2MASS catalogue located at
     ($18\hour48\min17\secp2, -03\adeg10\amin16\asecp5$) with
     magnitudes R=19.26, I=15.32, J=10.74, H=9.29 and K=8.46.  The
     source exhibits a column density of $\nh = 9 \times10^{22}\
     \cmmoinsdeux$, a $\Gamma=1.4$ and an energy cutoff at
     $E_{cut}=22$\,keV \citep{sguera:2007}.  Timing analysis of {\it
       Rossi-XTE}/ASM light curve have allowed investigators to derive
     an orbital period of $18.55\pm0.05$\,days \citep{levine:2006}.
     \cite{sguera:2007} also report that the source contains a pulsar
     with a spin period of $21.0526\pm0.0005$\,s with a pulse fraction
     of $65\pm10$\%, while the highly reddened optical counterpart
     suggests by analogy with other such systems that the source is an
     HMXB.  They further argue that it is a likely Be system due to
     its localisation in a Corbet diagram \citep{corbet:1986}, but
     they cannot rule out an SFXT, although the typical ratio between
     maximum and minimum luminosity they observed is at least 10 times
     lower than that of typical SFXT.

     We performed accurate astrometry of this source (rms of
     fit=$0\asecp67$), shown Figure \ref{figure:smallfields}.  There
     is a bright 2MASS counterpart 2MASSJ\,18481720-0310168 inside the
     {\it Swift} error circle, which is the candidate counterpart
     proposed by \cite{sguera:2007}.  However this source seems to be
     blended with another fainter source, at 2.5$\asec$ from the
     centre of the error circle.  We give the NIR magnitudes of the
     bright candidate counterpart in Table \ref{table:infrared}.

     Fitting the SED of this counterpart with the model described in
     Section \ref{section:seds} allows us to derive a stellar
     temperature of $T_\ast = 22500\ \mathrm{K}$ at the limit of 90\%
     of $\chi^2$, which is typical of a B star. The other parameters
     are given in Table \ref{table:fitsseds}, and the SED is shown in
     Figure \ref{figure:SEDs}. The stellar temperature is consistent
     with a B spectral type companion star. Since the $R_\ast/D_\ast$
     ratio minimizing $\chi^2$ is $2.15\times10^{-10}$, the distance
     of this source would then be 0.9\,kpc if the companion star is a
     main sequence star (with a typical stellar radius of
     $R_\ast=3\Rsol$), 1.5 kpc for a sub-giant and 2.7 kpc for a
     supergiant star (with a typical stellar radius of
     $R_\ast=20\Rsol$). This source exhibits a strong NIR excess,
     which might indicate the presence of a disk/wind such as in
     massive stars. 
Furthermore, its position in the Corbet diagram is 
between Be and wind accretor-supergiant  
X-ray binary systems. Although we cannot
firmly conclude on the spectral type and class, this SED
shows that this source is an HMXB system.

%
%

     \subsection{IGR\,J19140+0951}

IGR\,J19140+0951 was discovered on 2003 March 6-7 at the position
(RA DEC J2000.0) = ($19\hour13\min55\sec, +9\adeg51\aminp6$),
uncertainty 1$\amin$) by \cite{hannikainen:2003}. {\it Chandra}
observations performed on 2004 May 11 localized the source at
($19\hour14\min4\secp232, +09\adeg52\amin58\asecp29$) with
$0\asecp6$ accuracy \citep{intzand:2006}.  Fitting the hard X-ray
spectrum from {\it Rossi-XTE} observations allowed investigators to derive a
$\Gamma\sim1.6$ and $\nh\sim6\times10^{22} \cmmoinsdeux$
\citep{swank:2003} with variations of $\nh$ up to
$\sim10^{23}\cmmoinsdeux$ \citep{rodriguez:2005}.  An orbital period
of $13.55\ \mathrm{days}$ was found from timing analysis of {\it
Rossi-XTE} data, with an X-ray activity detected with {\it
Rossi-XTE}/ASM as early as 1996 \citep{corbet:2004}, confirming the
binary nature of the source.  \cite{rodriguez:2005}, after a
comprehensive analysis of \textit{INTEGRAL} and {\it Rossi-XTE} data,
showed that the source was spending most of its time in a faint state but
reported high variations of luminosity and absorption column density.
It is a persistent HMXB with evidence for the compact object being a
neutron star rather than a black hole, exhibiting a variable
absorption column density, and a bright iron line
\citep{rodriguez:2005}.  This source has other names: IGR\,J19140+098
\citep{hannikainen:2003}, and EXO\,1912+097.  \cite{intzand:2006}
proposed the 2MASS source 2MASS\,J19140422+0952577 as the NIR
counterpart of this source (J=8.55, H=7.67 and K$_{\rm S}$$=7.06\mag$).

We performed accurate astrometry of the field (rms of
fit=$0\asecp35$), shown in Figure \ref{figure:smallfields}.  The
accurate error circle allows us to confirm the 2MASS source
2MASS\,19140422+0952577 as the candidate counterpart, located at
$4\asecp6$ North of a bright 2MASS source (2MASS J19140417+0952538,
K$_{\rm S}$$=6.27\mag$). We give the NIR magnitudes of the deblended
candidate counterpart in Table \ref{table:infrared}. These 
deblended magnitudes we derive are different from the
magnitudes given in the 2MASS catalogue, because both 2MASS sources
are spatially very close in the 2MASS data.

\begin{figure*}
\setlength{\unitlength}{1.0cm}
\includegraphics[angle=0,width=9.5cm]{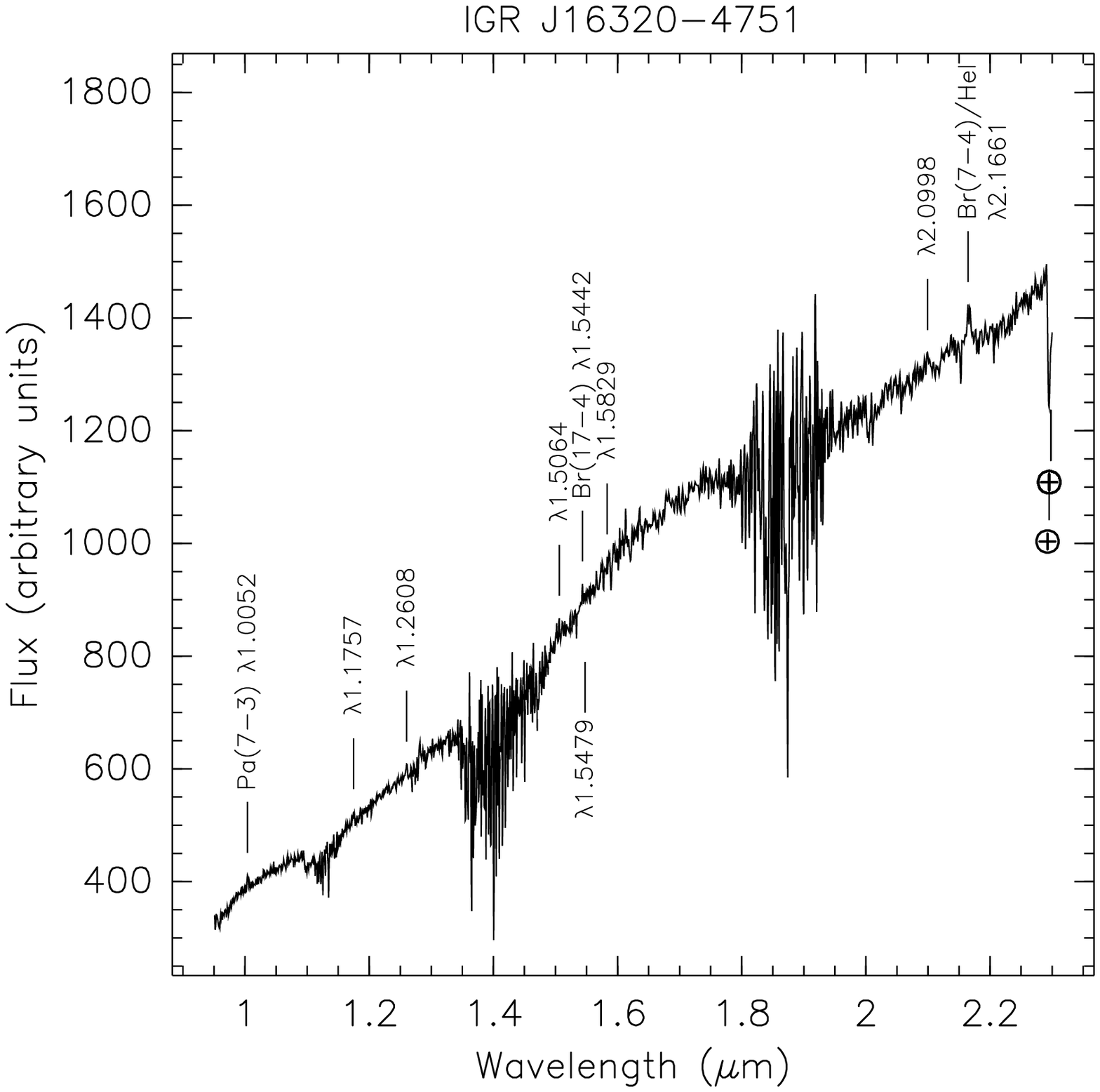}
%
\includegraphics[angle=0,width=9.5cm]{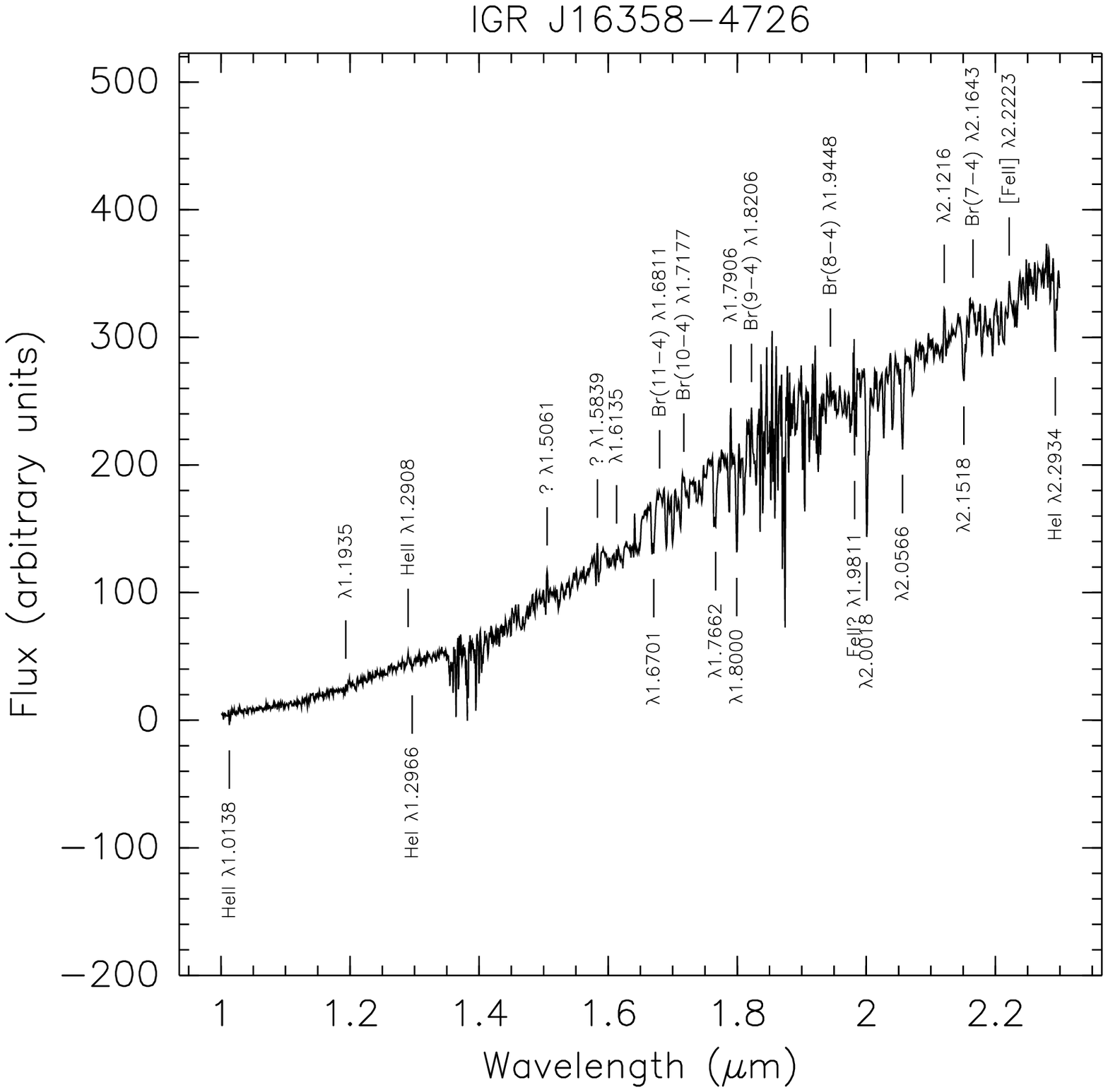}
%
\includegraphics[angle=0,width=9.5cm]{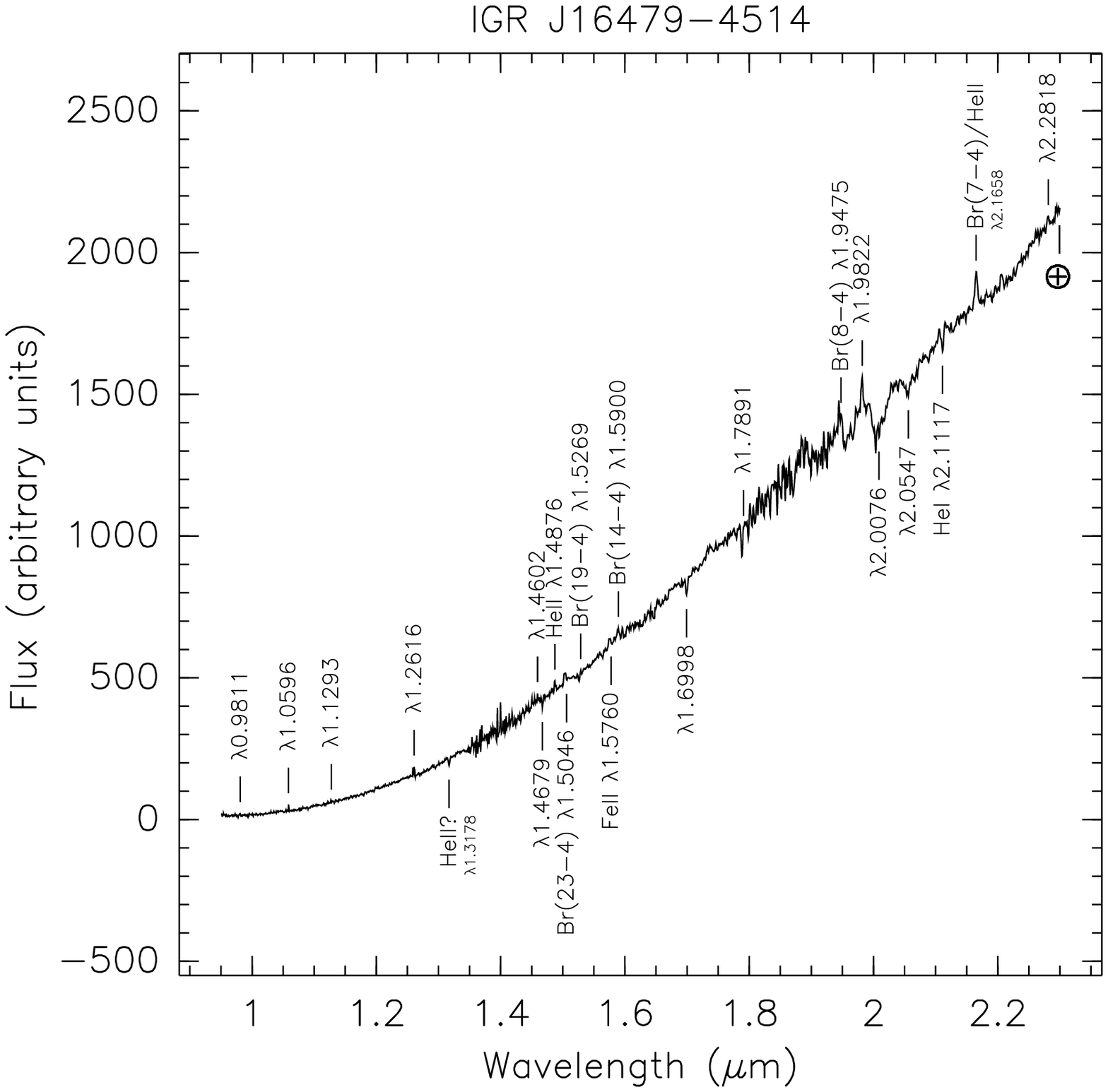}
%
\includegraphics[angle=0,width=9.5cm]{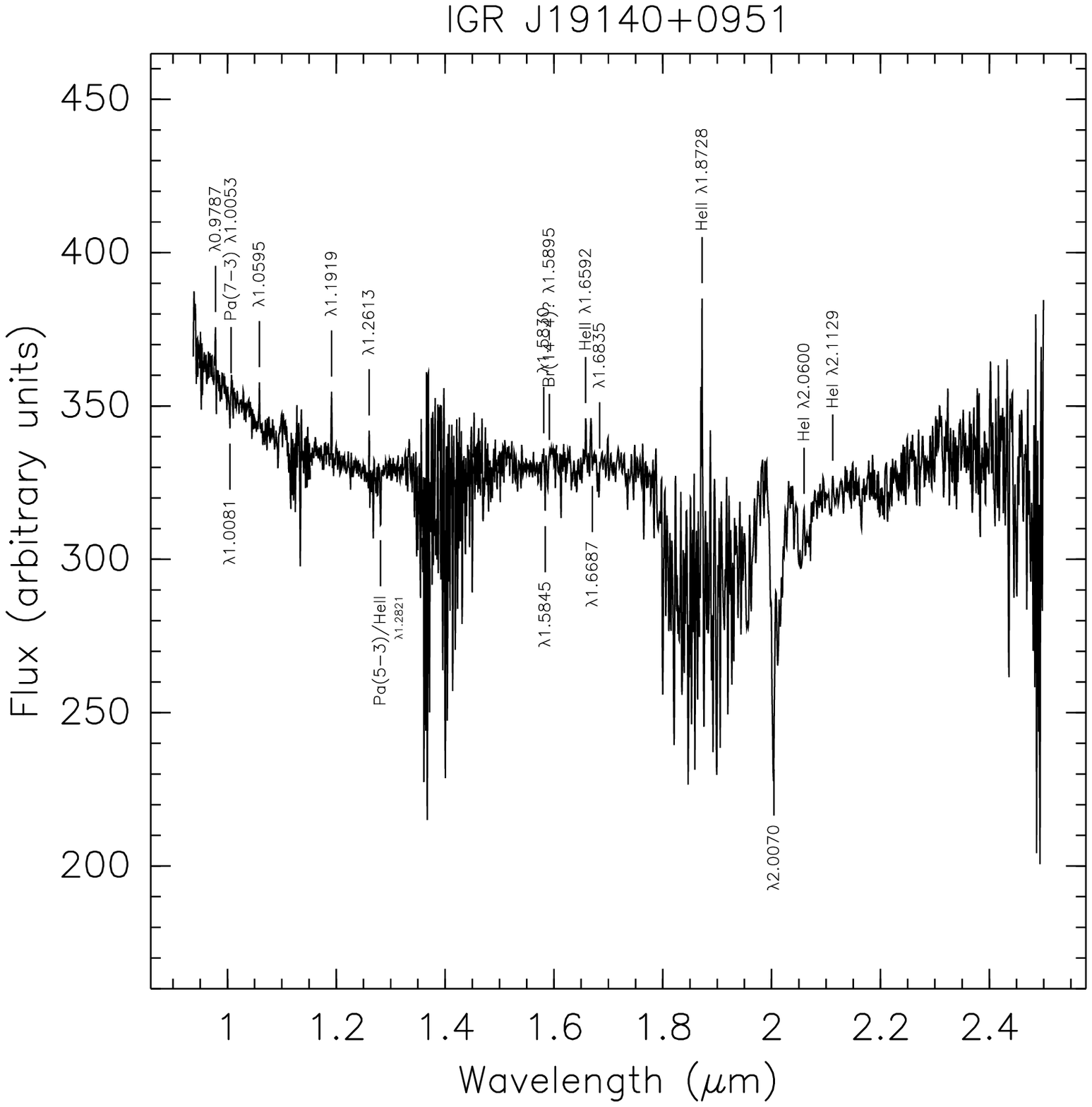}
%
\caption[]{\label{figure:nirspec1} From top to bottom, and left to
right: combined blue and red grism NIR spectra of IGR\,J16320-4751 (Candidate
1), IGR\,J16358-4726, IGR\,J16479-4514 (Candidate 1) and
IGR\,J19140+0951 (the y axis is in arbitrary units).}
\end{figure*}

We show the NIR spectra of this source in Figure
\ref{figure:nirspec1}. We report the detected lines in Table
\ref{table:spectralines}.  The spectrum is dominated by H (Paschen and
Brackett series), and $\ion{He}{i}$ and $\ion{He}{ii}$ in
emission.  The NIR spectra are typical of an OB spectral type
companion star, and the narrowness of the lines suggests a supergiant
type, which is consistent with the results by \cite{nespoli:2007} for
a B1I stellar type companion, derived from spectra obtained at
ESO/NTT/SofI. This classification has been refined to B0.5I based on
spectra obtained at UKIRT \citep{hannikainen:2007}, making
IGR\,J19140+0951 a supergiant X-ray binary. Both classifications
confirm the supergiant HMXB nature of IGR\,J19140+0951, hosting a
neutron star. The SED, consistent with the HMXB nature of the system,
is shown in Figure \ref{figure:SEDs}.

\cite{intzand:2006} also suggest a MIR counterpart
at 8.3 $\mu$m found in the Midcourse Space Experiment ({\it MSX},
\citeauthor{mill:1994} \citeyear{mill:1994}), however
\cite{rahoui:2008} show that \cite{intzand:2006} are in fact reporting
the summed flux of a blended source, composed of the MIR counterparts
of both IGR\,J19140+0951 and the bright 2MASS source at the South,
which is unrelated to IGR\,J19140+0951.  \cite{rahoui:2008} give the MIR
fluxes of the counterpart of IGR\,J19140+0951.

\section{Discussion} \label{section:discussion}

We begin this Section by giving a summary of the results of all
individual sources.  We continue by showing the large scale
environment of these sources, and report the presence of absorption in
their environment.  We then recall the general characteristics of
HMXBs, before discussing the results obtained, in the context of the
{\it INTEGRAL} era.

     \subsection{Summary of results of the sample of studied sources}

We found by spectroscopy of the most likely candidate
counterparts the spectral types of IGR~J16320-4751, IGR~J16358-4726,
IGR~J16479-4514, IGR~J17252-3616 and IGR~J18027-2016: they are all of
supergiant OB types, IGR~J16358-4726 likely hosting a sgB[e] companion
star.  We also confirm the supergiant O nature of IGR~J17391-3021, and
the supergiant B nature of IGR~J19140+0951. By fitting the SED of the
most likely candidates we found that IGR~J16418-4532 is an HMXB,
probably hosting a supergiant OB spectral type companion star, that
IGR~J16393-4643 is an HMXB system probably hosting a BIV-V companion
star, and that IGR~J18483-0311 is very likely an 
HMXB system.  By accurate astrometry we rejected the counterpart proposed for
IGR~J17091-3624 and IGR~J17597-2201, we propose two new candidate
counterparts for each of them, which by SED fitting we found 
consistent with an LMXB nature. We confirm the AGN nature of
IGR\,J16558-5203.

We summarize in Table \ref{table:results} the results obtained both
from spectroscopy and SED fitting to the sample of studied {\it INTEGRAL}
sources. We give the results
of the column density in optical/IR, spectral type of the companion
stars, and type of sources for all individual sources of our
sample. We also give all the parameters known about these
sources, such as the spin and orbital period, and column
density derived from X-ray observations, in order to facilitate the
following discussion.  The interstellar column density and absorption
in the optical and NIR domain have been derived from our observations,
given in Table \ref{table:fitsseds} for the sources IGR~J16393-4643,
IGR~J16418-4532, IGR~J17091-3624, IGR~J17597-2201, IGR~J18027-2016 and
IGR~J18483-0535, and in \cite{rahoui:2008} for the remaining sources.
We then converted the column density into absorption in magnitudes,
using the relation given in \cite{cardelli:1989}:
$\nh/\Av = 1.87 \times 10^{21}\ \cmmoinsdeux/\mags$.

The results, concerning the spectral type of the companion star, given
by these SED fittings are in agreement with those directly
derived from spectroscopy, when available.  On the other hand,
they allow us to derive the likely spectral type of the companion
star, when there is no spectroscopic information.  We can therefore
conclude that most of these obscured sources host luminous, massive,
hot and early-type companion stars, i.e. of OB spectral type, most of
them being evolved stars of the supergiant spectral class.

From these SEDs, all reported in Figure \ref{figure:SEDs}, we can
directly compare the optical to MIR with the hard X-ray domain
integrated flux (corresponding to the energy output). All sources for
which both fluxes are comparable are supergiant HMXBs:
IGR\,J16418-4532, IGR\,J16479-4514, IGR\,J17252-3616,
IGR\,J17391-3021, IGR\,J18027-2016, IGR\,J19140+0951. These comparable
optical--MIR and hard X-ray fluxes are expected from the HMXB nature
of the systems. Most of the sources for which the hard X-ray flux is
much higher than the optical--MIR flux are HMXBs, hosting either a
supergiant or B type companion star: IGR\,J16320-4751,
IGR\,J16358-4726, IGR\,J16393-4643. IGR\,J17091-3624 and
IGR\,J17597-2201 are the only cases with a hard X-ray flux higher than
the optical--MIR flux, consistent with their LMXB nature. Finally,
there is only one case where the hard X-ray flux is much less than the
optical--MIR flux: the 
HMXB IGR\,J18483-0311. However we have to be cautious because
the multi-wavelength observations of these SEDs were
not taken simultaneously, and the hard X-ray fluxes are variable.

\begin{table*}
  \begin{center}
    \caption{Summary of results and characteristics of our studied sample 
of {\it INTEGRAL} sources.  We give the name of the sources, the
region of the Galaxy in the direction in which they are located, their
spin and orbital period, the interstellar column density ($\nh$I),
the absorption derived from optical to infrared observations ($\nh$IR), 
the absorption derived from X-ray observations ($\nh$X), the
spectral type (SpT) of their most likely candidate 
obtained or confirmed by spectroscopy (spec) or by
fitting the SED (sed), the nature and type of the binary system, 
and the reference (Ref) to the
spectral type.  More details on each source are given in Section
\ref{section:results}. Details on how we obtained $\nh$I,
$\nh$IR and $\nh$X are given in Section \ref{section:seds}.
Type abbreviations: 
AGN = Active Galactic Nucleus,
B = Burster (neutron star),
BHC = Black Hole Candidate,
D = Dipping source,
LMXB = Low-Mass X-ray Binary,
HMXB = High-Mass X-ray Binary System, 
OBS = obscured source,
SFXT = Supergiant Fast X-ray Transient,
T: Transient source,
XP: X-ray Pulsar.
The classification as SFXT is still subjective, since we lack some
accurate observations on a long-term scale for most of the sources. 
The spectral types come from optical/NIR spectroscopy,
reported in the references given in the column Ref. 
c: this paper,
h: \cite{hannikainen:2007},
m: \cite{masetti:2006},
neg: \cite{negueruela:2006b}, 
nes: \cite{nespoli:2007}.
}
\label{table:results}
\vspace{1em}
    \renewcommand{\arraystretch}{1.2}
    \begin{tabular}{cccccccccc} 
\hline
Source & Region & Pspin & Porb & $\nh$I & $\nh$IR & $\nh$X & SpT & Nature & Ref \\
 & & (s) & (d) & $10^{22}\cmmoinsdeux$ & $10^{22}\cmmoinsdeux$ & $10^{22}\cmmoinsdeux$ & & & \\
\hline
IGR\,J16320-4751 & Norma & 1250 & 8.96(1) & 2.14 & 6.6 & 21 & spec: sgO & HMXB/XP/T/OBS & c \\
\hline
IGR\,J16358-4726 & Norma & 5880 & & 2.20 & 3.3 & 33 & spec: sgB[e]? & HMXB/XP/T/OBS & c \\
\hline
IGR\,J16393-4643 & Norma & 912 & 3.6875(6) & 2.19 & 2.19 & 24.98 & sed: BIV-V? & HMXB/XP/T & c \\
\hline 
IGR\,J16418-4532 & Norma & 1246 & 3.753(4) & 1.88 & 2.7 & 10 & sed: sgOB? & HMXB/XP/SFXT & c \\ 
\hline
IGR\,J16479-4514 & Norma & & & 2.14 & 3.4 & 7.7 & spec: sgOB & HMXB/SFXT? & c \\
\hline
IGR\,J16558-5203 & - & - & - & - & - & - &  AGN & Seyfert 1.2 & m \\
\hline 
IGR\,J17091-3624 & GC & & & 0.77 & 1.03 & 1.0 & sed: LMXB & BHC & c \\ 
\hline  
IGR\,J17252-3616 & GC & 413 & 9.74(4) & 1.56 & 3.8 & 15 & spec: sgB & HMXB/XP/OBS & c \\
\hline
IGR\,J17391-3021 & GC & & & 1.37 & 1.7 & 29.98 & spec: O8Iab(f) & HMXB/SFXT/OBS & neg \\
\hline
IGR\,J17597-2201 & GC & & & 1.17 & 2.84 & 4.50 & sed: LMXB & LMXB/B/D/P & c \\
\hline
IGR\,J18027-2016 & GC & 139 & 4.5696(9) & 1.04 & 1.53 & 9.05 & spec: sgOB & HMXB/XP/T & c \\
\hline
IGR\,J18483-0311 & GC & $21.05$ & 18.55 & 1.62 & 2.45 & 27.69 & sed: HMXB? & HMXB/XP & c \\ 
\hline 
IGR\,J19140+0951 &    & & 13.558(4) & 1.68 & 2.9 & 6 & spec: B0.5I & HMXB/OBS & h, nes \\
\hline
\end{tabular}
  \end{center}
\end{table*}

     \subsection{Environment} \label{environment}

\begin{figure*}
\centering
\subfigure[IGR\,J16358-4726]{\label{fig-abs:igrj16358}
  \includegraphics[angle=0,width=8.5cm]{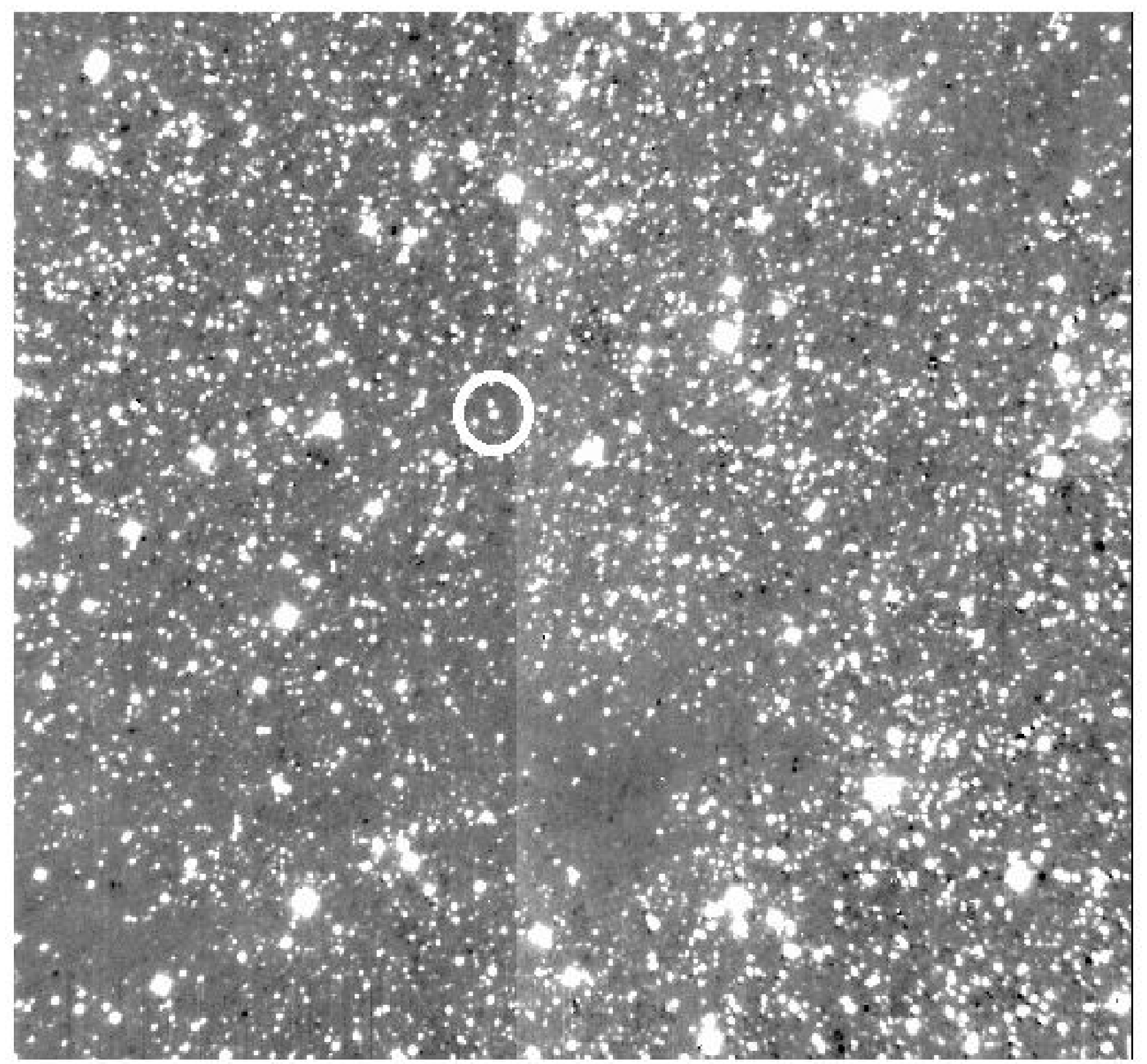}}
\subfigure[IGR\,J16418-4532]{\label{fig-abs:igrj16418}
\includegraphics[angle=0,width=8.5cm]{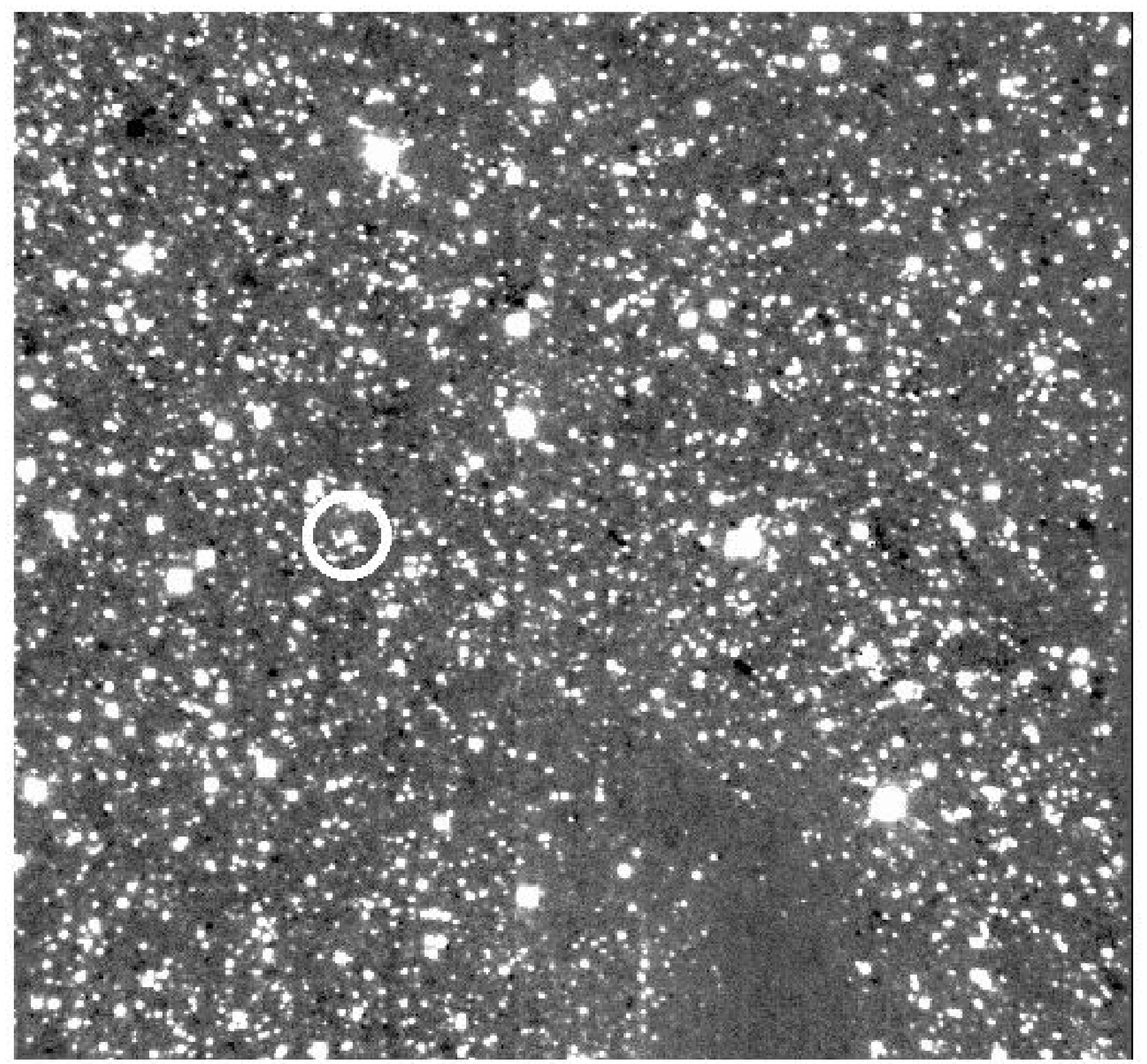}}
\subfigure[IGR\,J16479-4514]{\label{fig-abs:igrj16479}
\includegraphics[angle=0,width=8.5cm]{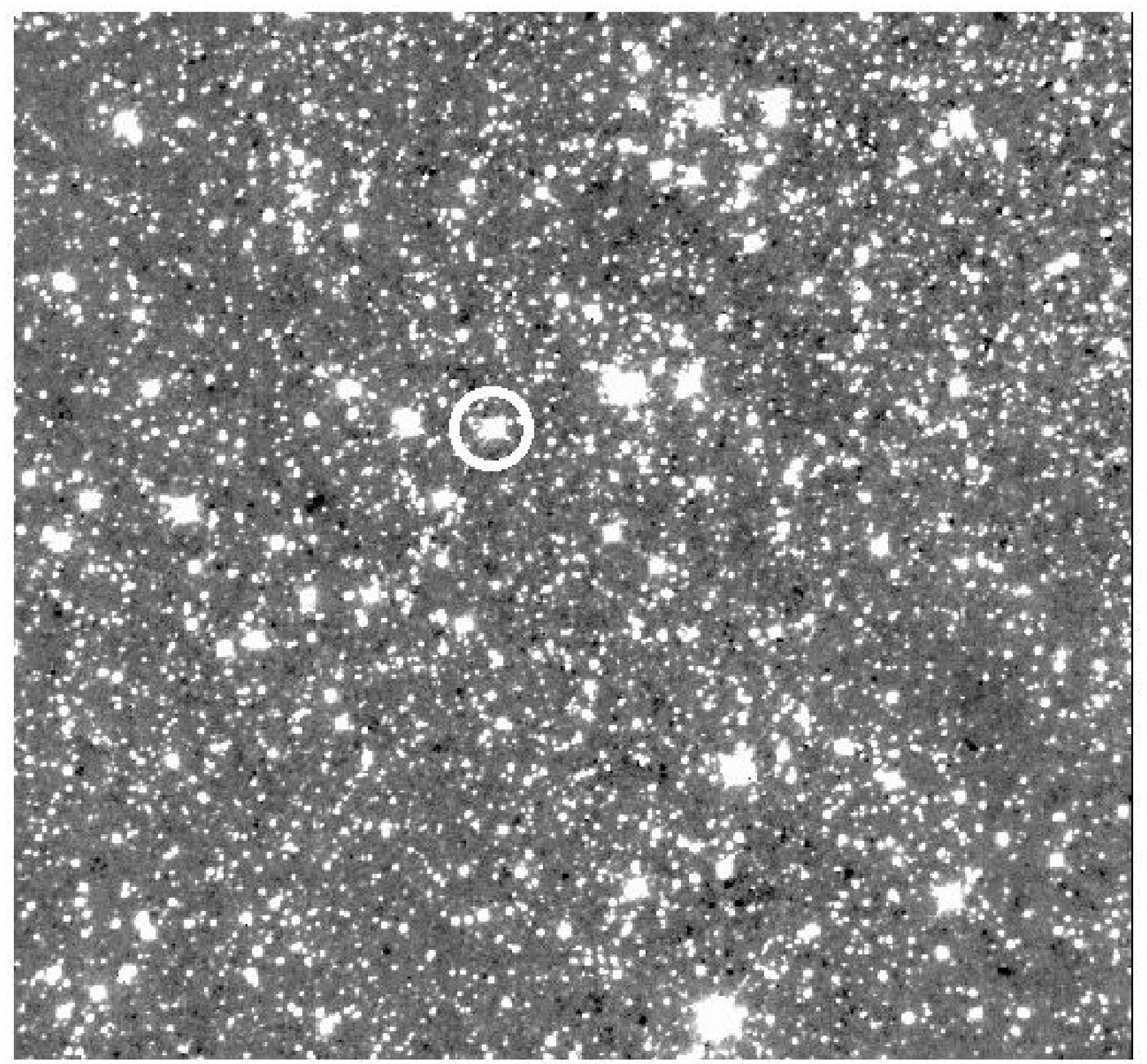}}
\subfigure[IGR\,J17391-3021]{\label{fig-abs:igrj17391}
\includegraphics[angle=0,width=8.5cm]{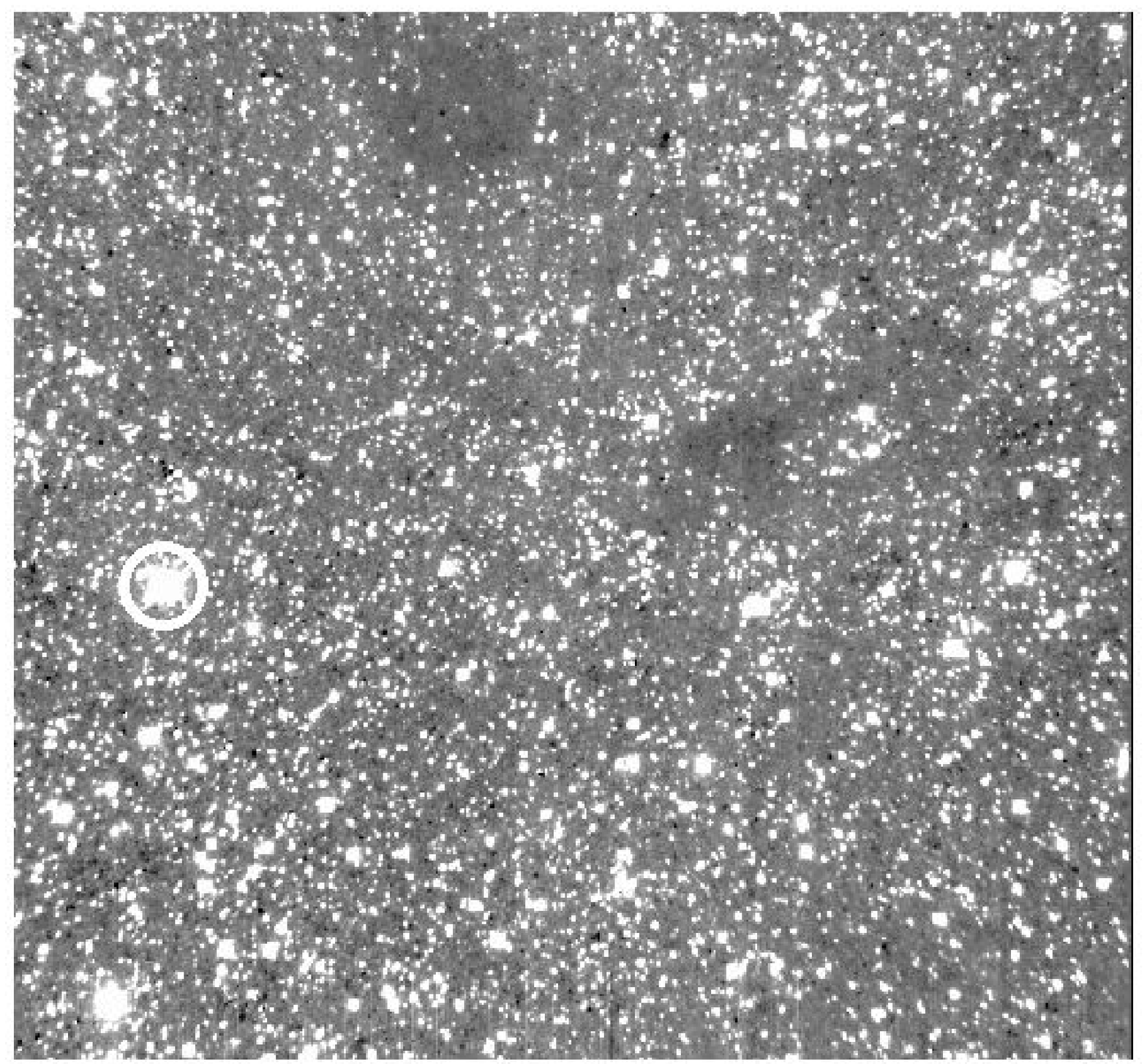}}
\caption[]{\label{figure:largefields} Large fields of 4 studied sources
exhibiting high absorption along their line of sight. Size:
$5.5\amin\times5.5\amin$; North is to the top and East to the left.  
We overplot a white
circle to indicate the position of the {\it INTEGRAL} sources.}
\end{figure*}

Four sources of the studied sample exhibit large-scale regions of
absorption in the NIR images, very close to the line of sight of the
hard X-ray source: IGR\,J16358-4726, IGR\,J16418-4532,
IGR\,J16479-4514 and IGR\,J17391-3021.  We show in
Figure~\ref{figure:largefields} the large field J band image of these
sources (except for IGR\,J16479-4514 for which we show the large field
K$_{\rm S}$ band image).  This absorption region might be due to the presence
of extended molecular clouds and/or $\hdeuxromain$ regions.  The
presence of such highly absorbed regions is not surprising close to
these peculiar hard X-ray sources, since they must be strongly linked
to their formation. For instance, the Norma arm region is one of the
richest star-forming regions of our Galaxy, where many high-mass stars
form and evolve \citep{bronfman:1996}. There exists therefore a high
probability for binary systems made up of high-mass stars to form,
and evolve into HMXBs (see for instance \citeauthor{tauris:2006}
\citeyear{tauris:2006}).  This might explain why so many
binary systems with high-mass companion stars and progenitors of
compact objects such as black holes or neutron stars have formed in
this region of our Galaxy.

\cite{bodaghee:2007} have compared the spatial distribution of HMXBs
discovered by {\it INTEGRAL} --for which the distance is known-- and
of star-forming complexes, mainly OB regions reported by
\cite{russeil:2003}. They have shown that their spatial distribution
is similar, suggesting that HMXBs are associated with these
complexes. The discovery of large-scale absorption regions in the
direction of these sources is therefore not surprising, since the
formation sites of HMXBs are closely linked to rich star forming
regions. Indeed, the short life of HMXBs prevents these systems from
migrating far away from their birthplace. Characterisation of these
large-scale absorption regions, and measurement of the metallicity of
stars hosted by these regions are required to clarify this situation.

     \subsection{HMXBs in the {\it INTEGRAL} era}

HMXBs are separated in two distinct groups.  The first group contains
the majority  of the HMXB systems, constituted of known
or suspected Be/X-ray Binary systems (BeXBs), called Be/X-ray
transients.  In Be systems, the donor is a Be star and the compact
object is a neutron star typically in a wide, moderately
eccentric orbit, spending little time in close proximity to the dense
circumstellar disk surrounding the Be companion 
(\citeauthor{coe:2000} \citeyear{coe:2000};
\citeauthor{negueruela:2004} \citeyear{negueruela:2004}).  
X-ray outbursts occur when the compact object
passes through the Be-star disk, accreting from the low-velocity and
high-density wind around Be stars, and exhibiting hard X-ray spectra.

The second group of HMXB systems contains the
Supergiant/X-ray Binaries (SXBs), where the compact object orbits deep inside
the highly supersonic wind of a supergiant early-type star, which
plays the role of the donor star \citep{kaper:2004}.  The X-ray
luminosity is powered either by accretion from the strong stellar wind
of the optical companion, or by Roche-lobe overflow.  In a wind-fed
system, accretion from the stellar wind results in a persistent X-ray
luminosity of $10^{35-36}$ erg/s, while in a Roche-lobe overflow
system, matter flows via the inner Lagrangian point to form an
accretion disc. In this case, a much higher X-ray luminosity ($\sim
10^{38}$ erg/s) is then produced during the outbursts.

In the pre-{\it INTEGRAL} era, known HMXBs were mostly BeXBs
systems.  For instance, in the catalogue of HMXBs of \cite{liu:2000},
there were 54 BeXBs and 7 SXBs identified, out of 130 HMXBs,
representing a proportion of 42\% and 5\% respectively.  Then, between
the two last editions of HMXB catalogues (\citeauthor{liu:2000}
\citeyear{liu:2000} and \citeauthor{liu:2006} \citeyear{liu:2006}),
the proportion of SXBs compared to BeXBs has increased,
with the first HMXBs identified in the {\it INTEGRAL} data.  The third
IBIS/ISGRI soft $\gamma$-ray survey catalogue \citep{bird:2007},
spanning nearly 3.5 years of operations, contains 421 sources detected
with the {\it INTEGRAL} observatory, of which 214 ($\sim 50$\,\%) 
were discovered by this satellite. This catalogue, extending up to 100
keV, includes 118 AGNs, 147 X-ray binaries (79 LMXBs and 68 HMXBs), 23
Cataclysmic Variables, 23 other objects, and 115 still unidentified
objects.  Among the 68 HMXBs, 24 have been identified as BeXBs and 19
as SXBs, representing a proportion of 35\% and 28\% respectively.
The proportion of BeXBs, relative to the total of HMXBs,
 has decreased by a factor 1.2 while the
proportion of SXBs has increased by a factor 5.6 between the
catalogue of \cite{liu:2000} and the one of \cite{bird:2007}.  Related
to this increasing proportion of SXBs, the other highlight
of the {\it INTEGRAL} catalogue is the emergence of the SFXT class,
with 12 {\it INTEGRAL} sources being firm or possible candidates.

Our studied sample of {\it INTEGRAL} sources allows us to add four
newly identified SXBs which were not classified as such in
\cite{bird:2007}: IGR\,J16320-4751, IGR\,J16358-4726, IGR\,J17252-3616
and IGR\,J18027-2016. The other sources that we have identified in
this paper as supergiants were already considered SXBs, based either on
spectral classification --IGR\,J17391-3021 and IGR\,J19140+0951-- or
on X-ray properties --the SFXT candidates IGR\,J16418-4532 and
IGR\,J16479-4514--.  With these new SXBs, the proportion of SXBs
has reached that of BeXBs: 35\% of all HMXBs for each population.
This clearly shows that the launch of {\it INTEGRAL} has drastically
changed the statistical situation concerning the nature of HMXBs, by
revealing a new dominant population of supergiant X-ray binaries,
which are purely wind accretor systems.  

Although this came as a surprise, it is {\it a posteriori} consistent
with the fact that these hard X-ray emitters are sources
ideally detected by {\it INTEGRAL}, as discussed in
\cite{lutovinov:2005a}, \cite{dean:2005} and \cite{bodaghee:2007}.
{\it INTEGRAL}, observing at energies higher than the threshold
above which photoelectric absorption becomes negligible in most
matter, can easily detect bright sources above a few tens of keV,
while they are not detectable below, and therefore had remained hidden
up to now. HMXBs accreting by stellar wind create a naturally dense
and highly absorbing circumstellar wind compared to Roche lobe
overflow in LMXBs, hiding the X-ray emission in a similar way to
Seyfert\,2 AGNs (\citeauthor{dean:2005} \citeyear{dean:2005};
\citeauthor{malizia:2003} \citeyear{malizia:2003}).

     \subsection{The nature of {\it INTEGRAL} HMXBs}

Let us now consider our results, summarized and put together with
other characteristics of these sources in Table \ref{table:results}.
In this sample of 13 {\it INTEGRAL} sources, concentrated both towards
the direction of the Norma arm and the Galactic centre, we classified
10 HMXBs (8 sources hosting sgOB and 2 sources hosting BIV-V companion
stars), 2 LMXBs and 1 AGN. We clearly confirm the predominance of
HMXBs hosting supergiants, as opposed to those hosting Be companion
stars: in our sample, 80\% of HMXBs host compact objects (probably
neutron stars) orbiting around OB supergiant secondaries. As already
discussed, this result is in agreement with the increase of SXBs in
the HMXBs population.  However, quantitatively, the proportion of
SXBs in our studied sample is higher than in the \cite{bird:2007}
catalogue. There are two reasons: first, our sample is much more
limited; second, most of the sources of our sample are in the
direction of the Norma arm, which is associated with rich star-forming
regions, the natural birthplace of massive stars, as discussed in Section
\ref{environment}.
%
Most of these new {\it INTEGRAL} sources are wind accretors,
consistent with their location in the Corbet diagram
\citep{corbet:1986}: nearly all of the HMXBs discovered by {\it
INTEGRAL}, for which both spin and orbital periods have been measured,
are located in the upper part of this diagram, among other wind
accretors, typical of supergiant HMXBs.  These systems exhibit 
extra absorption by a factor of $\sim4$ compared to the average of all
HMXBs, and they are X-ray pulsars, with longer pulsation periods than
previously known HMXBs \citep{bodaghee:2007}.

In the companion paper by \cite{rahoui:2008} it is shown that only 3
sources --IGR\,J16318-4848, IGR\,J16358-4726 and IGR\,J16195-4945--
of a sample of 12 {\it INTEGRAL} sources exhibit a MIR excess,
due to an absorbing component enshrouding the whole binary system. 
The MIR emission of the remaining sources comes from the supergiant
companion star, showing that absorbing material is enshrouding
the compact object for most of the sources (preliminary results
of this paper were described in \citeauthor{chaty:2006c}
\citeyear{chaty:2006c}).
We are therefore observing two classes of {\it INTEGRAL} sources, i.e. highly
absorbed sources such as IGR~J16358-4726, for which the extreme
representant is IGR~J16318-4848, and SFXTs such as IGR\,J17391-3021,
for which the archetype is IGR~J17544-2619. 

As shown in Table
\ref{table:results}, these classes share similar properties --for
instance they both host supergiant companion stars-- however they 
do not seem to have exactly the same configuration, and one way
to explain their different characteristics can be found in their excess in
absorption, which does not seem to have the same origin.
It is caused by two different phenomena in the case of the highly absorbed
sources: the observations from hard X-ray to MIR domains suggest the
presence of absorbing material concentrated around the compact object,
and also some dust and/or cold gas, perhaps forming a cocoon,
enshrouding the whole binary system (\citeauthor{filliatre:2004}
\citeyear{filliatre:2004}; \citeauthor{rahoui:2008}
\citeyear{rahoui:2008}).  Their characteristics might be explained by
the presence of a compact object (neutron star or black hole) orbiting
within the dense wind surrounding the companion star.
On the other hand, in the case of SFXTs, characterised by fast X-ray
outbursts, the presence of the absorbing material seems concentrated
around the compact object only, and MIR observations show that there
is no need for any other absorbing material around the whole system
\citep{rahoui:2008}. Their characteristics might be explained by the
presence of a compact object (neutron star or black hole) located on
an eccentric orbit around the companion star, and it would then be
when the compact object penetrates the clumpy circumstellar envelope
that outbursts are caused. However the situation is more complicated,
since some SFXT sources are also highly obscured, and the intrinsic
absorption derived from X-ray observations vary for some {\it
INTEGRAL} sources (see e.g. the case of IGR~J19140+0951;
\citeauthor{prat:2008} \citeyear{prat:2008}).

The X-ray transient or persistent nature of these sources might
therefore be related to the geometry of these systems, and  it is
possible to distinguish the nature of both classes by
speculating on the geometry of these systems (see also
\citeauthor{chaty:2006c} \citeyear{chaty:2006c}). However only the
knowledge of their orbital periods, which is currently unknown for
most of these sources, will allow us to confirm or not this view.  Many
questions are still open, most of them related to the
interaction between the compact object and the supergiant star, and to
the nature of the absorbing material, which seems to be present in the
environment either of the compact object, or of the whole binary
system, and even both. To better understand the general properties of
these systems, we will need to better characterise the dust, its
temperature, composition, geometry, extension around the system, and
to investigate where this dust or cold gas comes from. Finally,
probably the most important question to solve is whether this unusual
circumstellar environment is due to stellar evolution OR to the binary
system itself. To answer this question, this dominant population
of supergiant HMXBs now needs to be taken into account in
population synthesis models.

\section{Conclusions} \label{section:conclusions}

We have performed an extensive study of a
sample of 13 {\it INTEGRAL} sources, through optical and NIR
photometry and spectroscopy, performing for each source accurate
astrometry, identifying candidate counterparts, deriving their optical
and NIR magnitudes, and obtaining spectra for most of them. In
addition, we built and fitted the SED of these sources, from MIR to
hard X-rays.  We thus identified the nature of the companion stars and
of the binary systems by spectroscopy for 7 of these sources, and by
fitting their SED for 5 of them.  We finally reported NIR fields of
four of these sources, which exhibit large-scale regions of
absorption, probably linked to their formation process. We then
discussed the existence of this dominant population of supergiant
HMXBs in our Galaxy, born with two very massive components: a
population which only recently has been revealed by the {\it INTEGRAL}
observatory.

Thus, it clearly appears that a study of this new population of
supergiant HMXBs, constraining the nature and
orbital parameters of these systems, and linking them to population
synthesis models, will provide a better understanding of the evolution
of HMXBs.  These systems are probably the primary progenitors of
neutron star/neutron star or neutron star/black hole mergers.  There
is therefore the possibility that they are related to short/hard
$\gamma$-ray bursts, and also that they might be good candidates for
gravitational wave emitters. Because
most of these sources are obscured, the ''Norma arm''-like sources can
only be studied in the hard X-ray and infrared domains.  A joint study
of these sources with multi-wavelength hard X-ray, optical, NIR, MIR 
(and radio) observations is therefore necessary.


\begin{acknowledgements}
  Based on observations collected at the European Organisation for
  Astronomical Research in the Southern Hemisphere, Chile (ESO
  Programme 073.D-0339). We acknowledge Jorge Melnick for special DDT
  time at 3.6 telescope on EFOSC2.  SC thanks the ESO staff and
  especially Valenti Ivanov and Emanuela Pompei for their invaluable
  assistance during the run when we performed these optical and NIR
  observations.  JAT acknowledges partial support from {\em Chandra}
  award number GO5-6037X issued by the {\em Chandra X-Ray Observatory
    Center}, which is operated by the Smithsonian Astrophysical
  Observatory for and on behalf of the National Aeronautics and Space
  Administration (NASA), under contract NAS8-03060.  This research has
  made use of NASA's Astrophysics Data System, of the SIMBAD database,
  operated at CDS, Strasbourg, France, and of data products from the
  Two Micron All Sky Survey, which is a joint project of the
  University of Massachusetts and the Infrared Processing and Analysis
  Center/California Institute of Technology, funded by the National
  Aeronautics and Space Administration and the National Science
  Foundation.
\end{acknowledgements}

\bibliographystyle{aa} 
\bibliography{science}

\input{table_spectroscopy}

\end{document}

%% file: symbols.tex
\def\cmmoinsdeux{\mbox{ cm}^{-2}}

\def\microns{\mbox{ } \mu \mbox{m}}

\def\kpc{\mbox{ kpc}}

\def\Msol{\mbox{ }M_{\odot}}
\def\Rsol{\mbox{ }R_{\odot}}

\def\mag{\mbox{ magnitude}}
\def\mags{\mbox{ magnitudes}}

\def\ergcms{\mbox{ erg s}^{-1} \mbox{cm}^{-2}}

\def\ergcmsang{\mbox{ erg cm}^{-2} \mbox{s}^{-1} \AA}

\def\hour{^{h}}

\def\min{^{m}}
\def\minp{{\rlap.}^{m}}
\def\sec{^{s}}
\def\secp{{\rlap.}^{s}}

\def\adeg{^{\circ}}

\def\amin{^\prime}
\def\aminp{{\rlap.}^{\prime}}
\def\asec{^{\prime \prime}}
\def\asecp{{\rlap.}^{\prime \prime}}

\def\Av{A_{\rm v}}

\def\hdeuxromain{\mbox{H\,{\sc ii}}}
\def\nh{N_{\rm H}}
\def\nhone{N_{\rm HI}}

\def\ltsima{\; \buildrel < \over \sim \;}
\def\simlt{\lower.5ex\hbox{\ltsima}}            
\def\gtsima{\; \buildrel > \over \sim \;}
\def\simgt{\lower.5ex\hbox{\gtsima}}            


%% file: table_spectroscopy.tex
\onecolumn 
%
\begin{longtable}{c l l l l l l}     
\caption{\label{table:spectralines} Spectroscopy results. 
We indicate the name of the sources, the identification of the lines,
the rest wavelength ($\mu m$), the fitted central wavelength
($\mu m$), the flux (in $\ergcms$ for optical spectra and in arbitrary
units for NIR spectra), the equivalent width (EQW in $\AA$), and Full
Width Half Maximum (FWHM in $\AA$). T stands for telluric.} \\
\hline\hline       
Source & Identification   & $\lambda$ & $\lambda_{fit}$ & Flux & EQW & FWHM \\
\hline                    
\endfirsthead
\caption{continued.}\\
\hline\hline
Source & Identification   & $\lambda$ & $\lambda_{fit}$ & Flux & EQW & FWHM \\
\hline
\endhead
\hline
\endfoot
IGR\,J16320-4751 & & &     &        &      &    \\
& Pa(7-3)  & 1.005 & 1.0052  & 1296.88 & -3.773$\pm$1.3 & 55.6  \\
&          & & 1.1757 & 855.62  & -1.889  & 51.35 \\
&         & & 1.2609 & 673.3   & -1.275  & 29.4  \\
&         & & 1.5064 & 689.659 & -0.9421 & 8.044 \\
&Br(17-4) & 1.544 & 1.5442  & 1214.05 & -1.529$\pm$0.4  & 12.76 \\
&         & & 1.5479 & 2708.57 & -3.044  & 112.4 \\
&         & & 1.5829  & 0.0120243 & -0.175554 & 14.6  \\
 &  & & 2.0998  & 2160.39   & -1.656    & 88.75 \\
&Br(7-4)/HeI  & 2.166/2.166 & 2.1661 & 7327.22 & -5.454$\pm$1.0  & 114.6  \\ 
&T & & 2.2962 & -10695.5 &    7.402 &     51.66  \\
&T & & 2.3248 & -8645.31 &    6.087 &    38.76   \\
%
IGR\,J16358-4726 & & &     &        &      &  \\
&HeII & 1.0133 & 1.0138 & -331.65  & 24.01  & 7.408 \\
& & & 1.1935 &  803.177 & -9.219 & 2.876 \\ 
&HeII & 1.2914 & 1.2908 &  878.037 & -5.591 & 25.71 \\
&HeI & 1.2971 & 1.2966 & -285.973 &  1.77  & 12.26 \\
&? & & 1.5061 & 5060.16  & -14.96 & 25.61 \\
&? & & 1.5839 & 3229.4   & -7.776 & 25.37 \\
& & &  1.6135 & 2407.31  & -5.477 & 21.55 \\
& & & 1.6701 & -1731.86 & 10.56  & 46.58 \\
&Br(11-4)& 1.681 & 1.6811 & 5716.71 & -40.69 & 133.3   \\
&Br(10-4)& 1.737 & 1.7177 & 1292.73 & -7.716 & 46.55  \\ 
& & & 1.7662 & -2008.36 & 10.33  & 40.9  \\ 
& & & 1.7906 & 902.755  & -4.666 & 13.32 \\
& & & 1.8000 & -1825.61 & 9.3    & 23.42 \\
&Br(9-4) & 1.818 & 1.8206 & 3066.46 & -15.68 & 92.93  \\
&Br(8-4) & 1.945 & 1.9448 & 561.476 & -2.306 &  14.81 \\ 
&FeII? & 1.9746 & 1.9811 &  1281.53 & -5.106 &  2.16 \\
& & & 2.0018 & -4330.84 & 16.86  & 38.31 \\
&HeI? & 2.058 & 2.0566 & -2011.82 &  7.386 & 30.4  \\
& & & 2.1216 & 944.583 & -3.257 & 23.41 \\
& & & 2.1518 & -2205.66 &  7.095 & 47.87 \\
&Br(7-4) & 2.166 & 2.1643 & 1585.54 & -5.174 &  70.49 \\
&$[$FeII$]$ & 2.224 & 2.2223 & 3877.44 & -12.85 & 126.6 \\
&HeI & 2.3069 & 2.2934 & -1053.51 &  3.128 & 17.71 \\ \hline
%
%
%
%
IGR\,J16479-4514 && &     &        &      &  \\
& & &  0.9811 &   417.44 &   -39.77 &      28.98 \\
& & & 1.0597 &    282.616 &   -8.639 &        5.383 \\
& & &  1.1293 &   661.497 &   -10.13 &      69.46 \\
& & & 1.2616 &    376.308 &   -2.184 &        5.55 \\
&HeII? & 1.3150 & 1.3178 & -489.818 &    2.034 &      17.33 \\
& & &  1.4602 &    336.183 &  -0.7073 &       7.066 \\
& & &  1.4679 &   -280.531 &   0.5932 &      7.727 \\
&HeII & 1.4882 &  1.4876 &    595.672 &   -1.141 &       15.05 \\
&Br(23-4) & 1.504 & 1.5046 &    2512.46 &   -4.622 &       39.73 \\
&Br(19-4) & 1.526 & 1.5269 &   -1037.44 &    1.767 &      18.67 \\
&FeII & 1.576 & 1.5760 & 1858.42 &   -2.719 &       35.1 \\
&Br(14-4) & 1.588 & 1.5900 &    3130.59 &   -4.995 &        70.96 \\
& HeI & 1.70 & 1.6998 &   -1075.65 &     1.29 &      22.05 \\
& & & 1.7891 &  -2371.63  &    2.32  &      20.62 \\
&Br(8-4) & 1.945 & 1.9475 &    8758.25 &   -6.548 &      92.96 \\
& & & 1.9822 &    4953.06 &   -3.442 &      38.83 \\
& & & 2.0076 &   -22693.7 &    15.46 &      192.7 \\
& & & 2.0547  &  -4648.77  &   2.992  &      90.7 \\
&HeI & 2.1126 & 2.1117 &    -2376.81 &    1.379 &      30.27 \\
&Br(7-4)/HeII & 2.166/2.166 & 2.1658 &    6605.93 &   -3.652 &       51.56 \\ 
& & & 2.2818 &    2197.39 &   -1.054 &       47.44 \\
&T & & 2.3294 &   -2606.04 &    1.184 &      22.37 \\
%
%
IGR\,J17252-3616 && & &    &        &       \\
&HeI & & 0.7268 & 9.590E-17 & -61.37 & 32.97 \\
& & & 0.7375 & 6.308E-17 &   -34.66 &     21.14 \\
& & & 0.7622 & -1.29E-16 & 48.04  & 66.65 \\
& & & 0.7719 & 1.136E-16 & -39.71 & 37.44 \\
& & & 0.7788  & -1.06E-16 & 22.64  & 31.12 \\
&T & & 0.8277 & -1.30E-16 & 31.32  & 26.38 \\
& & & 0.8341 & 9.047E-16 & -195.9 & 39.92 \\
&Pa18 & 0.8437 & 0.8440 & 6.046E-16 & -106.5 & 44.48 \\
& & & 0.8819 & 1.176E-15 & -169.9 & 58.22 \\
& & & 0.8921 & 6.742E-16 & -91.68 & 46.02 \\
& & & 0.9301 & 5.538E-16 & -50.68 & 33.24 \\ 
& & & 0.9397 & 1.435E-16 &   -15.17 &      21.22 \\
& & & 0.9783 & 5.476E-16 & -48.98 & 57.03 \\ \hline
& & & 1.1933 & 17258.4   & -1.55  & 6.047 \\
&HeII & 1.2719 & 1.2717 &   -205225. &    12.25  &     116.1 \\
& & & 1.3652 & -1927788. & 80.62  & 151.6 \\
&FeII? & 1.576 & 1.5774 & 170625.   & -4.023 & 63.2  \\
&Br(13-4) & 1.611 & 1.6200 &    167968. &   -3.916 &       95.54 \\
&PCyg- & & 2.0555 &   -277866. &    5.281 &     63.73 \\
&HeI PCyg+ & 2.0587 & 2.0597 & 163020. & -3.08 & 23.96 \\
&HeI & 2.1126 & 2.1129 &   -105139. &    2.004 &       29.24 \\
&HeII & 2.1891 & 2.1894 & 158546.   & -3.309 & 63.66 \\
& & & 2.3190 &  -203026. &    3.913 &      25.96 \\
%
%
\hline                    
IGR~J17391-3021&  & &     &        &      & \\
&NaD & 0.5780 & 0.5786 & -9.84E-15  &   2.512 &     29.57 \\
&NaD & 0.5890 & 0.5894 & -5.74E-15  &   1.316 &     16.76 \\
& & &  0.6141 & 7.903E-15 &   -1.462 &     28.81 \\
&    & &  0.6183 & -9.74E-15  &   1.736 &     31.93 \\
&HeII & & 0.6286 & -1.18E-14 &     2.01 &     15.24 \\
&H$\alpha$ & 0.6562 & 0.6568 & 2.207E-14 &   -2.804  &    23.84 \\
&T & &  0.6888 & -7.97E-14 &    7.934  &    52.38 \\
& & &  0.7068 & -7.77E-15 &   0.6783 &     16.38 \\
&T & & 0.7189 & -2.15E-14 &    1.767 &      24.4 \\
& & &  0.7395 & 9.707E-15  & -0.7171 & 32.32 \\
& & &  0.7458 & 3.928E-15 &  -0.2813 & 11.34 \\
&T & & 0.7629 & -6.50E-13 &    41.87 &     78.06 \\
& & &  0.7931 & -1.15E-14  &  0.6468 &      22.8 \\
& & & 0.8171 & -5.01E-14 &    2.624 &       41. \\
& & & 0.8237 &-2.64E-14  &   1.349  &    18.94 \\
&T & &  0.8290 & -4.75E-15 &    0.241 &     22.14 \\
&Pa20 & 0.8392 & 0.8388 & -1.62E-15  & 0.07962  &    8.758 \\
&Pa16 & 0.8502 & 0.8505 & -1.11E-14 &   0.5035 &     13.02 \\
&Pa15 & 0.8545 & 0.8552 & -1.23E-14 &   0.5532 &     18.17 \\
&NI & 0.8703 & 0.8712 & 1.028E-13 &   -4.347 &     50.04 \\
&Pa12 & 0.8750 & 0.8756 & -3.44E-14 &    1.348 &     22.97 \\
& & & 0.8868 & -4.96E-14 &    1.887 &     25.83 \\
& & & 0.9006 & -1.66E-13 &    6.293 &     49.69 \\
& & & 0.9118 & 7.353E-15 &  -0.2933 &      17.46 \\
& & & 0.9201 & 4.781E-14 &   -1.887 &     28.79 \\
& & & 0.9230 & -4.13E-14 &    1.545 &     22.38 \\
& & & 0.9349 & -5.66E-13 &    22.48 &     64.58 \\
& & & 0.9423 & -1.49E-12 &     57.2 &     309.8 \\
& & & 0.9447  &-2.00E-12  &   72.96  &    274.8 \\
& & & 0.9504 & -8.41E-13 &    32.56 &     178.7 \\
&Pa(8-3) & 0.954 & 0.9550 & -1.19E-13 &    4.949 &     44.28 \\
& & & 0.9712 & 1.268E-13 &   -4.463 &     36.45 \\
& & & 0.9768 & -2.08E-13 &    6.634 &     73.77 \\
& & & 0.9930 & 7.416E-13 &   -25.57 &     218.4 \\
& & & 1.0111 & 1.372E-14 &  -0.4479 &     13.33 \\
& & & 1.0093 & 5.657E-14 &   -1.853 &     34.91 \\
&Pa(7-3)/HeII & 1.005/1.0049 & 1.0053 & -6.23E-14 &    1.959 &     25.14 \\
&HeII & 1.0133 & 1.0133 & -1.24E-14 &   0.4036 &     12.24 \\ \hline
& & & 0.9710 & 13487.8  & -2.658 & 24.21 \\
&Pa(7-3)/HeII & 1.005/1.0049 & 1.0047 & -3997.88 & 0.7305 & 9.054 \\
& & &   1.0123 & -4141.89 & 0.7398 & 9.734 \\
&HeI-D & 1.0832-3 & 1.0834 & 45732.6  & -7.06  & 42.2 \\
&Pa(6-3)/HeII & 1.094/1.0938 & 1.0937 & -9981.06 &  1.472 & 15.88 \\
&HeI & 1.1972 & 1.1971 & -11849.5 &  1.522 & 15.93 \\
& & & 1.2275 &    8130.68 &   -1.014  &       27.2 \\
& & &  1.2609 & -7058.26 &  0.851 & 1.225 \\
&HeI & 1.2789 & 1.2787 & -7781.74 & 0.9251 & 9.955 \\
&Pa(5-3)/HeII & 1.282/1.2817 & 1.2812 & 24041.7  & -2.847 & 69.16 \\
&OI & 1.3165 & 1.3178 & -9810.79 &  1.124 & 10.69 \\
&Br(23-4) & 1.504 & 1.5034 & 33968.9  & -3.427 & 39.53 \\
&Br(19-4) PCyg- & 1.526 & 1.5269 & -10867.7 &  1.074 & 11.76 \\
&PCyg+ & & 1.5296 &     2753.08 &  -0.2737 &       7.687 \\
&Br(17-4) & 1.544 & 1.5449 & -5302.73 & 0.5161 & 12.05 \\
&Br(15-4)/HeII PCyg- & 1.571/1.5719 & 1.5720 & -8690.81 &  0.816 & 11.44 \\
&PCyg+ & & 1.5763  &   28064.4 &    -2.625   &     34.17 \\
&HeII? & 1.6241 & 1.6286 & -14464.6 &  2.554 & 83.49 \\
&HeII & 1.6584 & 1.6591 & -8012.45 &  1.407 & 59.07 \\
&HeI PCyg- & 1.7007 & 1.7001 & -9883.17 &  1.721 & 25.47 \\
&PCyg+ & & 1.7108  &   7183.83  &  -1.248    &    46.05 \\
&Br(10-4)/HeII & 1.737/1.7355 & 1.7345 & 18147.7  & -3.105 & 82.78 \\
&? & & 2.0023 & 48773.3  & -8.054 & 53.72 \\
&? & & 2.0227 & -8537.24 &  1.394 & 54.94 \\
&HeI & 2.0587 & 2.0583 & -19765.2 &  3.163 & 32.95 \\
&NIII+CIII & 2.116 & 2.1157 & 11898.8  & -1.87  & 28.88 \\
&Br(7-4)/HeI & 2.166/2.166 & 2.1652 & 14485.3  & -2.267 & 48.56 \\
&HeII & 2.1891 & 2.1887 & -5196.59 & 0.8194 & 24.93 \\ \hline
%
IGR\,J18027-2016 && &    &      &      &  \\
&H$\zeta$ & 0.3889 & 0.3872 & -1.64E-16 &  85.36 &  13.75 \\
&H$\delta$ & 0.4101 & 0.4109 & 1.403E-15 &  INDEF &   13.79 \\
&HeII & 0.4200 & 0.4201 & -7.06E-16 & 2460.  &  8.549 \\
& & & 0.4614 & -5.94E-16 &  79.65 &   7.906 \\
&HeII & 0.4686 & 0.4696 & 2.762E-16 & -28.62 &   20.08 \\
&H$\beta$? & 0.4861 & 0.4901 & 9.769E-16 & -75.71 &   10.23 \\
& & & 0.5180 & -7.98E-16 &  51.37 &   9.967 \\
& & & 0.5460 & 2.099E-16 & -7.693 &   12.53 \\
&NaD & 0.5890 & 0.5884 & -1.99E-16 & 3.885  &   20.76 \\
& & & 0.6251 & 3.172E-16 & -5.072 &  42.17 \\
&H$\alpha$ & 0.6562 & 0.6570 & 6.033E-16 & -6.788 &  44.4 \\
&T & & 0.6885 & -1.21E-15 &  10.03 &  59.22 \\
& & & 0.7111 & -6.80E-17 & 0.4671 &   7.604 \\
& & & 0.7187 & -1.65E-16 &    1.096 &     21.06 \\
& & & 0.7380 & 3.866E-16 & -2.221 &   43.12 \\
& & & 0.7415 & -3.43E-16 &    1.845 &       31. \\
& & & 0.7621 & -8.49E-15 &  40.47 &   66.51 \\
& & & 0.8077 & 3.062E-16 & -1.061 &   20.52 \\
&T & & 0.8240 & -4.18E-16 &  1.409 &   10.03 \\
& & & 0.8345 & 1.376E-15 & -4.163 &   33.41 \\
& & & 0.8493 & -1.38E-15 &  3.771 &   22.71 \\
& & & 0.8819 & 2.979E-15 & -7.014 &   50.99 \\
& & & 0.8914  & 2.231E-15 &  -5.194&   42.9 \\
& & & 0.9001 & -1.79E-15 &  4.164 &  40.1 \\
& & & 0.9348 & -5.29E-15 &  13.54 &   50.03 \\
& & & 0.9441 & -1.93E-15 &   5.08 &   34.31 \\
&Pa(8-3) & 0.9549 & 0.9551 & -4.02E-15 &  9.739 &   83.46 \\ \hline
& & & 1.0193 & 169299.   & -16.81 &   67.81 \\
&HeI & 1.1016 & 1.1026  & -188368.  & 13.04  &   111.4 \\
& & & 1.2747 & -267133.  &  11.87 &  89.39 \\
&Br(17-4)? & 1.544 & 1.5484 &  257952.  & -8.298 &  116.5 \\
& & & 1.5813 & -137910.  &  4.266 &  59.41 \\
&PCyg- & & 1.6010 &  -87404.2  &   2.733  &   35.72 \\
&PCyg+? & & 1.6060 &  46540.9  & -1.455 &  7.072 \\
&HeII? & 1.6241 & 1.6220 & -105641.  &  3.331 &   41.99 \\
&Br(12-4)? PCyg- & 1.641 & 1.6439 & -340789.  &  10.27 &  82.88 \\
&PCyg+ & & 1.6493 &   -14621.9 &    0.4325 &     1.403 \\
&PCyg- & & 2.0033 & -644298.  &  25.66 &   59.55 \\
&PCyg+ & & 2.0103 &    123941.&    -4.991 &       29.17 \\
& & & 2.2983 & -299051.  &  14.65 &   55.77 \\
&T & & 2.3262 & -103670.  &   5.96 &   25.2 \\
&T & & 2.3563 & -226137.  &  14.08 &  59.82 \\
&T & & 2.3867 & -113159.  &  7.989 &  29.05 \\
&T & &   2.4189 &   -141757. &    9.892  &    37.36 \\ \hline
%
%
IGR\,J19140+0951 &&     &        &      &     \\
& & & 0.9787 &    254.752 &  -0.5518 &       8.255 \\
&Pa(7-3) & 1.005 & 1.0053 &   -179.359 &    0.397 &      10.52 \\
&- & & 1.0081 &     113.557&   -0.2515&         8.25 \\
&- & & 1.0595 &     193.383&   -0.4397&        6.494 \\
&- & & 1.1919 &       554. &    -1.31 &        15.17 \\
&- & & 1.2613 &    247.094 &  -0.5914 &        7.01 \\
&Pa(5-3)/HeII & 1.282/1.2817 & 1.2821 &   -359.765 &     0.86 &      13.84 \\
&- & & 1.5830 & 1002.7   & -3.077   & 83.78 \\
&- & & 1.5845 &   -414.155 &   0.9797 &       7.817 \\
&Br(14-4)? & 1.588 & 1.5895 &   -517.017 &    1.205 &         29.69 \\
&HeII & 1.6584 & 1.6592 & 324.516  &  -0.9892 & 12.57 \\
& & & 1.6687 & 1036.81  & -3.168   & 95.57 \\
&  & & 1.6834 &   -149.388 &   0.4517  &     7.682 \\
&HeII & 1.8725 & 1.8728&     1335.31 &   -4.457  &       11.47 \\
& & & 2.0070 & -12037.2 &  37.72   & 187.7 \\
&HeI & 2.0587 & 2.0600 & 563.242 & -1.878 & 30.69 \\
&HeI & 2.1126 & 2.1129 &  169.253 & -0.5294 & 18.93 \\ \hline
%
\hline                  
\end{longtable}

%% file: integral.bbl
\begin{thebibliography}{85}
\expandafter\ifx\csname natexlab\endcsname\relax\def\natexlab#1{#1}\fi

\bibitem[{{Bird} {et~al.}(2006){Bird}, {Barlow}, {Bassani}, {Bazzano},
  {B{\'e}langer}, {Bodaghee}, {Capitanio}, {Dean}, {Fiocchi}, {Hill}, {Lebrun},
  {Malizia}, {Mas-Hesse}, {Molina}, {Moran}, {Renaud}, {Sguera}, {Shaw},
  {Stephen}, {Terrier}, {Ubertini}, {Walter}, {Willis}, \&
  {Winkler}}]{bird:2006}
{Bird}, A.~J., {Barlow}, E.~J., {Bassani}, L., {et~al.} 2006, \apj, 636, 765

\bibitem[{{Bird} {et~al.}(2007){Bird}, {Malizia}, {Bazzano}, {Barlow},
  {Bassani}, {Hill}, {B{\'e}langer}, {Capitanio}, {Clark}, {Dean}, {Fiocchi},
  {G{\"o}tz}, {Lebrun}, {Molina}, {Produit}, {Renaud}, {Sguera}, {Stephen},
  {Terrier}, {Ubertini}, {Walter}, {Winkler}, \& {Zurita}}]{bird:2007}
{Bird}, A.~J., {Malizia}, A., {Bazzano}, A., {et~al.} 2007, \apjs, 170, 175

\bibitem[{{Bodaghee} {et~al.}(2007){Bodaghee}, {Courvoisier}, {Rodriguez},
  {Beckmann}, {Produit}, {Hannikainen}, {Kuulkers}, {Willis}, \&
  {Wendt}}]{bodaghee:2007}
{Bodaghee}, A., {Courvoisier}, T.~J.-L., {Rodriguez}, J., {et~al.} 2007, \aap,
  467, 585

\bibitem[{{Bodaghee} {et~al.}(2006){Bodaghee}, {Walter}, {Zurita Heras},
  {Bird}, {Courvoisier}, {Malizia}, {Terrier}, \& {Ubertini}}]{bodaghee:2006}
{Bodaghee}, A., {Walter}, R., {Zurita Heras}, J.~A., {et~al.} 2006, \aap, 447,
  1027

\bibitem[{{Bohlin} {et~al.}(1978){Bohlin}, {Savage}, \& {Drake}}]{bohlin:1978}
{Bohlin}, R.~C., {Savage}, B.~D., \& {Drake}, J.~F. 1978, \apj, 224, 132

\bibitem[{{Bronfman} {et~al.}(1996){Bronfman}, {Nyman}, \&
  {May}}]{bronfman:1996}
{Bronfman}, L., {Nyman}, L.-A., \& {May}, J. 1996, \aaps, 115, 81

\bibitem[{{Capitanio} {et~al.}(2006){Capitanio}, {Bazzano}, {Ubertini},
  {Zdziarski}, {Bird}, {De Cesare}, {Dean}, {Stephen}, \&
  {Tarana}}]{capitanio:2006}
{Capitanio}, F., {Bazzano}, A., {Ubertini}, P., {et~al.} 2006, \apj, 643, 376

\bibitem[{{Cardelli} {et~al.}(1989){Cardelli}, {Clayton}, \&
  {Mathis}}]{cardelli:1989}
{Cardelli}, J.~A., {Clayton}, G.~C., \& {Mathis}, J.~S. 1989, \apj, 345, 245

\bibitem[{{Caron} {et~al.}(2003){Caron}, {Moffat}, {St-Louis}, {Wade}, \&
  {Lester}}]{caron:2003}
{Caron}, G., {Moffat}, A.~F.~J., {St-Louis}, N., {Wade}, G.~A., \& {Lester},
  J.~B. 2003, \aj, 126, 1415

\bibitem[{{Chaty} \& {Filliatre}(2005)}]{chaty:2005a}
{Chaty}, S. \& {Filliatre}, P. 2005, \apss, 297, 235

\bibitem[{{Chaty} \& {Rahoui}(2006)}]{chaty:2006c}
{Chaty}, S. \& {Rahoui}, F. 2006, in The Obscured Universe, Procs. of 6th
  INTEGRAL workshop, Moscow, Russia, July 2-8, 2006, to be published by ESA's
  Publications Division in December 2006 as Special Publication SP-622, in
  press (astro-ph/0609474)

\bibitem[{{Chernyakova} {et~al.}(2003){Chernyakova}, {Lutovinov}, {Capitanio},
  {Lund}, \& {Gehrels}}]{chernyakova:2003}
{Chernyakova}, M., {Lutovinov}, A., {Capitanio}, F., {Lund}, N., \& {Gehrels},
  N. 2003, The Astronomer's Telegram, 157, 1

\bibitem[{{Coe}(2000)}]{coe:2000}
{Coe}, M.~J. 2000, in ASP Conf. Ser. 214: IAU Colloq. 175: The Be Phenomenon in
  Early-Type Stars, ed. M.~A. {Smith}, H.~F. {Henrichs}, \& J.~{Fabregat},
  656--+

\bibitem[{{Corbet} {et~al.}(2006){Corbet}, {Barbier}, {Barthelmy}, {Cummings},
  {Fenimore}, {Gehrels}, {Hullinger}, {Krimm}, {Markwardt}, {Palmer},
  {Parsons}, {Sakamoto}, {Sato}, {Tueller}, \& {Remillard}}]{corbet:2006}
{Corbet}, R., {Barbier}, L., {Barthelmy}, S., {et~al.} 2006, The Astronomer's
  Telegram, 779, 1

\bibitem[{{Corbet} {et~al.}(2005){Corbet}, {Barbier}, {Barthelmy}, {Cummings},
  {Fenimore}, {Gehrels}, {Hullinger}, {Krimm}, {Markwardt}, {Palmer},
  {Parsons}, {Sakamoto}, {Sato}, {Tueller}, \& {The Swift-Survey
  Team}}]{corbet:2005}
{Corbet}, R., {Barbier}, L., {Barthelmy}, S., {et~al.} 2005, The Astronomer's
  Telegram, 649, 1

\bibitem[{{Corbet}(1986)}]{corbet:1986}
{Corbet}, R.~H.~D. 1986, \mnras, 220, 1047

\bibitem[{{Corbet} {et~al.}(2004){Corbet}, {Hannikainen}, \&
  {Remillard}}]{corbet:2004}
{Corbet}, R.~H.~D., {Hannikainen}, D.~C., \& {Remillard}, R. 2004, The
  Astronomer's Telegram, 269, 1

\bibitem[{{Dean} {et~al.}(2005){Dean}, {Bazzano}, {Hill}, {Stephen}, {Bassani},
  {Barlow}, {Bird}, {Lebrun}, {Sguera}, {Shaw}, {Ubertini}, {Walter}, \&
  {Willis}}]{dean:2005}
{Dean}, A.~J., {Bazzano}, A., {Hill}, A.~B., {et~al.} 2005, \aap, 443, 485

\bibitem[{{Dickey} \& {Lockman}(1990)}]{dickey:1990}
{Dickey}, J.~M. \& {Lockman}, F.~J. 1990, \araa, 28, 215

\bibitem[{{Filliatre} \& {Chaty}(2004)}]{filliatre:2004}
{Filliatre}, P. \& {Chaty}, S. 2004, \apj, 616, 469

\bibitem[{{Hannikainen} {et~al.}(2007){Hannikainen}, {Rawlings}, {Muhli},
  {Vilhu}, {Schultz}, \& {Rodriguez}}]{hannikainen:2007}
{Hannikainen}, D.~C., {Rawlings}, M.~G., {Muhli}, P., {et~al.} 2007, \mnras,
  380, 665

\bibitem[{{Hannikainen} {et~al.}(2003){Hannikainen}, {Rodriguez}, \&
  {Pottschmidt}}]{hannikainen:2003}
{Hannikainen}, D.~C., {Rodriguez}, J., \& {Pottschmidt}, K. 2003, \iaucirc,
  8088, 4

\bibitem[{{Hanson} {et~al.}(2005){Hanson}, {Kudritzki}, {Kenworthy}, {Puls}, \&
  {Tokunaga}}]{hanson:2005}
{Hanson}, M.~M., {Kudritzki}, R.-P., {Kenworthy}, M.~A., {Puls}, J., \&
  {Tokunaga}, A.~T. 2005, \apjs, 161, 154

\bibitem[{{Hill} {et~al.}(2005){Hill}, {Walter}, {Knigge}, {Bazzano},
  {B{\'e}langer}, {Bird}, {Dean}, {Galache}, {Malizia}, {Renaud}, {Stephen}, \&
  {Ubertini}}]{hill:2005}
{Hill}, A.~B., {Walter}, R., {Knigge}, C., {et~al.} 2005, \aap, 439, 255

\bibitem[{{in't Zand} {et~al.}(2006){in't Zand}, {Jonker}, {Nelemans},
  {Steeghs}, \& {O'Brien}}]{intzand:2006}
{in't Zand}, J.~J.~M., {Jonker}, P.~G., {Nelemans}, G., {Steeghs}, D., \&
  {O'Brien}, K. 2006, \aap, 448, 1101

\bibitem[{{Kaper} {et~al.}(2004){Kaper}, {van der Meer}, \&
  {Tijani}}]{kaper:2004}
{Kaper}, L., {van der Meer}, A., \& {Tijani}, A.~H. 2004, in Revista Mexicana
  de Astronomia y Astrofisica Conference Series, ed. C.~{Allen} \& C.~{Scarfe},
  128--131

\bibitem[{{Kennea} \& {Capitanio}(2007)}]{kennea:2007}
{Kennea}, J.~A. \& {Capitanio}, F. 2007, The Astronomer's Telegram, 1140, 1

\bibitem[{{Kouveliotou} {et~al.}(2003){Kouveliotou}, {Patel}, {Tennant},
  {Woods}, {Finger}, \& {Wachter}}]{kouveliotou:2003}
{Kouveliotou}, C., {Patel}, S., {Tennant}, A., {et~al.} 2003, \iaucirc, 8109, 2

\bibitem[{{Kuulkers} {et~al.}(2003){Kuulkers}, {Lutovinov}, {Parmar},
  {Capitanio}, {Mowlavi}, \& {Hermsen}}]{kuulkers:2003}
{Kuulkers}, E., {Lutovinov}, A., {Parmar}, A., {et~al.} 2003, The Astronomer's
  Telegram, 149, 1

\bibitem[{{Landolt}(1992)}]{landolt:1992}
{Landolt}, A.~U. 1992, \aj, 104, 340

\bibitem[{{Lebrun} {et~al.}(2003){Lebrun}, {Leray}, {Lavocat}, {Cr{\'e}tolle},
  {Arqu{\`e}s}, {Blondel}, {Bonnin}, {Bou{\`e}re}, {Cara}, {Chaleil}, {Daly},
  {Desages}, {Dzitko}, {Horeau}, {Laurent}, {Limousin}, {Mathy}, {Mauguen},
  {Meignier}, {Molini{\'e}}, {Poindron}, {Rouger}, {Sauvageon}, \&
  {Tourrette}}]{lebrun:2003}
{Lebrun}, F., {Leray}, J.~P., {Lavocat}, P., {et~al.} 2003, \aap, 411, L141

\bibitem[{{Levine} \& {Corbet}(2006)}]{levine:2006}
{Levine}, A.~M. \& {Corbet}, R. 2006, The Astronomer's Telegram, 940, 1

\bibitem[{{Liu} {et~al.}(2000){Liu}, {van Paradijs}, \& {van den
  Heuvel}}]{liu:2000}
{Liu}, Q.~Z., {van Paradijs}, J., \& {van den Heuvel}, E.~P.~J. 2000, {Astron.
  Astrophys. Suppl. Ser.}, 147, 25

\bibitem[{{Liu} {et~al.}(2006){Liu}, {van Paradijs}, \& {van den
  Heuvel}}]{liu:2006}
{Liu}, Q.~Z., {van Paradijs}, J., \& {van den Heuvel}, E.~P.~J. 2006, \aap,
  455, 1165

\bibitem[{{Lutovinov} {et~al.}(2005{\natexlab{a}}){Lutovinov}, {Revnivtsev},
  {Gilfanov}, {Shtykovskiy}, {Molkov}, \& {Sunyaev}}]{lutovinov:2005a}
{Lutovinov}, A., {Revnivtsev}, M., {Gilfanov}, M., {et~al.} 2005{\natexlab{a}},
  \aap, 444, 821

\bibitem[{{Lutovinov} {et~al.}(2005{\natexlab{b}}){Lutovinov}, {Rodriguez},
  {Revnivtsev}, \& {Shtykovskiy}}]{lutovinov:2005b}
{Lutovinov}, A., {Rodriguez}, J., {Revnivtsev}, M., \& {Shtykovskiy}, P.
  2005{\natexlab{b}}, \aap, 433, L41

\bibitem[{{Lutovinov} {et~al.}(2003){Lutovinov}, {Walter}, {Belanger}, {Lund},
  {Grebenev}, \& {Winkler}}]{lutovinov:2003c}
{Lutovinov}, A., {Walter}, R., {Belanger}, G., {et~al.} 2003, The Astronomer's
  Telegram, 155, 1

\bibitem[{{Malizia} {et~al.}(2004){Malizia}, {Bassani}, {di Cocco}, {Stephen},
  {Walter}, {Bodaghee}, \& {Bazzano}}]{malizia:2004}
{Malizia}, A., {Bassani}, L., {di Cocco}, G., {et~al.} 2004, The Astronomer's
  Telegram, 227, 1

\bibitem[{{Malizia} {et~al.}(2003){Malizia}, {Bassani}, {Stephen}, {Di Cocco},
  {Fiore}, \& {Dean}}]{malizia:2003}
{Malizia}, A., {Bassani}, L., {Stephen}, J.~B., {et~al.} 2003, \apjl, 589, L17

\bibitem[{{Malizia} {et~al.}(2007){Malizia}, {Landi}, {Bassani}, {Bird},
  {Molina}, {De Rosa}, {Fiocchi}, {Gehrels}, {Kennea}, \&
  {Perri}}]{malizia:2007}
{Malizia}, A., {Landi}, R., {Bassani}, L., {et~al.} 2007, \apj, 668, 81

\bibitem[{{Markwardt} \& {Swank}(2003)}]{markwardt:2003c}
{Markwardt}, C.~B. \& {Swank}, J.~H. 2003, The Astronomer's Telegram, 156, 1

\bibitem[{{Masetti} {et~al.}(2006){Masetti}, {Morelli}, {Palazzi}, {Galaz},
  {Bassani}, {Bazzano}, {Bird}, {Dean}, {Israel}, {Landi}, {Malizia},
  {Minniti}, {Schiavone}, {Stephen}, {Ubertini}, \& {Walter}}]{masetti:2006}
{Masetti}, N., {Morelli}, L., {Palazzi}, E., {et~al.} 2006, \aap, 459, 21

\bibitem[{{Mill}(1994)}]{mill:1994}
{Mill}, J.~D. 1994, in Proc. SPIE Vol. 2232, p. 200-216, Signal Processing,
  Sensor Fusion, and Target Recognition III, Ivan Kadar; Vibeke Libby; Eds.,
  ed. I.~{Kadar} \& V.~{Libby}, 200--216

\bibitem[{{Molkov} {et~al.}(2003){Molkov}, {Mowlavi}, {Goldwurm}, {Strong},
  {Lund}, {Paul}, \& {Oosterbroek}}]{molkov:2003}
{Molkov}, S., {Mowlavi}, N., {Goldwurm}, A., {et~al.} 2003, The Astronomer's
  Telegram, 176, 1

\bibitem[{{Munari} \& {Tomasella}(1999)}]{munari:1999}
{Munari}, U. \& {Tomasella}, L. 1999, \aaps, 137, 521

\bibitem[{{Munari} \& {Zwitter}(1997)}]{munari:1997}
{Munari}, U. \& {Zwitter}, T. 1997, \aap, 318, 269

\bibitem[{{Negueruela}(2004)}]{negueruela:2004}
{Negueruela}, I. 2004, in Revista Mexicana de Astronomia y Astrofisica
  Conference Series, ed. G.~{Tovmassian} \& E.~{Sion}, 55--56

\bibitem[{{Negueruela} \& {Schurch}(2007)}]{negueruela:2007}
{Negueruela}, I. \& {Schurch}, M.~P.~E. 2007, \aap, 461, 631

\bibitem[{{Negueruela} {et~al.}(2006{\natexlab{a}}){Negueruela}, {Smith},
  {Harrison}, \& {Torrej{\'o}n}}]{negueruela:2006b}
{Negueruela}, I., {Smith}, D.~M., {Harrison}, T.~E., \& {Torrej{\'o}n}, J.~M.
  2006{\natexlab{a}}, \apj, 638, 982

\bibitem[{{Negueruela} {et~al.}(2006{\natexlab{b}}){Negueruela}, {Smith},
  {Reig}, {Chaty}, \& {Torrej{\'o}n}}]{negueruela:2006a}
{Negueruela}, I., {Smith}, D.~M., {Reig}, P., {Chaty}, S., \& {Torrej{\'o}n},
  J.~M. 2006{\natexlab{b}}, in ESA Special Publication, Vol. 604, ESA Special
  Publication, ed. A.~{Wilson}, 165--170

\bibitem[{{Nespoli} {et~al.}(2007){Nespoli}, {Fabregat}, \&
  {Mennickent}}]{nespoli:2007}
{Nespoli}, E., {Fabregat}, J., \& {Mennickent}, R. 2007, The Astronomer's
  Telegram, 982, 1

\bibitem[{{Patel} {et~al.}(2007){Patel}, {Zurita}, {Del Santo}, {Finger},
  {Kouveliotou}, {Eichler}, {G{\"o}{\u g}{\"u}{\c s}}, {Ubertini}, {Walter},
  {Woods}, {Wilson}, {Wachter}, \& {Bazzano}}]{patel:2007}
{Patel}, S.~K., {Zurita}, J., {Del Santo}, M., {et~al.} 2007, \apj, 657, 994

\bibitem[{{Pellizza} {et~al.}(2006){Pellizza}, {Chaty}, \&
  {Negueruela}}]{pellizza:2006}
{Pellizza}, L.~J., {Chaty}, S., \& {Negueruela}, I. 2006, \aap, 455, 653

\bibitem[{{Persson} {et~al.}(1998){Persson}, {Murphy}, {Krzeminski}, {Roth}, \&
  {Rieke}}]{persson:1998}
{Persson}, S.~E., {Murphy}, D.~C., {Krzeminski}, W., {Roth}, M., \& {Rieke},
  M.~J. 1998, \aj, 116, 2475

\bibitem[{{Prat} {et~al.}(2008){Prat}, {Rodriguez}, \&
  {Hannikainen}}]{prat:2008}
{Prat}, L., {Rodriguez}, J., \& {Hannikainen}, D.~C. 2008, ArXiv e-prints, 801

\bibitem[{{Predehl} \& {Schmitt}(1995)}]{predehl:1995}
{Predehl}, P. \& {Schmitt}, J. 1995, \aap, 293, 889

\bibitem[{{Rahoui} {et~al.}(2008){Rahoui}, {Chaty}, {Lagage}, \&
  {Pantin}}]{rahoui:2008}
{Rahoui}, F., {Chaty}, S., {Lagage}, P.-O., \& {Pantin}, E. 2008, \aap, in
  press

\bibitem[{{Revnivtsev} {et~al.}(2003{\natexlab{a}}){Revnivtsev}, {Gilfanov},
  {Churazov}, \& {Sunyaev}}]{revnivtsev:2003b}
{Revnivtsev}, M., {Gilfanov}, M., {Churazov}, E., \& {Sunyaev}, R.
  2003{\natexlab{a}}, The Astronomer's Telegram, 150, 1

\bibitem[{{Revnivtsev} {et~al.}(2003{\natexlab{b}}){Revnivtsev}, {Tuerler},
  {Del Santo}, {Westergaard}, {Gehrels}, \& {Winkler}}]{revnivtsev:2003a}
{Revnivtsev}, M., {Tuerler}, M., {Del Santo}, M., {et~al.} 2003{\natexlab{b}},
  \iaucirc, 8097, 2

\bibitem[{{Revnivtsev} {et~al.}(2004){Revnivtsev}, {Sunyaev}, {Varshalovich},
  {Zheleznyakov}, {Cherepashchuk}, {Lutovinov}, {Churazov}, {Grebenev}, \&
  {Gilfanov}}]{revnivtsev:2004}
{Revnivtsev}, M.~G., {Sunyaev}, R.~A., {Varshalovich}, D.~A., {et~al.} 2004,
  Astronomy Letters, 30, 382

\bibitem[{{Rodriguez} {et~al.}(2006){Rodriguez}, {Bodaghee}, {Kaaret},
  {Tomsick}, {Kuulkers}, {Malaguti}, {Petrucci}, {Cabanac}, {Chernyakova},
  {Corbel}, {Deluit}, {Di Cocco}, {Ebisawa}, {Goldwurm}, {Henri}, {Lebrun},
  {Paizis}, {Walter}, \& {Foschini}}]{rodriguez:2006}
{Rodriguez}, J., {Bodaghee}, A., {Kaaret}, P., {et~al.} 2006, \mnras, 366, 274

\bibitem[{{Rodriguez} {et~al.}(2005){Rodriguez}, {Cabanac}, {Hannikainen},
  {Beckmann}, {Shaw}, \& {Schultz}}]{rodriguez:2005}
{Rodriguez}, J., {Cabanac}, C., {Hannikainen}, D.~C., {et~al.} 2005, \aap, 432,
  235

\bibitem[{{Rodriguez} {et~al.}(2003){Rodriguez}, {Tomsick}, {Foschini},
  {Walter}, {Goldwurm}, {Corbel}, \& {Kaaret}}]{rodriguez:2003}
{Rodriguez}, J., {Tomsick}, J.~A., {Foschini}, L., {et~al.} 2003, \aap, 407,
  L41

\bibitem[{{Rupen} {et~al.}(2003){Rupen}, {Mioduszewski}, \&
  {Dhawan}}]{rupen:2003a}
{Rupen}, M.~P., {Mioduszewski}, A.~J., \& {Dhawan}, V. 2003, The Astronomer's
  Telegram, 152, 1

\bibitem[{{Russeil}(2003)}]{russeil:2003}
{Russeil}, D. 2003, \aap, 397, 133

\bibitem[{{Sguera} {et~al.}(2005){Sguera}, {Barlow}, {Bird}, {Clark}, {Dean},
  {Hill}, {Moran}, {Shaw}, {Willis}, {Bazzano}, {Ubertini}, \&
  {Malizia}}]{sguera:2005}
{Sguera}, V., {Barlow}, E.~J., {Bird}, A.~J., {et~al.} 2005, \aap, 444, 221

\bibitem[{{Sguera} {et~al.}(2006){Sguera}, {Bazzano}, {Bird}, {Dean},
  {Ubertini}, {Barlow}, {Bassani}, {Clark}, {Hill}, {Malizia}, {Molina}, \&
  {Stephen}}]{sguera:2006}
{Sguera}, V., {Bazzano}, A., {Bird}, A.~J., {et~al.} 2006, \apj, 646, 452

\bibitem[{{Sguera} {et~al.}(2007){Sguera}, {Hill}, {Bird}, {Dean}, {Bazzano},
  {Ubertini}, {Masetti}, {Landi}, {Malizia}, {Clark}, \&
  {Molina}}]{sguera:2007}
{Sguera}, V., {Hill}, A.~B., {Bird}, A.~J., {et~al.} 2007, \aap, 467, 249

\bibitem[{{Smith}(2004)}]{smith:2004}
{Smith}, D.~M. 2004, The Astronomer's Telegram, 338, 1

\bibitem[{{Smith} {et~al.}(2006){Smith}, {Heindl}, {Markwardt}, {Swank},
  {Negueruela}, {Harrison}, \& {Huss}}]{smith:2006}
{Smith}, D.~M., {Heindl}, W.~A., {Markwardt}, C.~B., {et~al.} 2006, \apj, 638,
  974

\bibitem[{{Stephen} {et~al.}(2005){Stephen}, {Bassani}, {Molina}, {Malizia},
  {Bazzano}, {Ubertini}, {Dean}, {Bird}, {Lebrun}, {Much}, \&
  {Walter}}]{stephen:2005}
{Stephen}, J.~B., {Bassani}, L., {Molina}, M., {et~al.} 2005, \aap, 432, L49

\bibitem[{{Sunyaev} {et~al.}(2003){Sunyaev}, {Lutovinov}, {Molkov}, \&
  {Deluit}}]{sunyaev:2003a}
{Sunyaev}, R., {Lutovinov}, A., {Molkov}, S., \& {Deluit}, S. 2003, The
  Astronomer's Telegram, 181, 1

\bibitem[{{Swank} \& {Markwardt}(2003)}]{swank:2003}
{Swank}, J.~H. \& {Markwardt}, C.~B. 2003, The Astronomer's Telegram, 128, 1

\bibitem[{{Tauris} \& {van den Heuvel}(2006)}]{tauris:2006}
{Tauris}, T.~M. \& {van den Heuvel}, E.~P.~J. 2006, {Formation and evolution of
  compact stellar X-ray sources} (Compact stellar X-ray sources), 623--665

\bibitem[{{Thompson} {et~al.}(2006){Thompson}, {Tomsick}, {Rothschild}, {in't
  Zand}, \& {Walter}}]{thompson:2006}
{Thompson}, T.~W.~J., {Tomsick}, J.~A., {Rothschild}, R.~E., {in't Zand},
  J.~J.~M., \& {Walter}, R. 2006, \apj, 649, 373

\bibitem[{{Thompson} {et~al.}(2007){Thompson}, {Tomsick}, {Zand}, {Rothschild},
  \& {Walter}}]{thompson:2007}
{Thompson}, T.~W.~J., {Tomsick}, J.~A., {Zand}, J.~J.~M.~i., {Rothschild},
  R.~E., \& {Walter}, R. 2007, \apj, 661, 447

\bibitem[{{Tomsick} {et~al.}(2006){Tomsick}, {Chaty}, {Rodriguez}, {Foschini},
  {Walter}, \& {Kaaret}}]{tomsick:2006a}
{Tomsick}, J.~A., {Chaty}, S., {Rodriguez}, J., {et~al.} 2006, \apj, 647, 1309

\bibitem[{{Tomsick} {et~al.}(2004{\natexlab{a}}){Tomsick}, {Lingenfelter},
  {Corbel}, {Goldwurm}, \& {Kaaret}}]{tomsick:2004a}
{Tomsick}, J.~A., {Lingenfelter}, R., {Corbel}, S., {Goldwurm}, A., \&
  {Kaaret}, P. 2004{\natexlab{a}}, in ESA SP-552: 5th INTEGRAL Workshop on the
  INTEGRAL Universe, ed. V.~{Schoenfelder}, G.~{Lichti}, \& C.~{Winkler},
  413--416

\bibitem[{{Tomsick} {et~al.}(2004{\natexlab{b}}){Tomsick}, {Lingenfelter},
  {Corbel}, {Goldwurm}, \& {Kaaret}}]{tomsick:2004b}
{Tomsick}, J.~A., {Lingenfelter}, R., {Corbel}, S., {Goldwurm}, A., \&
  {Kaaret}, P. 2004{\natexlab{b}}, The Astronomer's Telegram, 224, 1

\bibitem[{{Tomsick} {et~al.}(2003){Tomsick}, {Lingenfelter}, {Walter},
  {Rodriguez}, {Goldwurm}, {Corbel}, \& {Kaaret}}]{tomsick:2003}
{Tomsick}, J.~A., {Lingenfelter}, R., {Walter}, R., {et~al.} 2003, \iaucirc,
  8076, 1

\bibitem[{{T{\"u}rler} {et~al.}(2007){T{\"u}rler}, {Balman}, {Bazzano},
  {Beckmann}, {Belloni}, {Boggs}, {Capitanio}, {Chenevez}, {Del Santo},
  {Diehl}, {Donnarumma}, {Eckert}, {Goldoni}, {Gotz}, {Leyder}, {Mereghetti},
  {Paizis}, {Pottschmidt}, {Sidoli}, {Tarana}, {Tueller}, {Walter}, {Watanabe},
  \& {Weidenspointner}}]{turler:2007}
{T{\"u}rler}, M., {Balman}, S., {Bazzano}, A., {et~al.} 2007, The Astronomer's
  Telegram, 1019, 1

\bibitem[{{Walter} {et~al.}(2004{\natexlab{a}}){Walter}, {Bodaghee}, {Barlow},
  {Bird}, {Dean}, {Hill}, {Shaw}, {Bazzano}, {Ubertini}, {Bassani}, {Malizia},
  {Stephen}, {Belanger}, {Lebrun}, \& {Terrier}}]{walter:2004a}
{Walter}, R., {Bodaghee}, A., {Barlow}, E.~J., {et~al.} 2004{\natexlab{a}}, The
  Astronomer's Telegram, 229, 1

\bibitem[{{Walter} {et~al.}(2004{\natexlab{b}}){Walter}, {Courvoisier},
  {Foschini}, {Lebrun}, {Lund}, {Parmar}, {Rodriguez}, {Tomsick}, \&
  {Ubertini}}]{walter:2004b}
{Walter}, R., {Courvoisier}, T.~J.-L., {Foschini}, L., {et~al.}
  2004{\natexlab{b}}, in ESA Special Publication, Vol. 552, 5th INTEGRAL
  Workshop on the INTEGRAL Universe, ed. V.~{Schoenfelder}, G.~{Lichti}, \&
  C.~{Winkler}, 417--422

\bibitem[{{Walter} {et~al.}(2006){Walter}, {Zurita Heras}, {Bassani},
  {Bazzano}, {Bodaghee}, {Dean}, {Dubath}, {Parmar}, {Renaud}, \&
  {Ubertini}}]{walter:2006}
{Walter}, R., {Zurita Heras}, J., {Bassani}, L., {et~al.} 2006, \aap, 453, 133

\bibitem[{{Zurita Heras} {et~al.}(2006){Zurita Heras}, {de Cesare}, {Walter},
  {Bodaghee}, {B{\'e}langer}, {Courvoisier}, {Shaw}, \&
  {Stephen}}]{zurita-heras:2006}
{Zurita Heras}, J.~A., {de Cesare}, G., {Walter}, R., {et~al.} 2006, \aap, 448,
  261

\end{thebibliography}
